\documentclass[10pt,aps,showpacs,nofootinbib,prd,aps,epsf,floats,
               amsmath,amssymb,amsfonts,axodraw,floatfix,graphicx,twocolumn]{revtex4-1}
\usepackage{amsmath, amssymb}
\usepackage{paralist}
\newcommand{\mathsym}[1]{{}}

\usepackage{graphicx}
\usepackage{amsmath}
\usepackage{amssymb}
\usepackage{bm}
\newcommand{\be}{\begin{equation}}
\newcommand{\ee}{\end{equation}}
\newcommand{\bea}{\begin{eqnarray}}
\newcommand{\eea}{\end{eqnarray}}

\newsavebox{\PSLASH}
 \sbox{\PSLASH}{$p$\hspace{-1.8mm}/}
 
\renewcommand{\theequation}{\thesection.\arabic{equation}}
\newcounter{saveeqn}
\newcommand{\add}{\addtocounter{equation}{1}}
\newcommand{\alpheqn}{\setcounter{saveeqn}{\value{equation}}%
\setcounter{equation}{0}%
\renewcommand{\theequation}{\mbox{\thesection.\arabic{saveeqn}{\alph{equation}}}}}
\newcommand{\reseteqn}{\setcounter{equation}{\value{saveeqn}}%
\renewcommand{\theequation}{\thesection.\arabic{equation}}}

 \newsavebox{\notrightarrow}
 \sbox{\notrightarrow}{$\to$\hspace{-4mm}/}
 
 \newsavebox{\PARTIALSLASH}
 \sbox{\PARTIALSLASH}{$\partial$\hspace{-1.6mm}/}
 
 \newsavebox{\ASLASH}
 \sbox{\ASLASH}{$A$\hspace{-2.1mm}/}
 
 \newsavebox{\KSLASH}
 \sbox{\KSLASH}{$k$\hspace{-1.8mm}/}
 
 \newsavebox{\LSLASH}
 \sbox{\LSLASH}{$\ell$\hspace{-1.8mm}/}
 
 \newsavebox{\QSLASH}
 \sbox{\QSLASH}{$q$\hspace{-1.8mm}/}
 
 \newsavebox{\DSLASH}
 \sbox{\DSLASH}{$D$\hspace{-2.2mm}/}
 
 \newsavebox{\DbfSLASH}
 \sbox{\DbfSLASH}{${\mathbf D}$\hspace{-2.8mm}/}
 
 \newsavebox{\DELVECRIGHT}
 \sbox{\DELVECRIGHT}{$\stackrel{\rightarrow}{\partial}$}
 
 \newcommand{\blue}{\IfColor{\textCadetBlue}{}}
\newcommand{\black}{\IfColor{\textBlack}{}}
\newcommand{\red}{\IfColor{\textRed}{}}
\newcommand{\green}{\IfColor{\textOliveGreen}{}}
\newcommand{\lila}{\IfColor{\textRedViolet}{}}

\newcommand{\dprime}{\prime\prime}

\begin{document}
\title{Rotating solutions of nonideal transverse Chern-Simons magnetohydrodynamics}
\author{N. Sadooghi}\email{sadooghi@physics.sharif.ir}
\author{M. Shokri}\email{m\_shokri@physics.sharif.ir}
\affiliation{Department of Physics, Sharif University of Technology,
P.O. Box 11155-9161, Tehran, Iran}

\begin{abstract}
In order to gain deeper insight into the physics of the novel rotating solution of nonideal transverse magnetohydrodynamics (MHD), presented in one of our recent works, we replace the previously considered Maxwell theory with the ${\cal{CP}}$ violating Maxwell-Chern-Simons (MCS) theory. In this way, dissipationless chiral magnetic (CM) and anomalous Hall (AH) currents appear in the MCS equation of motion, that, together with equations of relativistic hydrodynamics, builds the set of constitutive equations of the nonideal transverse Chern-Simons magnetohydrodynamics (CSMHD). We are, in particular, interested in the effect of these currents on the evolution of electromagnetic fields in a uniformly and longitudinally expanding quark-gluon plasma with chirality imbalance. Combining the constitutive equations of CSMHD under these assumptions, we arrive, as expected, at two distinct rotating and nonrotating solutions for electromagnetic fields. The rotation occurs with increasing rapidity and a constant angular velocity $\omega_{0}$. Remarkably, the relative angle between the electric and magnetic fields, $\delta$, turns out to be given by the coefficient of AH current $\kappa_{E}$ and the electric conductivity  of the medium $\sigma$, as $\delta=\tan^{-1}(\kappa_{E}/\sigma)$. Whereas the nonrotating solution implies the AH coefficient to be vanishing, and thus nonrotating electric and magnetic fields to be either parallel or antiparallel, the relative orientation of rotating electric and magnetic fields and the evolution of the CM conductivity $\kappa_{B}$ are strongly affected by nonvanishing $\kappa_{E}$. We explore the effect of positive and negative $\omega_{0}$ on the evolution of the  CM current, and show, in particular, that a rotation of electromagnetic fields with negative $\omega_{0}$ implies a sign flip of the CM current in a chiral fluid with nonvanishing AH current.
\end{abstract}
\pacs{12.38.Mh, 25.75.-q, 47.75.+f, 47.65.-d, 52.27.Ny, 52.30.Cv }
\maketitle
\section{Introduction}\label{Introduction}\label{sec1}
Because of the formation of a plasma of deconfined quarks and gluons, modern heavy ion collision (HIC) experiments open up a unique possibility to study the topological sector of quantum chromodynamics (QCD). One of the most striking effects in this sector is the quantum chiral anomaly, arising, in particular, by an imbalance in the number of right-handed and left-handed quarks. The latter is produced through a transition between vacua of different Chern-Simons (CS) numbers, and leads to a local ${\cal{P}}$ and ${\cal{CP}}$ violation in chiral media. At high temperatures and energy densities, these transitions are mediated by unstable, spatially localized classical gauge field configurations of finite energy, called sphalerons \cite{mace2016}. Sphaleron transitions generate significant amount of axial charge by the mechanism of axial anomaly \cite{tanji2018}.
The induced axial charge asymmetry is then converted into an electric current along strong $U(1)$ magnetic fields, which are believed to be created in off-central HICs  \cite{magnetic}. This is known as the chiral magnetic effect (CME) \cite{cmeffect,magnetic}. Other effects associated with the presence of chiral fermions in a hot and dense quark matter include, among others, chiral vortical effect \cite{cveffect}, chiral separation effect \cite{cseffect} and chiral vortical separation effect \cite{cvseffect}.  Over the past few years, a growing number of theoretical studies have concentrated on various transport phenomena arisen from these effects, not only in the quark-gluon plasma (QGP) created in HICs (for a review, see \cite{cvseffect, hattori2016}), but also in relation to Weyl and Dirac semimetals (for a review see \cite{weylsemimetal}).
Experimental evidences of CME are reported  in \cite{cme-hic-exp} from HIC experiments at the Relativistic Heavy Ion Collider (RHIC) and Large Hadron collider (LHC) as well as in \cite{cme-cond-exp} from condensed matter experiments.
\par
Chiral kinetic theory \cite{stephanov2012} and anomalous (chiral) magnetohydrodynamics (MHD) \cite{son2009, tabatabaee2017,  hirono2017, sadofyev2014} are two main tools to describe the anomaly-induced transport phenomena in a chiral medium. In the present paper, we focus on chiral MHD, however, in a slightly different approach from \cite{tabatabaee2017} and \cite{hirono2017}, where, in particular, the MHD constitutive equations consist of homogeneous and inhomogeneous Maxwell equations, energy-momentum and vector current conservation laws  \textit{as well as} the axial anomaly equation. In the Chern-Simons MHD (CSMHD) setup, used in this paper, we start, in contrast, with a Lagrangian density of Maxwell-Chern-Simons (MCS) theory, also known as axion electrodynamics, which includes apart from the Maxwell  $F_{\mu\nu}F^{\mu\nu}$ term, a ${\cal{CP}}$ violating Chern-Simons (CS) $\vartheta F_{\mu\nu}\tilde{F}^{\mu\nu}$ term. Here, $\vartheta$ is an axionlike field and $\tilde{F}^{\mu\nu}\equiv \frac{1}{2}\epsilon^{\mu\nu\rho\sigma}F_{\rho\sigma}$. In this way, the CM and another anomalous currents, the AH current, appear automatically in the MCS equation of motion. As it is  shown in \cite{kharzeev-QED}, in the nonrelativistic limit, while the CM current is proportional to the magnetic field, and includes the time derivative of $\vartheta$, the AH current is perpendicular to the electric and magnetic fields, and includes the spatial gradient of $\vartheta$.
The combination of relativistic MCS equations of motion with the equation of relativistic hydrodynamics leads eventually to relativistic CSMHD. We are, in particular, interested in the effects of these currents on the evolution of the electric and magnetic fields in a chiral fluid with finite electric conductivity, that expands uniformly \cite{bjorken} in the direction transverse to electromagnetic fields (hereafter nonideal transverse CSMHD).
\par
Same assumptions were also made in our previous work \cite{shokri-1}, where we extended the method of self-similar solutions of relativistic hydrodynamics \cite{csorgo2002} to the case of nonconserved currents of a nonideal fluid, and presented two novel sets of solution to the transverse MHD \cite{rischke2015, rischke2016}. To do this, we used, as in the Bjorken solution to $1+1$ dimensional relativistic hydrodynamics \cite{bjorken}, the Milne coordinates $\tau$ and $\eta$, with the proper time $\tau\equiv \sqrt{t^{2}-z^{2}}$ and the rapidity  $\eta\equiv \frac{1}{2}\ln\left(\frac{t+z}{t-z}\right)$. We also parametrized the corresponding differential equations with two parameters $\zeta$ and $\phi$, that correspond to the angles of electric and magnetic fields with respect to a certain $x$-axis in the local rest frame (LRF) of the fluid. The solutions are shown to be characterized by parallel or antiparallel electric and magnetic fields, whose magnitudes were assumed to be boost-invariant. However, whereas electric and magnetic vectors in the first set of solutions are shown to be fixed in a $\tau$-$\eta$ plane, they rotate in the second set of solutions. The rotation occurs in the same $\tau$-$\eta$ plane with a constant angular velocity $\omega_{0}\equiv \frac{\partial\zeta}{\partial\eta}=\frac{\partial\phi}{\partial\eta}$. These two sets of solutions were referred to as nonrotating and rotating solutions of the nonideal transverse MHD. We showed that for both solutions, the relative angle between the electric and magnetic fields does not evolve with $\tau$, and  is boost invariant ($\eta$-independent). In both cases the evolution of the magnitude of the magnetic field $B$ is given by $B\propto \tau^{-1}\exp({\cal{M}})$. For the nonrotating solution, ${\cal{M}}$ arises from the solution of  $\frac{d{\cal{M}}}{du}=0$, with $u\equiv\ln\left(\tau/\tau_{0}\right)$. For the rotating solution, however, ${\cal{M}}$ satisfies a second order nonlinear differential equation, that arises by combining the constitutive equations of the transverse MHD. We set, without loss of generality, ${\cal{M}}=0$ for the nonrotating solution, and concluded that the frozen flux theorem is also valid in the nonideal transverse MHD, as in the ideal case. Solving the aforementioned nonlinear differential equation, we numerically determined the corresponding ${\cal{M}}$ to the rotating solution. Once ${\cal{M}}$ was determined, the evolution of the magnitude of the electric field $E$ and the temperature $T$ could also be separately determined in nonrotating and rotating cases. We defined  a parameter $\Omega_{0}\equiv \ell\omega_{0}$, where $\ell\equiv \pm 1$ corresponds to parallel ($\ell=+1$) and antiparallel ($\ell=-1$) electric and magnetic fields, and explored, in particular, the effect of $\Omega_{0}$ on the evolution of $B,E$ and $T$. We showed that the lifetime of the magnetic field increases with increasing negative values of $\Omega_{0}$. In \cite{shokri-1},  $\Omega_{0}$ remained as a free parameter, among other free parameters such as the finite electric conductivity of the medium.
\par
The main purpose of the present paper is to gain deeper insight into the physics of nonrotating and rotating solutions of the nonideal transverse MHD. To do this, we replace, the previously considered Maxwell theory with the ${\cal{CP}}$ violating MCS theory, and  explore the effect of aforementioned currents on the evolution of electromagnetic and hydrodynamic fields. We essentially make the same assumptions as in \cite{shokri-1}, and arrive at a number of novel results concerning the solutions of nonideal transverse CSMHD, where, in particular, anomalous CM and AH currents are created by nonvanishing temporal and spatial gradients of the axionlike $\vartheta$ field.
\par
One of the most remarkable results is that the relative angle between $\boldsymbol{E}$ and $\boldsymbol{B}$ fields, $\delta$, is given by the coefficient of the AH current $\kappa_{E}$ and the electric conductivity of the medium $\sigma$, as $\delta=\tan^{-1}\alpha_{E}$ with $\alpha_{E}\equiv\kappa_{E}/\sigma$.  For vanishing $\alpha_{E}$, we arrive at parallel or antiparallel electric and magnetic fields, as expected from \cite{shokri-1}. For nonvanishing $\tan\delta$, we show that $\kappa_{E}\propto \tau^{-1}$, and, in this way, $\delta$ evolves with the proper time $\tau$. This is in contrast to our findings in \cite{shokri-1}, where $\delta$ was a constant in $\tau$ and $\eta$. We also determine the $(\tau,\eta)$-dependence of the $\vartheta$-field, and show that it depends on the coefficients of CM and AH currents, $\kappa_{B}$ and $\kappa_{E}$.  We also show that for nonvanishing $\alpha_{E}$, in contrast to \cite{shokri-1}, where the angular velocity $\omega_{0}$ was a free parameter, it is possible to determine $\omega_{0}$ as a function of $\alpha_{0}\equiv \alpha_{E}(\tau_{0})$ and a number of other free parameters, among others, the electric conductivity of the medium, $\sigma_{0}$, and the ratio of the electric and magnetic field magnitudes at a certain initial proper time $\tau_{0}$, $\beta_{0}\equiv E_{0}/B_{0}$. As concerns the evolution of $E$ and $B$ fields, it is shown, that for $\alpha_{E}\neq 0$, the function ${\cal{M}}$ in $B\propto\tau^{-1}\exp({\cal{M}})$ can be analytically determined as a function of the same free parameters  $\sigma_{0}, \beta_{0},\omega_{0}$ and $\alpha_{0}$. In the case of vanishing $\alpha_{E}$, however, ${\cal{M}}$ is shown to arise,  from the solution of either $\frac{d{\cal{M}}}{du}=0$ (nonrotating solution) or a second order nonlinear differential equation (rotating solution), as in \cite{shokri-1}. Once ${\cal{M}}$ is determined, the evolution of $E$ can also be analytically determined in the $\alpha_{E}\neq 0$ case.
For nonvanishing $\alpha_{E}$, the evolution of the CM conductivity $\kappa_{B}$ can also be analytically determined in terms of the aforementioned set of free parameters. Being proportional to the axial chemical potential $\mu_{5}$, the evolution of $\kappa_{B}$ leads automatically to the $\tau$-dependence of $\mu_{5}$. The latter turns out to be important in relation to the production of axial charge in a hot QGP \cite{tanji2018}.
\par
All the above results show the significant role played by nonvanishing AH current in the QGP produced at the RHIC and LHC.  Let us notice that the relation of this anomalous and dissipationless current to nonlocal chiral condensates is recently demonstrated in \cite{ferrer2015, ferrer2016}. Here, it is shown that in the presence of a magnetic field, the axion electrodynamics, or equivalently the MCS theory, is realized within the dual chiral density wave phase of dense quark matter, characterized by nonlocal condensates, and that it exhibits an anomalous Hall current perpendicular to the magnetic field and an anomalous electric charge density. A large number of papers discuss the effect of this dissipationless current on the properties of Weyl and Dirac semimetals (see e.g. \cite{shovkovy2018}, for one of the most recent ones).
\par
The organization of this paper is as follows: In Sec. \ref{sec2}, we formulate the nonideal transverse CSMHD by presenting a number of definitions and useful relations as well as important properties of the additional anomalous currents induced by the ${\cal{CP}}$ violating CS term $\vartheta F_{\mu\nu}\tilde{F}^{\mu\nu}$. In Sec. \ref{sec3}, we present the constitutive equations of the transverse CSMHD, and present formal results for the evolution of electromagnetic and hydrodynamic fields. In Sec. \ref{sec4},
we summarize the above mentioned analytical results, and prove them. In Sec. \ref{sec5}, we focus on nonrotating solution of the electromagnetic fields, and show that for the electric and CM conductivities, $\sigma$ and $\kappa_{B}$, to be constant, the initial value of $\alpha_{E}$ at $\tau_{0}$ vanishes identically, and the evolution of the electric and magnetic fields are explicitly determined inter alia in terms of $\sigma$ and $\kappa_{B}$. In  Sec. \ref{sec6}, we first focus on the relation between $\omega_{0}$ and $\alpha_{0}$. The latter is related to the initial value of the angle between electric and magnetic fields, $\delta_{0}$. Choosing a number of consistent $\omega_{0}$, and determining their corresponding $\alpha_{0}$, we then plot the $\tau$-dependence of the electric, magnetic fields and the temperature as well as axial chemical potential $\mu_{5}$. The latter is known to be related to $\kappa_{B}$, whose $\tau$-dependence can be determined once $\alpha_{E}$ is nonvanishing. We show that  for positive $\omega_{0}$, $\mu_{5}$ increases during the evolution of the fluid and for negative $\omega_{0}$, $\mu_{5}$ changes its sign from positive to negative.  The latter indicates a sign flip in the current induced by the CME.  We also explore the effect of the initial electric conductivity of the medium on these properties. A number of concluding remarks is then presented in Sec. \ref{sec7}.
\section{Maxwell-Chern-Simons Theory and relativistic transverse magnetohydrodynamics}\label{sec2}
\setcounter{equation}{0}
\par\noindent
\textbf{\textit{Definitions and useful relations:}}
Let us start with the Lagrangian density of the MCS theory
\begin{eqnarray}\label{N1}
{\cal{L}}=-\frac{1}{4}F_{\mu\nu}F^{\mu\nu}-A_{\mu}J^{\mu}-\frac{c}{4}\vartheta F_{\mu\nu}\tilde{F}^{\mu\nu},
\end{eqnarray}
with $c\equiv\sum_{f}q_{f}^{2}\frac{e^{2}}{2\pi^{2}}$  and $\tilde{F}^{\mu\nu}=\frac{1}{2}\epsilon^{\mu\nu\rho\sigma}F_{\rho\sigma}$ \cite{kharzeev-QED}. Here, we assume $f=u,d$ quarks with $\left(q_{u},d_{d}\right)=\left(\frac{2}{3},-\frac{1}{3}\right)$. In this model, $\vartheta=\vartheta(t,\boldsymbol{x})$ plays the role of an axionlike field. Homogeneous and inhomogeneous MCS equations of motion are given by
\begin{eqnarray}\label{N2}
\partial_{\mu}\tilde{F}^{\mu\nu}=0,\qquad\mbox{and}\qquad
\partial_{\mu}F^{\mu\nu}={\cal{J}}^{\nu}.
\end{eqnarray}
Here, the modified current ${\cal{J}}^{\mu}$ is defined by ${\cal{J}}^{\mu}\equiv J^{\mu}-cP_{\nu}\tilde{F}^{\nu\mu}$, with $J^{\mu}$ being the electric current and $P_{\mu}\equiv \partial_{\mu}\vartheta$. The MCS energy-momentum tensor reads
\begin{eqnarray}\label{N3}
T^{\mu\nu}_{\mbox{\tiny{MCS}}}={\cal{F}}^{\mu\rho}F_{\rho}^{~\nu}+\frac{1}{4}{\cal{F}}_{\rho\sigma}F^{\rho\sigma}g^{\mu\nu},
\end{eqnarray}
with ${\cal{F}}_{\mu\nu}\equiv F_{\mu\nu}+c\vartheta\tilde{F}_{\mu\nu}$. It satisfies
\begin{eqnarray}\label{N4}
\partial_{\mu}T^{\mu\nu}_{\mbox{\tiny{MCS}}}=J_{\lambda}F^{\lambda\nu}+\frac{c}{4}P^{\nu}F_{\mu\lambda}\tilde{F}^{\mu\lambda},
\end{eqnarray}
[see Appendix \ref{appA} for the proof of (\ref{N3}) and (\ref{N4})].
In what follows, we study the effect of the additional ${\cal{CP}}$ violating term $\frac{c}{4}\vartheta F_{\mu\nu}\tilde{F}^{\mu\nu}$ in (\ref{N1}) on the evolution of electromagnetic and hydrodynamic fields in a $1+1$-dimensional relativistic fluid dynamical framework. This is characterized by a translational symmetry in a transverse $x$-$y$ plane. To do this, we use, in particular, the Bjorken flow \cite{bjorken}, characterized by the fluid four-velocity $u^{\mu}=\gamma(1,0,0,v_{z})$ with  $v_z=\frac{z}{t}$.\footnote{Here, $u^{\mu}=\frac{dx^{\mu}}{d\tau}$ satisfies $u\cdot u=1$, where, in general, $a\cdot b=a_{\mu}b^{\mu}$.} We combine the MCS equations of motion (\ref{N2}) with the corresponding conservation equations
\begin{eqnarray}\label{N5}
\hspace{-0.5cm}\partial_{\mu}T^{\mu\nu}=0,\qquad \partial_{\mu}{\cal{J}}^{\mu}=0,
\end{eqnarray}
where  $T^{\mu\nu}=T^{\mu\nu}_{\mbox{\tiny{MCS}}}+T^{\mu\nu}_{F}$ is the total energy-momentum tensor, including the fluid energy momentum tensor $T_{F}^{\mu\nu}$ and the MCS tensor $T_{\mbox{\tiny{MCS}}}^{\mu\nu}$ from (\ref{N3}).
The fluid tensor $T_{F}^{\mu\nu}$, expressed in terms of the energy density $\epsilon$, pressure $p$ and magnetization tensor $M^{\mu\nu}$ is given by
\begin{eqnarray}\label{N6}
T^{\mu\nu}_{F}=(\epsilon+p)u^{\mu}u^{\nu}-pg^{\mu\nu}-\frac{1}{2}\left(M^{\mu\lambda}F_{\lambda}^{~\nu}+M^{\nu\lambda}F_{\lambda}^{~\mu}
\right).\nonumber\\
\end{eqnarray}
Here, $g^{\mu\nu}=\mbox{diag}\left(1,-1,-1,-1\right)$ and
\begin{eqnarray}\label{N7}
M^{\mu\nu}=-\chi_{e}\left(E^{\mu}u^{\nu}-E^{\nu}u^{\mu}\right)-\chi_{m}B^{\mu\nu},
\end{eqnarray}
with constant  $\chi_{e}$ and $\chi_{m}$ the electric and magnetic susceptibilities, and $B^{\mu\nu}\equiv \epsilon^{\mu\nu\alpha\beta}B_{\alpha}u_{\beta}$. The electric and magnetic fields are defined by $E^{\mu}=F^{\mu\nu}u_{\nu}$, $B^{\mu}=\frac{1}{2}\epsilon^{\mu\nu\alpha\beta}F_{\nu\alpha}u_{\beta}$. They satisfy $E\cdot E=-E^{2}$ and  $B\cdot B=-B^{2}$. In the local rest frame (LRF) of the fluid, with $u^{\mu}=(1,\boldsymbol{0})$, we have $E^{\mu}=(0,\boldsymbol{E})$ and $B^{\mu}=(0,\boldsymbol{B})$. In terms of $E^{\mu},B^{\mu}$ and $u^{\mu}$, the antisymmetric field strength tensor $F^{\mu\nu}$ and its dual are given by
\begin{eqnarray}\label{N8}
F^{\mu\nu}&=&E^{\mu}u^{\nu}-E^{\nu}u^{\mu}-B^{\mu\nu},\nonumber\\
\tilde{F}^{\mu\nu}&=&B^{\mu}u^{\nu}-B^{\nu}u^{\mu}+E^{\mu\nu}.
\end{eqnarray}
Here, in analogy to $B^{\mu\nu}$, the antisymmetric tensor $E^{\mu\nu}$ is defined by $E^{\mu\nu}\equiv \epsilon^{\mu\nu\alpha\beta}E_{\alpha}u_{\beta}$. In a dissipative fluid with electric charge density $\rho_{e}$ and electric conductivity $\sigma$, we have
\begin{eqnarray}\label{N9}
J^{\mu}\equiv \rho_{e}u^{\mu}+\sigma E^{\mu}+\partial_{\rho}M^{\rho\mu}.
\end{eqnarray}
The modified current ${\cal{J}}^{\mu}$ appearing on the right-hand side (rhs) of (\ref{N2}) is thus given by
\begin{eqnarray}\label{N10}
{\cal{J}}^{\mu}&= &J^{\mu}-cP_{\nu}\tilde{F}^{\nu\mu},\nonumber\\
&=&\rho_{e}u^{\mu}+\sigma E^{\mu}+\chi_{e}\partial_{\nu}(E^{\mu}u^{\nu})-\chi_{m}\partial_{\nu}B^{\nu\mu}\nonumber\\
&&-c\left(P\cdot B\right)u^{\mu}+c\left(P\cdot u\right)B^{\mu}+c\epsilon^{\mu\nu\rho\sigma}P_{\nu}E_{\rho}u_{\sigma}.\nonumber\\
\end{eqnarray}
Here, the definitions of  $\tilde{F}^{\mu\nu}$ from (\ref{N8}) and $M^{\mu\nu}$ from (\ref{N7}) are used.
\\
\\
\textbf{\textit{Properties of the transverse MHD:}}
As aforementioned, the transverse MHD is mainly characterized by a translational symmetry in the transverse $x$-$y$ plane. The evolution of the fluid occurs in the longitudinal $z$-direction. Moreover, $\boldsymbol{v}\cdot \boldsymbol{E}=0$ and $\boldsymbol{v}\cdot \boldsymbol{B}=0$ are assumed. Together with $u\cdot E=0$ and $u\cdot B=0,$ that arise from the above definitions of $E^{\mu}$ and $B^{\mu}$, they lead to\footnote{For a generic four-vector $a^{\mu}$, the notation $a^{\mu}=(a^{0},a^{1},a^{2},a^{3})=(a_{0},a_x,a_y,a_z)$ is used.}
\begin{eqnarray}\label{N11}
E^{\mu}=(0,E_x,E_y,0), \qquad \mbox{and}\qquad B^{\mu}=(0,B_x,B_y,0). \nonumber\\
\end{eqnarray}
Because of the assumed translational invariance in the $x$-$y$ plane, the transverse components of $E^{\mu}$ and $B^{\mu}$ fields turn out to be independent of $x$ and $y$ variables.
Moreover, using the homogeneous and inhomogeneous Maxwell equations, and following the method also used in \cite{shokri-1} (for details, see Appendix \ref{appB}), it is easy to show that the longitudinal components of $E^{\mu}$ and $B^{\mu}$ do not evolve with $\tau$ and $\eta$,
\begin{eqnarray}\label{N12}
\frac{\partial E_{i}}{\partial\tau}=\frac{\partial E_{i}}{\partial\eta}=0,\qquad i=0,z,\nonumber\\
\frac{\partial B_{i}}{\partial\tau}=\frac{\partial B_{i}}{\partial\eta}=0,\qquad i=0,z.
\end{eqnarray}
Here, $\tau=\left(t^{2}-z^{2}\right)^{1/2}$ and $\eta=\frac{1}{2}\ln\frac{t+z}{t-z}$ are the proper time and the rapidity in the $1+1$-dimensional Milne parametrization, where in particular, $x^{\mu}$ is given by $x^{\mu}=(t,0,0,z)=(\tau\cosh\eta,0,0,\tau\sinh\eta)$. Choosing the Bjorken ansatz for $u^{\mu}$, we arrive at
\begin{eqnarray}\label{N13}
u^{\mu}=\left(\cosh\eta,0,0,\sinh\eta\right).
\end{eqnarray}
In terms of the above $\tau$ and $\eta$ parameters, the derivative $\partial_{\mu}=(\partial_t,0,0,\partial_z)$ is defined by
\begin{eqnarray}\label{N14}
\frac{\partial}{\partial t}&=&\cosh\eta\frac{\partial}{\partial\tau}-\frac{1}{\tau}\sinh\eta\frac{\partial}{\partial\eta},\nonumber\\
\frac{\partial}{\partial z}&=&-\sinh\eta\frac{\partial}{\partial\tau}+\frac{1}{\tau}\cosh\eta\frac{\partial}{\partial\eta}.
\end{eqnarray}
Using these relations, apart from $\partial\cdot E=0$ and $\partial\cdot B=0$, we have $E\cdot \partial=0$ and $B\cdot \partial=0$.
\par
In addition to the above properties, which are also discussed in \cite{shokri-1}, the transverse CSMHD is characterized by $\partial_x\vartheta=\partial_y\vartheta=0$. This is because of the assumed translational invariance in the transverse plane. We thus have
\begin{eqnarray}\label{N15}
P_{\mu}=\left(\partial_0\vartheta,0,0,\partial_z\vartheta\right).
\end{eqnarray}
\\
\textbf{\textit{Properties of anomalous terms in $\bm{\mathcal{J}}^{\boldsymbol{\mu}}$:}}
Let us first consider the inhomogeneous MCS equation of motion from (\ref{N2}). Multiplying it with $u_{\nu}$, and using the definitions of $F^{\mu\nu}$ from (\ref{N8}), $M^{\mu\nu}$ from (\ref{N7}) and ${\cal{J}}^{\mu}$ from (\ref{N10}), we arrive at
\begin{eqnarray}\label{N16}
2B\cdot \omega=\rho_{e}-c~P\cdot B,
\end{eqnarray}
where $\omega^{\mu}\equiv \frac{1}{2}\epsilon^{\mu\nu\alpha\beta}u_{\nu}\partial_{\alpha}u_{\beta}$ is the vorticity of the fluid. Bearing in mind that in transverse MHD only the longitudinal components of $u^{\mu}$ and $\partial_{\mu}$ are nonvanishing,  the vorticity of the fluid vanishes identically. We thus arrive at
\begin{eqnarray}\label{N17}
\rho_{e}=c\ P\cdot B.
\end{eqnarray}
In transverse CSMHD, this result is also consistent with the continuity equation $\partial_{\mu}{\cal{J}}^{\mu}=0$ from \eqref{N5} with ${\cal{J}}^{\mu}$ from \eqref{N10}.
Plugging $B^{\mu}$ and $P_{\mu}$ from (\ref{N11}) and (\ref{N15}) into (\ref{N17}), it turns out that the electric charge density arising from the gradient of the axionlike field $\vartheta$ vanishes. As described in \cite{kharzeev-QED}, (\ref{N17}) is a manifestation of the Witten effect, according to which, dyons are created in a system with  nonvanishing spatial gradient of $\vartheta$ \cite{witten,kharzeev-QED}. However, the fact that in the transverse MHD, $P\cdot B$ and hence $\rho_{e}$ vanish means that in a system with a translational invariance in two spatial $x$ and $y$ directions, even for nonvanishing $\partial_{3}\vartheta$, no dyons can be built.
\par
Let us now consider the second term proportional to $c$ in (\ref{N10}), $c(P\cdot u) B^{\mu}$. In the LRF of the fluid, the corresponding coefficient is given by
\begin{eqnarray}\label{N18}
c\left(P\cdot u\right)\stackrel{\mbox{\tiny{LRF}}}{=}\kappa_{B},
\end{eqnarray}
where $\kappa_{B}\equiv c\partial_{0}\vartheta=c\mu_{5}$ is the coefficient of Chiral Magnetic Effect (CME), and $\mu_{5}$ is the axial chemical potential.
Plugging, at this stage, $u^{\mu}$ from (\ref{N13}) and $P_{\mu}$ from (\ref{N15}) into (\ref{N18}), and bearing in mind that $P\cdot u$ is a Lorentz scalar, we obtain
\begin{eqnarray}\label{N19}
\kappa_{B}\equiv cP_{0}\cosh\eta+cP_{3}\sinh\eta.
\end{eqnarray}
The appearance of the CME current in ${\cal{J}}^{\mu}$ from (\ref{N10}) was previously indicated in \cite{kharzeev-QED}.
\par
Let us finally consider the third term $c\epsilon^{\mu\nu\rho\sigma}P_{\nu}E_{\rho}u_{\sigma}$ appearing on the rhs of ${\cal{J}}^{\mu}$ from (\ref{N10}). Plugging $u^{\mu}$ from (\ref{N13}) and $P_{\mu}$ from (\ref{N15}) into this expression, we obtain
\begin{eqnarray}\label{N20}
c\epsilon^{\mu\nu\rho\sigma}P_{\nu}E_{\rho}u_{\sigma}=\kappa_{E}\epsilon^{0\mu\nu 3}E_{\nu},
\end{eqnarray}
where $\kappa_{E}(\tau,\eta)$ is defined by\footnote{Later, for the sake of simplicity, we introduce a function $\alpha_{E}=\kappa_{E}/\sigma$, where $\sigma$ is the electric conductivity of the medium.}
\begin{eqnarray}\label{N21}
\kappa_{E}\equiv cP_{0}\sinh\eta+cP_{3}\cosh\eta.
\end{eqnarray}
Let us notice that in \cite{kharzeev-QED}, this term appears in the form $\boldsymbol{P}\times \boldsymbol{E}$ in the modified inhomogeneous MCS equation
$$
\boldsymbol{\nabla}\times \boldsymbol{B}-\frac{\partial \boldsymbol{E}}{\partial t}=\boldsymbol{J}+c\left(P_{0}\boldsymbol{B}-\boldsymbol{P}\times \boldsymbol{E}\right),
$$
where $P_{\mu}=\left(\partial_{0}\vartheta,\boldsymbol{\nabla}\vartheta\right)\equiv \left(P_{0},\boldsymbol{P}\right)$ is introduced. This  dissipationless anomalous Hall (AH) current is also known from \cite{ferrer2015,ferrer2016}, where its connection to topological insulators and its implications to heavy ion physics as well as neutron stars are outlined.
\par
Parametrizing, as in \cite{shokri-1}, the electric and magnetic four vectors from (\ref{N11}) in terms of the relative angles of $\boldsymbol{E}$ and $\boldsymbol{B}$  with respect to the $x$-axis in the LRF of the fluid, $\zeta$ and $\phi$, we arrive at
\begin{eqnarray}\label{N22}
E^{\mu}&=&\left(0,E\cos\zeta,E\sin\zeta,0\right),\nonumber\\
B^{\mu}&=&\left(0,B\cos\phi,B\sin\phi,0\right).
\end{eqnarray}
In Sec. \ref{sec4}, we show that $\kappa_{E}$ from (\ref{N21}) is related to $\tan\delta$, with $\delta\equiv \phi-\zeta$. Using the boost invariance ($\eta$-independence) of $\delta$, which is explicitly shown in Sec. \ref{sec4}, $\kappa_{E}$ turns out to be solely a function of $\tau$.
\par
Combining the above results, the modified current (\ref{N10}) thus reads
\begin{eqnarray}\label{N23}
{\cal{J}}^{\mu}&=&\sigma E^{\mu}+\chi_{e}\partial_{\nu}(E^{\mu}u^{\nu})-\chi_{m}\partial_{\nu}B^{\nu\mu}+\kappa_{B}B^{\mu}\nonumber\\
&&+\kappa_E
\epsilon^{0\mu\nu 3}E_{\nu}.
\end{eqnarray}
Let us notice, at this stage, that, according to definitions (\ref{N19}) and (\ref{N21}), $\kappa_{B}$ and $\kappa_{E}$ turn out to be the Lorentz boost transformed of $cP_{0}$ and $cP_{3}$ from the LRF of the fluid. In Sec. \ref{sec4}, we use the inverse transformation
\begin{eqnarray}\label{N24}
cP_{0}&=&+\kappa_{B}\cosh\eta-\kappa_{E}\sinh\eta,\nonumber\\
cP_{3}&=&-\kappa_{B}\sinh\eta+\kappa_{E}\cosh\eta,
\end{eqnarray}
and determine the evolution of the axionlike field $\vartheta$ as a function of $\tau$ and $\eta$.
\section{Constitutive equations of the CSMHD in $\boldsymbol{1+1}$ dimensions}\label{sec3}
\setcounter{equation}{0}
The constitutive equations of CSMHD include the homogeneous and inhomogeneous Maxwell equations from (\ref{N2}), the Euler equation arising from $\Delta_{\mu\nu}\partial_{\rho}T^{\rho\nu}=0$
with $T^{\mu\nu}$ the total energy-momentum tensor, and the equation arising from  $\Delta_{\mu\nu}\partial_{\rho}T^{\rho\nu}_{F}=-\Delta_{\mu\nu}
\left(J_{\lambda}F^{\lambda\nu}+\frac{c}{4}P^{\nu}F_{\rho\sigma}F^{\rho\sigma}\right)$. Here, $\Delta^{\mu\nu}\equiv g^{\mu\nu}-u^{\mu}u^{\nu}$. These equations and a number of other useful relations are presented in this section.
\par
Plugging the definitions of $\tilde{F}^{\mu\nu}$ from (\ref{N8}) into the homogeneous Maxwell equation (\ref{N2}), and combining the resulting expressions for $\nu=1$ and $\nu=2$ components of $\partial_{\mu}\tilde{F}^{\mu\nu}=0$, we arrive after some algebra at
\begin{eqnarray}\label{E1}
\partial_{\mu}(Bu^{\mu})-\frac{E}{\tau}\cos\delta\frac{\partial\zeta}{\partial\eta}=0,
\end{eqnarray}
and
\begin{eqnarray}\label{E2}
B\frac{\partial\phi}{\partial\tau}+\frac{E}{\tau}\sin\delta\frac{\partial\zeta}{\partial\eta}=0.
\end{eqnarray}
Here, the derivatives defined in (\ref{N14}) and the parametrization of $E^{\mu}$ and $B^{\mu}$ in terms of $\zeta$ and $\phi$ from (\ref{N22}) are used. Following same steps, the equations arising from the combination of $\nu=1$ and $\nu=2$ components of the inhomogeneous Maxwell equation $\partial_{\mu}F^{\mu\nu}={\cal{J}}^{\nu}$ with $F^{\mu\nu}$ from (\ref{N8}) and ${\cal{J}}^{\nu}$ from (\ref{N23}) reads
 \begin{eqnarray}
&&(1+\chi_{e})E\frac{\partial\zeta}{\partial\tau}+(1-\chi_{m})\sin\delta\frac{B}{\tau}\frac{\partial\phi}{\partial\eta}+\kappa_{E} E
\nonumber\\
&&~~+\kappa_{B}B\sin\delta=0,\label{E3}\\
&&(1+\chi_{e})\partial_{\mu}(Eu^{\mu})+(1-\chi_{m})\cos\delta\frac{B}{\tau}\frac{\partial\phi}{\partial\eta}+\sigma E\nonumber\\
&&~~+\kappa_{B}B\cos\delta=0,\label{E4}
\end{eqnarray}
where $\kappa_{B}$ and $\kappa_{E}$ are defined in (\ref{N19}) and (\ref{N21}). Let us notice that the additional terms including these two coefficients in (\ref{E3}) and (\ref{E4}), arise from additional terms of the modified MCS current ${\cal{J}}^{\mu}$ proportional to $c$, and are absent in a fluid with no chirality imbalance.
\par
In contrast to the above inhomogeneous MCS equations, the Euler equation arising from $\Delta_{\mu\nu}\partial_{\rho}T^{\rho\nu}=0$ does not receive any additional term proportional to $c$. It reads
\begin{eqnarray}\label{E5}
\hspace{-0.5cm}Du_{\mu}=\frac{\nabla_{\mu}p_{tot}-\chi {\cal{C}}_{\mu}}{\epsilon+p+\left(1-\chi_{m}\right)B^{2}+\left(1-\chi_{e}\right)E^{2}}.
\end{eqnarray}
Here, $D\equiv u^{\mu}\partial_{\mu}$ and $\nabla_{\mu}\equiv \Delta_{\mu\nu}\partial^{\nu}$. Moreover, the total pressure $p_{tot}$ and ${\cal{C}}_{\mu}$ are defined by
\begin{eqnarray}\label{E6}
p_{tot}\equiv p-\chi_{m}B^{2}+\frac{1}{2}\left(E^{2}+B^{2}\right),
\end{eqnarray}
and
\begin{eqnarray}\label{E7}
\hspace{-1cm}{\cal{C}}_{\mu}\equiv E^{\lambda}B_{\lambda\rho}\partial^{\rho}u_{\mu}+\theta E^{\lambda}B_{\lambda\mu}+\Delta_{\mu\nu}
D\left(E_{\lambda}B^{\lambda\nu}\right),
\end{eqnarray}
with $\theta=\partial_{\mu}u^{\mu}$. The coefficient $\chi$, appearing in (\ref{E5}), is defined by
$\chi\equiv \frac{1}{2}\big[\left(1+\chi_{e}\right)+\left(1-\chi_{m}\right)\big]$. For a uniformly expanding fluid with $Du_{\mu}=0$, (\ref{E5}) leads to $\nabla_{\mu}p_{tot}=\chi{\cal{C}}_{\mu}$. Bearing in mind that in the Milne coordinates, we have $\theta=\frac{1}{\tau}$, $D=\frac{\partial}{\partial\tau}$, $\nabla_{\mu}=-\frac{1}{\tau}\left(\sinh\eta,0,0,-\cosh\eta\right)\frac{\partial}{\partial\eta}$, the Euler equation (\ref{E5}) is given by
\begin{eqnarray}\label{E8}
\frac{1}{\tau}\frac{\partial p_{tot}}{\partial\eta}=-\chi\left(\frac{\partial}{\partial\tau}+\frac{2}{\tau}\right)\left(EB\sin\delta\right).
\end{eqnarray}
Assuming, as in \cite{shokri-1}, $p,E$ and $B$ to be $\eta$-independent, we arrive for $\chi\neq 0$ at ${\cal{C}}_{\mu}=0$. This leads to
\begin{eqnarray}\label{E9}
\left(\frac{\partial}{\partial\tau}+\frac{2}{\tau}\right)\left(EB\sin\delta\right)=0.
\end{eqnarray}
In contrast to \cite{shokri-1}, where the combination of constituent equations of MHD led to $\sin\delta=0$, for nonvanishing $\kappa_{E}$, $\sin\delta\neq 0$ turns out to be also possible. In this case (\ref{E9}) leads to an additional equation apart from (\ref{E1})-(\ref{E4}) and (\ref{E10}), that determines the evolution of electromagnetic  and thermodynamic fields as well as $\kappa_{B}$ and $\kappa_{E}$ [see Sec. \ref{sec4}].
\par
Using $Du_{\mu}=0$ and ${\cal{C}}_{\mu}=0$, and combining the expressions arising from $\mu=0$ and $\mu=3$ components of $\Delta_{\mu\nu}\partial_{\rho}T^{\rho\nu}_{F}=-\Delta_{\mu\nu}
\left(J_{\lambda}F^{\lambda\nu}+\frac{c}{4}P^{\nu}F_{\rho\sigma}F^{\rho\sigma}\right)$ with $T_{F}^{\mu\nu}$ given in (\ref{N6}), we arrive at
\begin{eqnarray}\label{E10}
\hspace{-1.2cm}\big[\sigma E+\chi_{e}\partial_{\mu}\left(Eu^{\mu}\right)\big]\tan\delta-E\chi_{e}\frac{\partial\zeta}{\partial\tau}=\alpha_{E}\sigma E.
\end{eqnarray}
Here, $\cos\delta\neq 0$ is used.\footnote{In the next section, we show that for $\sigma E\neq 0$, $\cos\delta$ is nonvanishing.} For the evolution of the temperature, we shall also  evaluate $u_{\nu}\partial_{\mu}T^{\mu\nu}_{F}=-u_{\nu}
\left(J_{\mu}F^{\mu\nu}+\frac{c}{4}P^{\nu}F_{\rho\sigma}F^{\rho\sigma}\right)$, which, upon using (\ref{N18}), yields
\begin{eqnarray}\label{E11}
\lefteqn{\hspace{-0.5cm}D\epsilon+\chi_{e}EDE+\theta\left(\epsilon+p-\chi_{m}B^{2}\right)+\frac{1}{2}(\chi_{e}-\chi_{m})
}\nonumber\\
&&\times\frac{EB}{\tau}\cos\delta\frac{\partial\delta}{\partial\eta}
=\sigma E^{2}-\chi_{m}\left(\frac{EB}{\tau}\cos\delta\frac{\partial\phi}{\partial\eta}\right)\nonumber\\
&&+\kappa_{B}EB\cos\delta.
\end{eqnarray}
Similar to the case of nonideal transverse MHD with no chirality imbalance, previously discussed in \cite{shokri-1}, the dynamics of nonideal CSMHD in $1+1$ dimensions is governed by a number of inhomogeneous differential equations
\begin{eqnarray}\label{E12}
&&\partial_{\mu}\left(T^{\kappa}u^{\mu}\right)=T^{\kappa}D{\cal{L}},\qquad \partial_{\mu}\left(Bu^{\mu}\right)=BD{\cal{M}},\nonumber\\
&&\partial_{\mu}\left(Eu^{\mu}\right)=ED{\cal{N}},
\end{eqnarray}
whose formal solutions are given by
\begin{eqnarray}\label{E13}
&&T=T_{0}\left(\frac{\tau_{0}}{\tau}\right)^{\frac{1}{\kappa}}e^{\frac{\cal{L}}{\kappa}},\qquad B=B_0 \left(\frac{\tau_{0}}{\tau}\right) e^{\cal{M}},\nonumber\\
&& E=E_{0}\left(\frac{\tau_{0}}{\tau}\right)e^{\cal{N}},
\end{eqnarray}
respectively. The aim is to use the constitutive equations, presented above, to determine the unknown functions ${\cal{L}},{\cal{M}}$ and ${\cal{N}}$. Another useful relation, which plays an essential role in determining the evolution of electromagnetic and thermodynamical fields in the case of nonvanishing AH coefficient, $\kappa_{E}$, arises by combining (\ref{E9}) with the formal solutions for $E$ and $B$ from (\ref{E13}),
\begin{eqnarray}\label{E14}
\frac{\partial\delta}{\partial\tau}\cos\delta+\left(\frac{d{\cal{M}}}{d\tau}+\frac{d{\cal{N}}}{d\tau}\right)\sin\delta=0.
\end{eqnarray}
For $\cos\delta\neq 0$, (\ref{E14}) turns out to be
\begin{eqnarray}\label{E15}
\frac{\partial\delta}{\partial\tau}=-\left(\frac{d{\cal{M}}}{d\tau}+\frac{d{\cal{N}}}{d\tau}\right)\tan\delta.
\end{eqnarray}
In the next section, we combine the constitutive equations (\ref{E1})-(\ref{E4}), (\ref{E10}) and (\ref{E15}), and determine the $\left(\tau,\eta\right)$-dependence of $\phi$ and $\zeta$, as well as the $\tau$-dependence of $E,B$ and $T$. To do this, we assume, as in \cite{shokri-1}, the  equation of state $\epsilon=\kappa p$, where $\kappa$ is related to the sound velocity $c_s$ in the fluid as $\kappa^{-1}=c_{s}^{2}=1/3$. Moreover, we set $p=nT$, where $n$ is the baryon number density, whose evolution is described by the conservation law
\begin{eqnarray}\label{E16}
\partial_{\mu}(nu^{\mu})=0.
\end{eqnarray}
This leads to a simple Bjorken scaling solution for $n$ in transverse MHD
\begin{eqnarray}\label{E17}
n(\tau)=n_{0}\left(\frac{\tau_{0}}{\tau}\right).
\end{eqnarray}
These kinds of assumptions are also made originally by Bjorken in order to present the most simple analytical solution for transverse hydrodynamics \cite{bjorken}. Taking the equation of state of an ultrarelativistic ideal gas $\epsilon =\kappa p$ with $\kappa=3$ is motivated by the fact that at high temperature $T\gg T_{c}$, the trace anomaly $\epsilon-3 p$ approximately vanishes (see e.g. the results arising from lattice QCD in \cite{bazanov2014}).\footnote{Equivalently, $c_{s}^{2}=\frac{dp}{d\epsilon}\approx 1/3$ is found in lattice QCD for $T\gg T_{c}$ (see e.g. Table 1 in \cite{bazanov2014}).} In the present work, we neglect, for simplicity, the effect of electric and magnetic susceptibilities on the pressure $p$ and energy density $\epsilon$, and use the same ideal gas equation of state $\epsilon=3 p$ as in \cite{ csorgo2002,shokri-1}, where extensions of Bjorken's solutions are presented. To make an analytical treatment possible, it is enough to have $\kappa$ = constant \cite{shokri-1}.\par
Assuming, apart from $\epsilon=3p$, the following empirical $\tau$-dependence for the electric conductivity $\sigma$,
\begin{eqnarray}\label{E18}
\sigma=\sigma_{0}\left(\frac{\tau_{0}}{\tau}\right)^{1/\kappa},
\end{eqnarray}
and combining the definitions of $\kappa_{B}$ and $\kappa_{E}$ from (\ref{N19}) and (\ref{N21}), we also determine the evolution of the axionlike field $\vartheta$.  For the case of nonvanishing $\sin\delta$, we also arrive at the $\tau$-dependence of $\kappa_{B}$ and $\kappa_{E}$. Let us notice that in order to write (\ref{E18}), we were inspired by the temperature dependence of the electric conductivity $\sigma$, which is computed in lattice QCD \cite{latticeconductivity} (see also \cite{mclerran2013}),\footnote{The determination of the $T$ dependence of the electric conductivity $\sigma$ is beyond the scope of the present paper. The most recent results for $\sigma(T)$ is presented in \cite{electricconduct}.}
\begin{eqnarray}\label{E19}
\sigma=\sigma_{c}\frac{T}{T_{c}}, ~~ \mbox{with}~~ \sigma_{c}=5.8\pm 2.9 \mbox{MeVc},
\end{eqnarray}
where $T_{c}$ is the critical temperature of the QCD phase transition.
Plugging the evolution of the temperature $T$ from (\ref{E13}) into (\ref{E19}), and neglecting $e^{\frac{{\cal{L}}}{\kappa}}$, we arrive at (\ref{E18}), with $\sigma_{0}$ defined by
\begin{eqnarray}\label{E20}
\sigma_{0}\equiv \sigma_{c}\frac{T_{0}}{T_{c}}.
\end{eqnarray}
Here, $T_{0}=T(\tau_{0})$ is the initial temperature at $\tau=\tau_{0}$. Here, as in the assumption concerning $\kappa$, we neglect, in the first approximation, the effect of electric and magnetic susceptibilities on the electric conductivity $\sigma$.
\section{The evolution in a uniformly expanding magnetized fluid with chirality imbalance}\label{sec4}
\setcounter{equation}{0}
Before presenting the $(\tau,\eta)$ dependence of $\zeta,\phi,E,B$ and $T$ as well as $\vartheta$, $\kappa_{E}$ and $\kappa_{B}$, let us emphasize that in the present paper, as in our previous work \cite{shokri-1}, our arguments are based on three main assumptions:
\begin{enumerate}
\setlength\itemsep{-0.05cm}
\item The system is translational invariant in the transverse $x$-$y$ plane, i.e. no quantity depends on $x$ and $y$ coordinates.
\item The system evolves uniformly, i.e. $Du_{\mu}=0$, $\forall t$.
\item The pressure $p$ and the magnitude of the electric and magnetic fields $E=|\boldsymbol{E}|$ and $B=|\boldsymbol{B}|$ are boost invariant, i.e.
$\frac{\partial p}{\partial\eta}=0, \frac{\partial E}{\partial\eta}=0$ and $\frac{\partial B}{\partial\eta}=0$.
\end{enumerate}
\subsection{Summary of results}\noindent
\textbf{\textit{i) Relative angle between $\boldsymbol{E}$ and $\boldsymbol{B}$}}\\
\textit{i.a)} Using the aforementioned constitutive equations, we show that in nonideal transverse MHD with nonvanishing $\sigma E$, the electric and magnetic fields cannot be perpendicular to each other, i.e. $\cos\delta\neq 0$.
\vspace{0.1cm}\par\noindent
\textit{i.b)} The combination of constitutive equations leads to
\begin{eqnarray}\label{S1}
\tan\delta=\alpha_{E},
\end{eqnarray}
where $\alpha_{E}=\kappa_{E}/\sigma$. According to (\ref{N21}), $\kappa_{E}$ is given as a linear combination of temporal and spatial gradients of the $\vartheta$-vacuum, $P_{0}$ and $P_{3}$, and vanishes in a fluid with no chirality imbalance. Hence, in a chiral fluid, within the aforementioned Bjorkenian
framework, the relative angle between the electric and magnetic fields, $\delta$, is solely determined by $\alpha_{E}$, and is thus related to the AH coefficient $\kappa_{E}$ and the electric conductivity of the chiral fluid $\sigma$. This generalizes the results from our previous work \cite{shokri-1}, where it was shown that in a nonchiral magnetized fluid, the electric and magnetic fields are either parallel or antiparallel.
\vspace{0.1cm}\par\noindent
\textit{i.c)} It turns out that the relative angle between $\boldsymbol{E}$ and $\boldsymbol{B}$ fields is boost invariant, i.e. $\frac{\partial\delta}{\partial\eta}=0$.  This leads immediately to the boost invariance of $\alpha_{E}$ through (\ref{S1}). We show that in a chiral fluid, the case of $\alpha_{E}=0$ is not generally excluded.
\\
In what follows, two cases of $\alpha_{E}\neq 0$ and $\alpha_{E}=0$ are separately considered.
\vspace{0.2cm}\par\noindent\textbf{\textit{ii) Evolution of $\boldsymbol{\kappa_{E}}$  and $\boldsymbol{\delta}$}}\\
\textit{ii.a)} In the case of $\tan\delta\neq 0$, we use (\ref{N24}) and the definitions of $cP_{0}$ and $cP_{3}$ in terms of the temporal and spatial derivatives of $\vartheta$ in the Milne parametrization (\ref{N14}), and arrive at the evolution of $\kappa_{E}$,
\begin{eqnarray}\label{S2}
\kappa_{E}(\tau)=\kappa_{E}^{(0)}\left(\frac{\tau_{0}}{\tau}\right),
\end{eqnarray}
where $\kappa_{E}^{(0)}\equiv \kappa_{E}(\tau_{0})$. Using (\ref{E18}), the evolution of $\alpha_{E}=\kappa_{E}/\sigma$ is thus given by
\begin{eqnarray}\label{S3}
\alpha_{E}=\alpha_{0}\left(\frac{\tau_{0}}{\tau}\right)^{1-\frac{1}{\kappa}}.
\end{eqnarray}
This fixes the evolution of the relative angle between the electric and magnetic fields as \begin{eqnarray}\label{S4}
\delta(\tau)=\tan^{-1}\left(\alpha_{0}\left(\frac{\tau_{0}}{\tau}\right)^{1-\frac{1}{\kappa}}\right).
\end{eqnarray}
\par\noindent
\textit{ii.b)} For $\tan\delta=0$, the electric and magnetic fields turn out to be either parallel or antiparallel.
\vspace{0.2cm}\par\noindent\textbf{\textit{iii) Evolution of the $\boldsymbol{\vartheta}$-vacuum}}\\
\textit{iii.a)} For $\tan\delta\neq 0$, the evolution of the axionlike field $\vartheta$ is given by
\begin{eqnarray}\label{S5}
\hspace{-0.5cm}\vartheta\left(\tau,\eta\right)=\frac{\tau_{0}\kappa_{E}^{(0)}}{c}\eta+\frac{1}{c}\int_{\tau_{0}}^{\tau}\kappa_{B}(\tau^{\prime})d\tau^{\prime}+\vartheta_{0}.
\end{eqnarray}
Here, $\kappa_{E}^{(0)}=\sigma_{0}\alpha_{0}$ and $\vartheta_{0}\equiv \vartheta(\tau_{0})$.
\par\noindent
\textit{iii.b)} For $\tan\delta=0$, the axionlike field $\vartheta$ is $\eta$-independent. For $\kappa_{B}=\mbox{const.}$, its evolution is simply given by
\begin{eqnarray}\label{S6}
\vartheta(\tau)=\frac{\kappa_{B}}{c}(\tau-\tau_0)+\vartheta_{0},
\end{eqnarray}
with $\vartheta_{0}=\mbox{const.}$
\vspace{0.2cm}\par\noindent\textbf{\textit{iv) Evolution of the angles $\boldsymbol{\zeta}$  and $\boldsymbol{\phi}$}}\\
\textit{iv.a)} The evolution of the angles $\zeta$ and $\phi$, appearing in (\ref{N22}) is  given by
\begin{eqnarray}\label{S7}
\hspace{-0.8cm}\zeta(\tau,\eta)&=&\omega_{0}\eta+\phi_{0}-\alpha_{0}\tau_{0}^{1-\frac{1}{\kappa}}\int_{\tau_{0}}^{\tau}
\frac{d\tau^{\prime}}{\tau^{\prime~1-\frac{1}{\kappa}}}
\frac{d{\cal{M}}}{d\tau^{\prime}}\nonumber\\
&&-\mbox{tan}^{-1}\left(\alpha_{E}(\tau)\right),
\nonumber\\
\hspace{-0.8cm}\phi(\tau,\eta)&=&\omega_{0}\eta+\phi_{0}-\alpha_{0}\tau_{0}^{1-\frac{1}{\kappa}}\int_{\tau_{0}}^{\tau}
\frac{d\tau^{\prime}}{\tau^{\prime~1-\frac{1}{\kappa}}}
\frac{d{\cal{M}}}{d\tau^{\prime}}.
\end{eqnarray}
Here, for $\tau$-dependent $\sigma$, $\alpha_{E}$ is given in (\ref{S3}), and ${\cal{M}}$, appearing in (\ref{E12}) and (\ref{E13}), describes the deviation from the frozen flux theorem in a nonideal fluid.
\par\noindent
\textit{iv.b)} For $\tan\delta=0$, the evolution of the angles $\zeta$ and $\phi$ is given by
\begin{eqnarray}\label{S8}
\zeta(\tau,\eta)&=&\omega_{0}\eta+\zeta_{0},\nonumber\\
\phi(\tau,\eta)&=&\omega_{0}\eta+\phi_{0}.
\end{eqnarray}
Same relations arise also in \cite{shokri-1}, where in the absence of any chirality imbalance $\tan\delta=0$.
\vspace{0.2cm}\par\noindent\textbf{\textit{v) Determination of $\boldsymbol{\alpha_{0}}$}}\\
\textit{v.a)} Evaluating (\ref{E4}) at $u=0$, and using (\ref{E15}), we show that for $\tan\delta\neq 0$,  $\alpha_{0}$ from (\ref{S3}) is given by
\begin{eqnarray}\label{S9}
\hspace{-0.5cm}\alpha_{0}=
\pm\left(-1+4\left({\cal{C}}_{1}\pm\sqrt{{\cal{C}}_{1}^{2}
-4{\cal{C}}_{2}}\right)^{-2}\right)^{1/2},
\end{eqnarray}
where
\begin{eqnarray}\label{S10}
{\cal{C}}_{1}\equiv\bigg\{\beta_{0}\omega_{0}-\frac{[\left(1-\chi_{m}\right)\omega_{0}+
\kappa_{B}^{(0)}\tau_{0}]}{\beta_{0}\left(1+\chi_{e}\right)}\bigg\}\left(\frac{\kappa}{\kappa-1}\right),\nonumber\\
\end{eqnarray}
and
\begin{eqnarray}\label{S11}
{\cal{C}}_{2}\equiv \left(\frac{\sigma_{0}\tau_{0}}{1+\chi_{e}}\right)\left(\frac{\kappa}{\kappa-1}\right).
\end{eqnarray}
\textit{v.b)} In the case of $\tan\delta=0$, $\alpha_{0}$ vanishes identically.
\vspace{0.2cm}\par\noindent \textbf{\textit{vi) Evolution of $\bm{\mathcal{M}}$, $\bm{\mathcal{N}}$ and $\boldsymbol{\kappa}_{\boldsymbol{B}}$}}\\
\textit{vi.a)} For $\tan\delta\neq 0$, it is possible to determine ${\cal{M}}$ and ${\cal{N}}$ analytically. They are given by
\begin{eqnarray}\label{S12}
{\cal{M}}=\frac{1}{2}\ln\left\{1-\frac{2\kappa\beta_{0}\omega_{0}\cos\delta_{0}}{\kappa-1}\bigg[1-\left(\frac{\tau_{0}}{\tau}\right)^{-\left(1-\frac{1}{\kappa}\right)}\bigg]\right\},\nonumber\\
\end{eqnarray}
and
\begin{eqnarray}\label{S13}
{\cal{N}}&=&
\frac{1}{2}\ln\left\{\cos^{2}\delta_{0}\bigg[\tan^{2}\delta_{0}+\left(\frac{\tau_{0}}{\tau}\right)^{-2\left(1-\frac{1}{\kappa}\right)}\bigg]\right\}
\nonumber\\
&&-\frac{1}{2}\ln\left\{1-\frac{2\kappa\beta_{0}\omega_{0}\cos\delta_{0}}{\kappa-1}\bigg[1-\left(\frac{\tau_{0}}{\tau}\right)^{-\left(1-\frac{1}{\kappa}\right)}\bigg]\right\},\nonumber\\
\end{eqnarray}
with $\beta_{0}\equiv \frac{E_{0}}{B_{0}}$ and $\cos\delta_{0}=\ell\left(1+\alpha_{0}^{2}\right)^{-1/2}$. Here, $\ell\equiv\pm 1$ and $\alpha_{0}$ is given in (\ref{S9}). Plugging ${\cal{M}}$ and ${\cal{N}}$ from (\ref{S12}) and (\ref{S13}) into formal solutions for $B$ and $E$ from (\ref{E13}), the evolution of these fields is completely determined in terms of free parameters $\kappa,\tau_{0},\beta_{0},\sigma_{0},\omega_{0}$ and $\kappa_{B}^{(0)}$ as well as $\chi_{e}$ and $\chi_{m}$.
\par
For $\tan\delta\neq 0$, we also have the possibility to determine the evolution of $\kappa_{B}$ using (\ref{E4}). It is given by
\begin{eqnarray}\label{S14}
\lefteqn{\hspace{-0.5cm}\kappa_{B}(u)=-\frac{1}{\beta_{0}\tau_{0}\cos\delta}\bigg[\left(1+\chi_{e}\right) \beta_{0}^{2}e^{{\cal{N}}-{\cal{M}}-u}\frac{d{\cal{N}}}{du}
}\nonumber\\
&&\hspace{-0.3cm}+\left(1-\chi_{m}\right)e^{{\cal{M}}-{\cal{N}}-u}\frac{d{\cal{M}}}{du}+\sigma\tau_{0}\beta_{0}^{2}e^{{\cal{N}}-{\cal{M}}}
\bigg],
\end{eqnarray}
with $u\equiv\ln\left(\frac{\tau}{\tau_{0}}\right)$. Plugging the corresponding expressions for ${\cal{M}}$ and ${\cal{N}}$ from (\ref{S12}) and (\ref{S13}) into (\ref{S14}) and using the time dependence of $\sigma$ from (\ref{E18}) as well as $\alpha_{E}$ from (\ref{S3}), the $\tau$-dependence of $\kappa_{B}$ is completely determined in terms of the aforementioned free parameters.
\par\noindent
\textit{vi.b)} For $\tan\delta=0$, we show that ${\cal{M}}$ either satisfies
\begin{eqnarray}\label{S15}
\frac{d{\cal{M}}}{du}=0,
\end{eqnarray}
or a second order nonlinear differential equation
\begin{eqnarray}\label{S16}
\lefteqn{\hspace{-2cm}\frac{d^{2}{\cal{M}}}{du^{2}}+\frac{d{\cal{M}}}{du}\bigg[\frac{d{\cal{M}}}{du}+\frac{\sigma\tau_{0}e^{u}}{(1+\chi_{e})}\bigg]+
\frac{\omega_{0}^{2}}{(1+\chi_{e})}
}\nonumber\\
&&\times\left(1-\chi_{m}+\frac{\kappa_{B}\tau_{0}e^{u}}{\omega_{0}}\right)
=0.
\end{eqnarray}
In contrast to the case of $\tan\delta\neq 0$, here $\kappa_{B}$ is assumed to be constant and part of the initial condition. As described in \cite{shokri-1}, (\ref{S15})  corresponds to $\omega_{0}=0$, which characterizes nonrotating $\boldsymbol{E}$ and $\boldsymbol{B}$ vectors. Moreover, for $\omega_{0}=0$, we have ${\cal{M}}=0$. This leads, according to (\ref{E13}), to $B=B_{0}\left(\frac{\tau_{0}}{\tau}\right)$, even in the nonideal fluid. For $\omega_{0}\neq 0$, which corresponds to rotating $\boldsymbol{E}$ and $\boldsymbol{B}$ vectors, the solution of (\ref{S16}) leads to nonvanishing ${\cal{M}}$, which describes a deviation from the frozen flux theorem.
\par
Once ${\cal{M}}$ is determined, ${\cal{N}}$ can also be given by
\begin{eqnarray}\label{S17}
e^{{\cal{N}}}=\frac{e^{{\cal{M}}}}{\beta_{0}\ell\omega_{0}}\frac{d{\cal{M}}}{du}.
\end{eqnarray}
\vspace{0.2cm}\par\noindent\textbf{\textit{vii) Evolution of $\bm{\mathcal{L}}$}}\\
\textit{vii.a)} For $\tan\delta\neq 0$, using (\ref{E11}), $\epsilon=\kappa p$ as well as $p=nT$, we show that $e^{{\cal{L}}/\kappa}$, appearing in the formal solution of $T(\tau)$ from (\ref{E13}), reads
\begin{widetext}
\begin{eqnarray}\label{S18}
e^{\frac{{\cal{L}}}{\kappa}}&=&1+\frac{E_{0}^{2}}{\epsilon_{0}}\int_{\tau_{0}}^{\tau}d\tau^{\prime}\sigma(\tau^{\prime})\left(\frac{\tau_{0}}{\tau^{\prime}}\right)^{1-\frac{1}{\kappa}}e^{2{\cal{N}}}+
\frac{E_{0}B_{0}}{\epsilon_{0}}\int_{\tau_{0}}^{\tau}d\tau^{\prime}\left(\frac{\tau_{0}}{\tau^{\prime}}\right)^{1-\frac{1}{\kappa}}e^{{\cal{N}}+{\cal{M}}}\kappa_{B}(\tau')\cos\delta(\tau')\nonumber\\
&&+\frac{\chi_{m} B_{0}^{2}}{\tau_{0}\epsilon_{0}}\int_{\tau_{0}}^{\tau}d\tau^{\prime}\left(\frac{\tau_{0}}{\tau^{\prime}}\right)^{2-\frac{1}{\kappa}}e^{2{\cal{M}}}+
\frac{\chi_{e} E_{0}^{2}}{\tau_{0}\epsilon_{0}}\int_{\tau_{0}}^{\tau}d\tau^{\prime}\left(\frac{\tau_{0}}{\tau^{\prime}}\right)^{2-\frac{1}{\kappa}}e^{2{\cal{N}}}\nonumber\\
&&-
\frac{\chi_{m} B_{0}^{2}}{\epsilon_{0}}\int_{\tau_{0}}^{\tau}d\tau^{\prime}\left(\frac{\tau_{0}}{\tau^{\prime}}\right)^{1-\frac{1}{\kappa}}e^{2{\cal{M}}}\frac{d{\cal{M}}}{d\tau^{\prime}}-
\frac{\chi_{e} E_{0}^{2}}{\epsilon_{0}}\int_{\tau_{0}}^{\tau}d\tau^{\prime}\left(\frac{\tau_{0}}{\tau^{\prime}}\right)^{1-\frac{1}{\kappa}}e^{2{\cal{N}}}\frac{d{\cal{N}}}{d\tau^{\prime}}.
\end{eqnarray}
\end{widetext}
Here, $\sigma(\tau), {\cal{M}}(\tau), {\cal{N}}(\tau),$ and $\kappa_{B}(\tau)$ are given in (\ref{S13}), (\ref{S12}),  (\ref{E18}) and (\ref{S14}), respectively. \\
\textit{vii.b)} For $\tan\delta=0$, ${\cal{L}}$ is determined by choosing a constant $\kappa_{B}$, and plugging ${\cal{M}}$ as well as ${\cal{N}}$ from the solution of master equations (\ref{S16}) and (\ref{S17}) into (\ref{S18}). In this case $\cos\delta=\ell=\pm 1$.
\subsection{Proofs}
\noindent
\textbf{\textit{i) Relative angle between $\boldsymbol{E}$ and $\boldsymbol{B}$}}\\
\textit{i.a)} We prove by contradiction that in a fluid with nonvanishing $\sigma E$, $\cos\delta$ does not vanish. Let us assume that $\cos\delta=0$, and plug this into (\ref{E14}). Assuming, without loss of generality, that ${\cal{M}}(0)={\cal{N}}(0)=0$, we obtain ${\cal{M}}(\tau)=-{\cal{N}}(\tau)$. Then, plugging  $\cos\delta=0$ into (\ref{E1}), and comparing the remaining $\partial_{\mu}B^{\mu}=0$ with the inhomogeneous differential equation $\partial_{\mu}(Bu^{\mu})=BD{\cal{M}}$ from (\ref{E12}), we obtain ${\cal{M}}={\cal{N}}=0,~\forall \tau$. Plugging these results into (\ref{E4}), we finally arrive at $\sigma E=0$, which is however assumed to be nonvanishing. This shows that for $\sigma E\neq 0$, we have $\cos\delta\neq 0$, i.e. in a nonideal fluid with finite electric conductivity, $\boldsymbol{E}$ and $\boldsymbol{B}$ cannot be perpendicular to each other. As aforementioned, to derive (\ref{E10}) and (\ref{E15}), we have used $\cos\delta\neq 0$.
\vspace{0.1cm}\par\noindent
\textit{i.b)} To show $\tan\delta=\alpha_{E}$, let us consider (\ref{E10}). Using $\partial_{\mu}(Eu^{\mu})=E\frac{d{\cal{N}}}{d\tau}$ from (\ref{E13}), we arrive first at
\begin{eqnarray}\label{S19}
\frac{\partial\zeta}{\partial\tau}=\left(\frac{\sigma}{\chi_{e}}+\frac{d{\cal{N}}}{d\tau}\right)\tan\delta-\frac{\sigma}{\chi_{e}}\alpha_{E}.
\end{eqnarray}
Plugging this expression into (\ref{E3}), and subtracting it from (\ref{E4}) multiplied with $\tan\delta$, we arrive at (\ref{S1}).
\vspace{0.1cm}\par\noindent
\textit{i.c)} To show the boost invariance of $\delta$, $\frac{\partial\delta}{\partial\eta}=0$, let us consider (\ref{N24}). Bearing in mind that $P_{i}=\partial_{i}\vartheta, i=0,3$ and using the definitions of longitudinal components of  $\partial_{\mu}$ from (\ref{N14}), we arrive first at
\begin{eqnarray}\label{S20}
\frac{\partial\vartheta\left(\tau,\eta\right)}{\partial\tau}=\frac{\kappa_{B}}{c},\hspace{0.3cm}
\frac{\partial\vartheta\left(\tau,\eta\right)}{\partial\eta}=\frac{\tau\kappa_{E}}{c}.
\end{eqnarray}
Then, using the $\eta$-independence of $\kappa_{B}$, we have $\frac{\partial^{2}\vartheta}{\partial\eta\partial\tau}=0$ from the first relation in (\ref{S20}). Differentiating then the second relation from  (\ref{S20}) with respect to $\tau$, we arrive at
\begin{eqnarray}\label{S21}
\tau\frac{\partial\kappa_{E}}{\partial \tau}+\kappa_{E}=0.
\end{eqnarray}
Using at this stage (\ref{E18}), (\ref{S21}) is equivalently given by
\begin{eqnarray}\label{S22}
\sigma\cos^{-2}\delta\frac{\partial\delta}{\partial u}+\left(1-\frac{1}{\kappa}\right)\sigma\tan\delta=0,
\end{eqnarray}
with $u=\ln\left(\frac{\tau}{\tau_{0}}\right)$. Here, (\ref{S1}) is also used.
Plugging $\frac{\partial\delta}{\partial u}$ from (\ref{E15}) into (\ref{S22}), we arrive at
\begin{eqnarray}\label{S23}
\hspace{-0.8cm}\sigma\bigg[1-\frac{1}{\kappa}-\frac{1}{\cos^{2}\delta}\left(\frac{d{\cal{M}}}{du}+\frac{d{\cal{N}}}{du}\right)\bigg]\tan\delta=0.
\end{eqnarray}
Since in a nonideal fluid $\sigma$ is nonvanishing, we are faced with two distinct equations
\begin{eqnarray}\label{S24}
\cos^{2}\delta=
\frac{\kappa}{\kappa-1}\left(\frac{d{\cal{M}}}{du}+\frac{d{\cal{N}}}{du}\right),
\end{eqnarray}
for nonvanishing $\alpha_{E}$, or
\begin{eqnarray}\label{S25}
\tan\delta=0,
\end{eqnarray}
for vanishing $\alpha_{E}$. Let us consider (\ref{S24}). Here, the $\eta$-independence of ${\cal{M}}$ and ${\cal{N}}$ leads immediately to the $\eta$-independence of $\delta$. Moreover, using (\ref{S1}), we obtain $\frac{\partial\alpha_{E}}{\partial\eta}=0$. The same is also true for $\tan\delta=0$ from (\ref{S25}), which leads also to $\frac{\partial\delta}{\partial\eta}=0$.
\vspace{0.2cm}\par\noindent\textbf{\textit{ii) Evolution of $\boldsymbol{\kappa_{E}}$  and $\boldsymbol{\delta}$}}\\
\textit{ii.a)} For $\tan\delta\neq 0$, (\ref{S1}) yields $\alpha_{E}\neq 0$. Bearing in mind that $\kappa_{E}=\sigma\alpha_{E}$, in a nonideal fluid with finite electric conductivity $\sigma$, the AH current $\kappa_{E}\epsilon^{0\mu\nu 3}E_{\nu}$ from (\ref{N23}), is therefore nonvanishing. To determine the evolution of $\kappa_{E}$, we simply solve (\ref{S21}), and arrive first at (\ref{S2}).
Then, plugging (\ref{S2}) and (\ref{E18}) into $\alpha_{E}=\kappa_{E}/\sigma$, it turns out that $\alpha_{E}$ evolves as (\ref{S3}). Using at this stage $\delta=\tan^{-1}\alpha_{E}$ from (\ref{S1}), the evolution of the relative angle between the electric and magnetic fields is thus given by (\ref{S4}).
\vspace{0.1cm}\par\noindent
\textit{ii.b)} For $\tan\delta=0$, the electric and magnetic fields are either parallel or antiparallel, and remain so during the evolution of the chiral fluid. Their relative angle is thus given by
\begin{eqnarray}\label{S26}
\delta=n\pi,\quad\mbox{with}\quad n=0,1,2,\cdots.
\end{eqnarray}
Let us notice that according to (\ref{S1}), $\tan\delta=0$ leads to $\alpha_{E}=0$, and hence to a vanishing AH current once $\boldsymbol{E}$ and $\boldsymbol{B}$ are parallel or antiparallel. This is also expected from \cite{ferrer2015,ferrer2016}.
\vspace{0.2cm}\par\noindent\textbf{\textit{iii) Evolution of the $\boldsymbol{\vartheta}$-vacuum}}\\
\textit{iii.a)} To determine the $\tau$- and $\eta$-dependence of the axionlike field $\vartheta$ for the $\tan\delta\neq 0$ case, let us differentiate the second equation in (\ref{S20}) with respect to $\eta$. Using the boost invariance of $\kappa_{E}$, we have $\frac{\partial^{2}\vartheta}{\partial\eta^{2}}=0$, which leads to the ansatz
\begin{eqnarray}\label{S27}
\vartheta\left(\tau,\eta\right)=\lambda_{\vartheta}(\tau)\eta+\vartheta_{0}(\tau).
\end{eqnarray}
Plugging (\ref{S27}) into (\ref{S20}), and using (\ref{S2}), we obtain
\begin{eqnarray}\label{S28}
\frac{\partial\vartheta\left(\tau,\eta\right)}{\partial\eta}=\lambda_{\vartheta}=\frac{\tau_{0}\kappa_{E}^{(0)}}{c}=\mbox{const.}
\end{eqnarray}
Here, $\kappa_{E}^{(0)}=\sigma_{0}\alpha_{0}$.  Differentiating (\ref{S27}) with constant $\lambda_{\vartheta}$ with respect to $\tau$, and using $\frac{\partial\vartheta}{\partial\tau}=\frac{\kappa_{B}}{c}$ from (\ref{S20}), $\vartheta_{0}(\tau)$ is given by
\begin{eqnarray}\label{S29}
\vartheta_{0}(\tau)=\frac{1}{c}\int_{\tau_{0}}^{\tau}\kappa_{B}(\tau^{\prime})d\tau^{\prime}+\vartheta_{0},
\end{eqnarray}
with $\vartheta_{0}=$ const. Plugging these results into (\ref{S27}), the evolution of the axionlike field $\vartheta$ is thus given by (\ref{S5}).
\vspace{0.1cm}\par\noindent
\textit{iii.b)} As concerns the $\tau$- and $\eta$-dependence of $\vartheta$ for $\tan\delta=0$, let us consider (\ref{S20}). In this case $\alpha_{E}=0$ leads to $\frac{\partial\vartheta}{\partial\eta}=0$, and hence to $\vartheta(\tau,\eta)=\vartheta_{0}(\tau)$. Plugging this into the first relation in (\ref{S20}), and assuming $\kappa_{B}=\mbox{const.}$, we arrive at (\ref{S6}).
\vspace{0.2cm}\par\noindent\textbf{\textit{iv) Evolution of the angles $\boldsymbol{\zeta}$  and $\boldsymbol{\phi}$}}\\
\textit{iv.a)} Let us consider the case $\tan\delta\neq 0$. To derive the evolution of the angles $\zeta$ and $\phi$ in this case, we use (\ref{S1}) and $\frac{\partial\delta}{\partial\eta}=0$ to obtain
\begin{eqnarray}
\frac{\partial\zeta}{\partial\tau}&=&\frac{\partial\phi}{\partial\tau}-\frac{1}{1+\alpha_{E}^{2}}\frac{d\alpha_{E}}{d\tau},\label{S30}\\
 \frac{\partial\zeta}{\partial\eta}&=&\frac{\partial\phi}{\partial\eta}.\label{S31}
\end{eqnarray}
Plugging $\partial_{\mu}(Bu^{\mu})=B\frac{d{\cal{M}}}{d\tau}$ from (\ref{E12}) into (\ref{E1}), we arrive at
\begin{eqnarray}\label{S32}
\frac{\partial\zeta\left(\tau,\eta\right)}{\partial\eta}=\frac{B}{E\cos\delta}\frac{d{\cal{M}}}{du},
\end{eqnarray}
where $u= \ln\left(\frac{\tau}{\tau_{0}}\right)$. Using the $\eta$-independence of $E, B,\delta$ and ${\cal{M}}$, we have
\begin{eqnarray}\label{S33}
\frac{\partial^{2}\zeta\left(\tau,\eta\right)}{\partial\eta^{2}}=0,
\end{eqnarray}
and, upon using (\ref{S31}),
\begin{eqnarray}\label{S34}
\frac{\partial^{2}\phi\left(\tau,\eta\right)}{\partial\eta^{2}}=0.
\end{eqnarray}
The last two equations lead to
\begin{eqnarray}\label{S35}
\zeta(\tau,\eta)&=&\omega_{\zeta}(\tau)\eta+\zeta_{0}(\tau),\nonumber\\
\phi(\tau,\eta)&=&\omega_{\phi}(\tau)\eta+\phi_{0}(\tau).
\end{eqnarray}
Plugging at this stage (\ref{S32}) into (\ref{E2}), we obtain
\begin{eqnarray}\label{S36}
\frac{\partial\phi\left(\tau,\eta\right)}{\partial\tau}=-\alpha_{E}\frac{d{\cal{M}}}{d\tau}.
\end{eqnarray}
Here, (\ref{S1}) is used. Then, using (\ref{S30}), we arrive at
\begin{eqnarray}\label{S37}
\frac{\partial\zeta\left(\tau,\eta\right)}{\partial\tau}=-\alpha_{E}\frac{d{\cal{M}}}{d\tau}-\frac{1}{1+\alpha_{E}^{2}}\frac{d\alpha_{E}}{d\tau}.
\end{eqnarray}
Bearing in mind that the rhs of (\ref{S36}) and (\ref{S37}) are independent of $\eta$, we have
\begin{eqnarray}\label{S38}
\frac{\partial}{\partial\eta}\left(\frac{\partial \phi}{\partial\tau}\right)=\frac{\partial}{\partial\eta}\left(\frac{\partial \zeta}{\partial\tau}\right)=0.
\end{eqnarray}
Plugging (\ref{S35}) into (\ref{S38}) leads immediately to
\begin{eqnarray}\label{S39}
\frac{\partial\omega_{\phi}}{\partial\tau}=0, \qquad \frac{\partial\omega_{\zeta}}{\partial\tau}=0,
\end{eqnarray}
and hence to $\omega_{\phi}=$ const and $\omega_{\zeta}=$ const. Using at this stage (\ref{S31}), we obtain
\begin{eqnarray}\label{S40}
\omega_{\phi}=\omega_{\zeta}\equiv \omega_{0}.
\end{eqnarray}
The relation (\ref{S32}) reduces therefore to
\begin{eqnarray}\label{S41}
\omega_{0}=\frac{B}{E\cos\delta}\frac{d{\cal{M}}}{du}.
\end{eqnarray}
Plugging $\zeta(\tau,\eta)=\omega_{0}\eta+\zeta_{0}(\tau)$ and $\phi(\tau,\eta)=\omega_{0}\eta+\phi_{0}(\tau)$ into (\ref{S37}) and (\ref{S36}), we arrive at the differential equations for $\zeta_{0}(\tau)$ and $\phi_{0}(\tau)$. Then, plugging (\ref{S2}) into these equations, and solving them lead to (\ref{S7}).
\vspace{0.1cm}\par\noindent
\textit{iv.b)} Following the same method as described above, we arrive for the case $\tan\delta=0$ at (\ref{S8}).
Let us notice that in this case,  the results from (\ref{S8}) coincide with those presented in \cite{shokri-1}.
\vspace{0.2cm}\par\noindent\textbf{\textit{v) Determination of $\boldsymbol{\alpha_{0}}$}}\\
\textit{v.a)}
To prove (\ref{S9}), which is only valid for the case $\tan\delta\neq 0$, let us consider (\ref{E4}). Using $\partial_{\mu}\left(Eu^{\mu}\right)=E\frac{d{\cal{N}}}{d\tau}$, multiplying (\ref{E4}) with $\frac{\tau}{E}$, and evaluating the resulting expression at $u=0$, we arrive first at
\begin{eqnarray}\label{S42}
\frac{d{\cal{N}}}{du}\bigg|_{u=0}+\frac{\big[(1-\chi_{m})\omega_{0}+\kappa_{B}\tau_{0}\big]}{\beta_{0}(1+\chi_{e})}
\cos\delta_{0}+\frac{\sigma \tau_{0}}{1+\chi_{e}}=0,\nonumber\\
\end{eqnarray}
where, according to (\ref{S3}), $\delta_{0}\equiv \delta(\tau=\tau_{0})=\tan^{-1}\alpha_{0}$.
To determine $\frac{d{\cal{N}}}{du}\bigg|_{u=0}$, let us then evaluate (\ref{S24}) at $u=0$. For $\alpha_{0}\neq 0$, we obtain
\begin{eqnarray}\label{S43}
\frac{d{\cal{N}}}{du}\bigg|_{u=0}=\left(\frac{\kappa-1}{\kappa}\right)\cos^{2}\delta_{0}-\frac{d{\cal{M}}}{du}\bigg|_{u=0}.
\end{eqnarray}
Plugging
\begin{eqnarray}\label{S44}
\frac{d{\cal{M}}}{du}\bigg|_{u=0}=\beta_{0}\omega_{0}\cos\delta_{0},
\end{eqnarray}
from (\ref{S41}) into (\ref{S43}) and the resulting expression into (\ref{S42}), we arrive at
\begin{eqnarray}\label{S45}
\cos^{2}\delta_{0}-{\cal{C}}_{1}\cos\delta_{0}+{\cal{C}}_{2}=0,
\end{eqnarray}
where ${\cal{C}}_{1}$ and ${\cal{C}}_{2}$ are defined in (\ref{S10}) and (\ref{S11}), respectively. The solution of the above equation reads
\begin{eqnarray}\label{S46}
\cos\delta_{0}=\frac{1}{2}\left({\cal{C}}_{1}\pm\sqrt{{\cal{C}}_{1}^{2}-4{\cal{C}}_{2}}\right).
\end{eqnarray}
Using at this stage $\cos\delta_{0}=\ell\left(1+\alpha_{0}^{2}\right)^{-1/2}$, we arrive at $\alpha_{0}$ from (\ref{S9}). In this way, the initial value of the relative angle between $\boldsymbol{E}$ and $\boldsymbol{B}$ fields, $\delta_{0}=\tan^{-1}\alpha_{0}$, is completely determined in terms of free parameters $\kappa,\tau_{0},\beta_{0},\sigma_{0},\omega_{0}$ and $\kappa_{B}^{(0)}$ as well as $\chi_{e}$ and $\chi_{m}$.
\vspace{0.3cm}\par\noindent\textbf{\textit{vi) Evolution of $\bm{\mathcal{M}}$, $\bm{\mathcal{N}}$ and $\boldsymbol{\kappa}_{\boldsymbol{B}}$}}\\
\textit{vi.a)} For $\tan\delta\neq 0$, the quantities ${\cal{M}}$, ${\cal{N}}$ and $\kappa_{B}$ can be determined using (\ref{E4}), (\ref{E9}) and (\ref{S41}). In what follows, we assume $\frac{d{\cal{M}}}{du}\neq 0$ (see below). To determine ${\cal{M}}$, let us first consider (\ref{E9}). Integrating this relation with respect to $\tau$, and using the formal solution of $E$ and $B$ from (\ref{E13}), we arrive first at
\begin{eqnarray}\label{S47}
e^{{\cal{N}}}\cos\delta=\frac{e^{-{\cal{M}}}\sin\delta_{0}}{\alpha_{E}}.
\end{eqnarray}
Here, (\ref{S1}) is also used. Plugging then the formal solution of $E$ and $B$ into (\ref{S41}), and comparing the resulting expression for $e^{{\cal{N}}}\cos\delta$,
\begin{eqnarray}\label{S48}
e^{{\cal{N}}}\cos\delta=\frac{e^{\cal{M}}}{\beta_{0}\omega_{0}}\frac{d{\cal{M}}}{du},
\end{eqnarray}
with (\ref{S47}), we arrive at the corresponding differential equation to ${\cal{M}}$,
\begin{eqnarray}\label{S49}
e^{2{\cal{M}}}\frac{d{\cal{M}}}{du}=\frac{\beta_{0}\omega_{0}\sin\delta_{0}}{\alpha_{E}},
\end{eqnarray}
with $\alpha_{E}$ from (\ref{S3}). Integrating (\ref{S49}) with respect to $u$, we arrive at ${\cal{M}}$ from (\ref{S12}).
Bearing in mind that  since $\alpha_{0}$ from (\ref{S9}) is solely a function of free parameters
$\{\kappa,\tau_{0},\beta_{0},\sigma_{0},\omega_{0},\kappa_{B}^{(0)},\chi_{e},\chi_{m}\},$ ${\cal{M}}$ turns also out to be a function of the same free parameters.
\par
There are many equivalent possibilities to determine ${\cal{N}}$ arising in the formal solution of $E$ from (\ref{E13}). One of the most simple ones is to solve the differential equation
\begin{eqnarray}\label{S50}
\frac{d{\cal{N}}}{du}=\left(\frac{\kappa-1}{\kappa}\right)\cos^{2}\delta-\frac{d{\cal{M}}}{du},
\end{eqnarray}
from (\ref{S24}) with $\cos^{2}\delta=(1+\alpha_{E}^{2})^{-1}$. Inserting $\alpha_{E}$ from (\ref{S3}) and ${\cal{M}}$ from (\ref{S12}) into (\ref{S50}), we arrive at ${\cal{N}}$ from (\ref{S13}).
Similar to ${\cal{M}}$, ${\cal{N}}$ is also a function of free parameters $\kappa,\tau_{0},\beta_{0},\sigma_{0},\omega_{0}$ and $\kappa_{B}^{(0)}$ as well as $\chi_{e}$ and $\chi_{m}$.
Plugging ${\cal{M}}$ and ${\cal{N}}$ from (\ref{S12}) and (\ref{S50}) into the formal solutions of $B$ and $E$ from (\ref{E13}), the magnitude of the magnetic
and electric fields are given in terms of these free parameters.
\par
To prove (\ref{S14}), let us finally consider (\ref{E4}). Using $\partial_{\mu}\left(Eu^{\mu}\right)=E\frac{d{\cal{N}}}{d\tau}$, and multiplying (\ref{E4}) with $\frac{\tau}{B}$, we arrive first at
\begin{eqnarray}\label{S51}
&&\left(1+\chi_{e}\right)\frac{E}{B}\frac{d{\cal{N}}}{du}+(1-\chi_{m})\frac{B}{E}\frac{d{\cal{M}}}{du}+\sigma\tau_{0}e^{u}\frac{E}{B}\nonumber\\
&&+\kappa_{B}\tau_{0}e^{u}\cos\delta=0.
\end{eqnarray}
This gives rise to $\kappa_{B}$, that in terms of ${\cal{M}}, {\cal{N}}$ and their derivatives with respect to $u$ is given by (\ref{S14}).
\vspace{0.1cm}\par\noindent
\textit{vi.b)} Let us now consider the case $\tan\delta=0$. In this case, the constraint relation (\ref{E9}) is automatically satisfied, and no explicit relation between $\frac{d{\cal{M}}}{du}$ and $\frac{d{\cal{N}}}{du}$ arises. To determine ${\cal{M}}$ and ${\cal{N}}$, we follow the same method as described in \cite{shokri-1}, where in a nonchiral fluid, $\tan\delta$ vanished.
\par
To derive the master equation (\ref{S16}) for ${\cal{M}}$, let us consider (\ref{S51}). Multiplying it with $\frac{E}{B}$, and using (\ref{S41}), we arrive at
\begin{eqnarray}\label{S52}
&&\frac{1}{\omega_0^{2}\cos^{2}\delta}\frac{d{\cal{M}}}{du}\bigg\{
(1+\chi_{e})\frac{d{\cal{M}}}{du}\frac{d{\cal{N}}}{du}+(1-\chi_{m})\omega_{0}^{2}\cos^{2}\delta\nonumber\\
&&+\sigma\tau_{0}e^{u}\frac{d{\cal{M}}}{du}+\kappa_{B}\omega_0\tau_0e^{u}\cos^{2}\delta
\bigg\}=0.
\end{eqnarray}
Two distinct differential equations for ${\cal{M}}$, (\ref{S15}) or (\ref{S16}), thus arise. The latter is derived using
\begin{eqnarray}\label{S53}
\frac{d{\cal{M}}}{du}\frac{d{\cal{N}}}{du}=\frac{d^{2}{\cal{M}}}{du^{2}}+\left(\frac{d{\cal{M}}}{du}\right)^{2},
\end{eqnarray}
which arises from (\ref{S41}). Once ${\cal{M}}$ is determined either analytically or numerically by solving (\ref{S15}) or (\ref{S16}), it is possible to determine ${\cal{N}}$ via (\ref{S48}), with $\cos\delta=\ell=\pm 1$.
\vspace{0.2cm}\par\noindent\textbf{\textit{vii) Evolution of $\bm{\mathcal{L}}$}}\\
\textit{vii.a)} To drive (\ref{S18}) in the case of $\tan\delta\neq 0$, let us consider (\ref{E11}), that is equivalently given by
\begin{eqnarray}\label{S54}
D\epsilon+\theta\left(\epsilon+p\right)+{\cal{O}}=0.
\end{eqnarray}
For nonvanishing $\frac{d{\cal{M}}}{d\tau}$, ${\cal{O}}$ is defined by
\begin{eqnarray}\label{S55}
{\cal{O}}\equiv \chi_{e}EDE+\chi_{m}BDB-\sigma E^{2}-\kappa_{B}EB\cos\delta.\nonumber\\
\end{eqnarray}
It arises from (\ref{E11}) with $\frac{\partial\delta}{\partial\eta}=0$ and $\frac{\partial\phi}{\partial\eta}=\omega_{0}$ with $\omega_{0}$ satisfying (\ref{S41}).\footnote{
For $\frac{d{\cal{M}}}{d\tau}=0$, or equivalently $\partial_{\mu}(Bu^{\mu})=0$, we have to replace $BDB$ in (\ref{S55}) by $B^{2}\theta$.}
Then, plugging the equation of state $\epsilon=\kappa p$ with constant $\kappa$ into (\ref{S54}) and using $p=nT$, we arrive at
\begin{eqnarray}\label{S56}
\hspace{-1cm}\partial_{\mu}\left(T^{\kappa}u^{\mu}\right)=T^{\kappa}D{\cal{L}}, \quad\mbox{with}\quad D{\cal{L}}=-\frac{{\cal{O}}}{p},
\end{eqnarray}
and ${\cal{O}}$ given in (\ref{S55}). Finally, plugging the formal solutions of $E$ and $B$ from (\ref{E13}) into (\ref{S55}), and using
\begin{eqnarray}\label{S57}
p=p_{0}\left(\frac{\tau_{0}}{\tau}\right)^{1+\frac{1}{\kappa}}e^{\frac{\cal{L}}{\kappa}},
\end{eqnarray}
which arises by combining the evolution of $n(\tau)$ from (\ref{E17}) and $T(\tau)$
from (\ref{E13}), we arrive at (\ref{S18}).
\vspace{0.1cm}\par\noindent
\textit{vii.b)} In the case of vanishing $\tan\delta$, the coefficient $e^{{\cal{L}}/\kappa}$ is given by (\ref{S18}) with constant $\kappa_{B}$ and $\cos\delta=\ell$. In this case, ${\cal{M}}$ and ${\cal{N}}$ arising from the master equations (\ref{S16}) and (\ref{S17}), are to be used.
\section{Nonrotating electric and magnetic fields; Analytical results}\label{sec5}
\setcounter{equation}{0}
Let us assume for simplicity that the electric and CM conductivity, $\sigma$ and  $\kappa_{B}$, are constant. In this case, the evolution of $\alpha_{E}$ is given by plugging (\ref{S2}) into $\kappa_{E}=\sigma\alpha_{E}$, and reads
\begin{eqnarray}\label{A0}
\alpha_{E}(\tau)=\alpha_{0}\left(\frac{\tau_{0}}{\tau}\right).
\end{eqnarray}
In what follows, we show that in the nonideal transverse CSMHD with nonvanishing electric field, the case $\frac{d{\cal{M}}}{du}=0$ leads to $\alpha_{0}=0$, and therefore to a vanishing AH current. The key point in this case is that, according to (\ref{S41}), $\frac{d{\cal{M}}}{d\tau}=0$ corresponds to vanishing angular velocity $\omega_{0}$. This is dubbed ``nonrotating electric and magnetic fields''. Using these assumptions, (\ref{S7}) leads to
\begin{eqnarray}\label{A1}
\zeta(\tau,\eta)&=&\phi_{0}-\tan^{-1}\left(\alpha_{E}\right),\nonumber\\
\phi(\tau,\eta)&=&\phi_{0},
\end{eqnarray}
with $\alpha_{E}$ given in (\ref{A0}). It is also possible to show that the evolution of $B$ and $E$ fields is given by
\begin{eqnarray}\label{A2}
B(\tau)&=&B_{0}\left(\frac{\tau_{0}}{\tau}\right),\nonumber\\
E(\tau)&=&E_{0}\left(\frac{\tau_{0}}{\tau}\right)\bigg[\left(1+\frac{\ell\kappa_{B}}{\beta_{0}\sigma}\right)e^{-\frac{\sigma\left(\tau-\tau_{0}\right)}{1+\chi_{e}}}-\frac{\ell\kappa_{B}}{\beta_{0}\sigma}\bigg], \nonumber\\
\end{eqnarray}
with $\beta_{0}=\frac{E_{0}}{B_{0}}$.
\par
To do this, we start with the formal solution of $B$ from (\ref{E13}). For $\frac{d{\cal{M}}}{du}=0$, we have ${\cal{M}}=0$. Choosing, without loss of generality, ${\cal{M}}_{0}=0$, the evolution of $B$ is given by $B=B_{0}\left(\frac{\tau_{0}}{\tau}\right)$ from (\ref{A2}). Hence, as in the case of the ideal MHD, the magnetic fluxes are frozen.
\par
To arrive at the evolution of the electric field from (\ref{A2}), let us consider (\ref{E4}). Using $\frac{\partial\phi}{\partial\eta}=\omega_{0}$, and bearing in mind that for $\frac{d{\cal{M}}}{d\tau}=0$, we have $\omega_{0}=0$, we arrive at a differential equation for ${\cal{U}}\equiv e^{\cal{N}}$
\begin{eqnarray}\label{A3}
\frac{d{\cal{U}}}{d\tau}+{\cal{C}}_{0}~{\cal{U}}+{\cal{A}}\left(\tau\right)=0,
\end{eqnarray}
where
\begin{eqnarray}\label{A4}
\hspace{-0.5cm}{\cal{C}}_{0}&\equiv& \frac{\sigma}{1+\chi_{e}},\nonumber\\
\hspace{-0.5cm}{\cal{A}}(\tau)&\equiv&\frac{{\cal{A}}_{0}}{\left(1+\alpha_{E}^{2}\right)^{1/2}},\hspace{0.2cm}\mbox{with}\hspace{0.2cm}{\cal{A}}_{0}\equiv \frac{\kappa_{B}\ell}{\beta_{0}\left(1+\chi_{e}\right)}.
\end{eqnarray}
The most general solution to (\ref{A3}) reads
\begin{eqnarray}\label{A5}
	e^{{\cal{N}}}=e^{-{\cal{C}}_{0}\left(\tau-\tau_{0}\right)} -e^{-{\cal{C}}_{0}\tau}\int_{\tau_{0}}^{\tau}d\tau^{\prime}~e^{{\cal{C}}_{0}\tau^{\prime}}
{\cal{A}}(\tau^{\prime}).
\end{eqnarray}
However, it turns out that in this nonrotating case $\alpha_{E}=0$, and $e^{{\cal{N}}}$ is therefore given by
\begin{eqnarray}\label{A6}
e^{{\cal{N}}}=\left(1+\frac{{\cal{A}}_{0}}{{\cal{C}}_{0}}\right)e^{-{\cal{C}}_{0}\left(\tau-\tau_{0}\right)}-\frac{{\cal{A}}_{0}}{{\cal{C}}_{0}}.
\end{eqnarray}
To show this, let us consider (\ref{A3}), that leads to
\begin{eqnarray}\label{A7}
\hspace{-0.3cm}{\cal{N}}=\ln\left(-\frac{
{\cal{A}}\left(\tau\right)}{\frac{d{\cal{N}}}{d\tau}+{\cal{C}}_{0}}\right).
\end{eqnarray}
Plugging
\begin{eqnarray}\label{A8}
\frac{d{\cal{N}}}{d\tau}=-\frac{1}{\alpha_{E}\left(1+\alpha_{E}^{2}\right)}\frac{d\alpha_{E}}{d\tau},
\end{eqnarray}
from (\ref{E15}) with $\frac{d{\cal{M}}}{d\tau}=0$ and $\tan\delta=\alpha_{E}$ from (\ref{S1}) into (\ref{A7}), and differentiating both sides of the resulting expression with respect to $\tau$, we arrive, after using (\ref{A8}) once again, at the following differential equation for $\alpha_{E}$
\begin{eqnarray}\label{A9}
\alpha_{E}\alpha_{E}^{\dprime}-{\cal{C}}_{0}\alpha_{E}\alpha_{E}^{\prime}
\left(\alpha_{E}^{2}-1\right)-2\alpha_{E}^{\prime 2}=0,
\end{eqnarray}
where the primes denote the derivation with respect to $\tau$. Plugging, at this stage,  $\alpha_{E}$ from (\ref{A0}) into (\ref{A9}), we obtain
\begin{eqnarray}\label{A10}
\frac{{\cal{C}}_{0}\alpha_{0}^{2}\tau_{0}^{2}\left(\alpha_{0}^{2}\tau_{0}^{2}-\tau^{2}\right)}{\tau^{5}}=0.
\end{eqnarray}
This leads immediately to $\alpha_{0}=0$, and thus to $\alpha_{E}=0$. As aforementioned, $e^{{\cal{N}}}$ in this case is given by (\ref{A6}). Plugging the definitions of ${\cal{A}}_{0}$ and ${\cal{C}}_{0}$ from (\ref{A4}) into (\ref{A6}), and using the formal solution of $E=E_{0}\left(\frac{\tau_{0}}{\tau}\right)e^{{\cal{N}}}$, we arrive at $E(\tau)$ from (\ref{A2}).
Let us notice that for parallel $E$ and $B$ fields with $\ell=+1$, $E(\tau)$ is always positive. For antiparallel $E$ and $B$ fields, the positiveness of $E$ sets certain constraint on the ratio $\frac{\kappa_{B}}{\beta_{0}\sigma}$.
\section{Rotating electric and magnetic fields; Numerical results}\label{sec6}
\setcounter{equation}{0}
\par\noindent
In this section, we focus on the evolution of electromagnetic and hydrodynamic fields in a fluid with finite electric conductivity $\sigma$. In particular, we separately study two cases of nonvanishing and vanishing AH coefficients $\kappa_{E}$ [see Secs. \ref{sec6A} and \ref{sec6B}]. We are interested in the effect of various free parameters $\{\kappa,\tau_{0},\beta_{0},\sigma_{0}, \omega_{0}, \kappa_{B}^{(0)},\chi_{e},\chi_{m}\}$ on the proper time dependence of $E,B$ and $T$. To be brief, we only use
\begin{eqnarray}\label{D1}
\{\kappa,\tau_{0},\beta_{0}\}=\{3,0.5~\mbox{fm/c},0.1\}.
\end{eqnarray}
As concerns $\sigma_{0}$, arising in (\ref{E18}), and defined in (\ref{E20}), we mainly work with two values, $\sigma_{0}\simeq 8.6,17.1$ MeVc, corresponding to $T_{0}=250,500$ MeV. Here, $\sigma_c= 6$ MeVc and $T_{c}=175$ MeV are chosen. In our numerical results and corresponding plots, these two cases are referred to as $T_{0}=250$ MeV and $T_{0}=500$ MeV cases.
Since the effect of electric and magnetic susceptibilities, $\chi_{e}$ and $\chi_{m}$, on the evolution of $E$ and $B$ for the case of $\tan\delta=\alpha_{E}=0$ is already studied in \cite{shokri-1}, we set $\chi_{e}=\chi_{m}=0$, and focus only on the interplay between the rest of these parameters, $\{\sigma_{0}, \omega_{0}, \kappa_{B}^{(0)}\}$, and their effect on the $\tau$-dependence of $E,B$ and $T$.
\par
In the case of $\tan\delta=\alpha_{E}\neq 0$, we first explain the method from which a valid range for the constant angular velocity $\omega_{0}$ is found by choosing a fixed initial value of $\kappa_{B}$. Because of the definition $\kappa_{B}=c\mu_{5}$ with $c=\sum_{f=\{u,d\}}q_{f}^{2}\frac{e^{2}}{2\pi^{2}}$, a fixed initial value of $\kappa_{B}$ corresponds to a fixed initial value of the axial chemical potential $\mu_{5}$ of the medium. In our plots different values of $\kappa_{B}$ are denoted by corresponding values of $\mu_{5}$. Bearing in mind that $\omega_{0}$ remains constant during the evolution of the fluid, we choose fixed values of $\omega_{0}$, and compute ${\cal{L}}$, ${\cal{M}}$ and ${\cal{N}}$ from which the evolution of $T,E$ and $B$ arises as a function of $\tau$ [see (\ref{E13})]. Apart from these quantities, the $\tau$-dependence of $\mu_{5}$ can also be determined in the case of nonvanishing $\alpha_{E}$.
\par
In the case of vanishing AH coefficient, i.e. for $\tan\delta=\alpha_{E}=0$, we have to work with fixed values of $\kappa_{B}$ (or equivalently $\mu_{5}$) during the evolution of the fluid. For the sake of comparison, we use the same values of $\omega_{0}$ as in the case of nonvanishing $\alpha_{E}$. Following the method, originally introduced in our previous work \cite{shokri-1}, we numerically solve the master equation (\ref{S16}) for ${\cal{M}}$ using different sets of free parameters. Once ${\cal{M}}$ is determined ${\cal{N}}$ and ${\cal{L}}$ can also be determined, using (\ref{S17}) and (\ref{S18}). This leads eventually to the proper time dependence of $B,E$ and $T$ via (\ref{E13}).
\subsection{Case 1: Nonvanishing AH coefficient}\label{sec6A}
\subsubsection{Determination of suitable values for $\omega_{0}$}\label{sec6A1}
\par
\begin{figure}[hbt]
	\includegraphics[width=8cm,height=6cm]{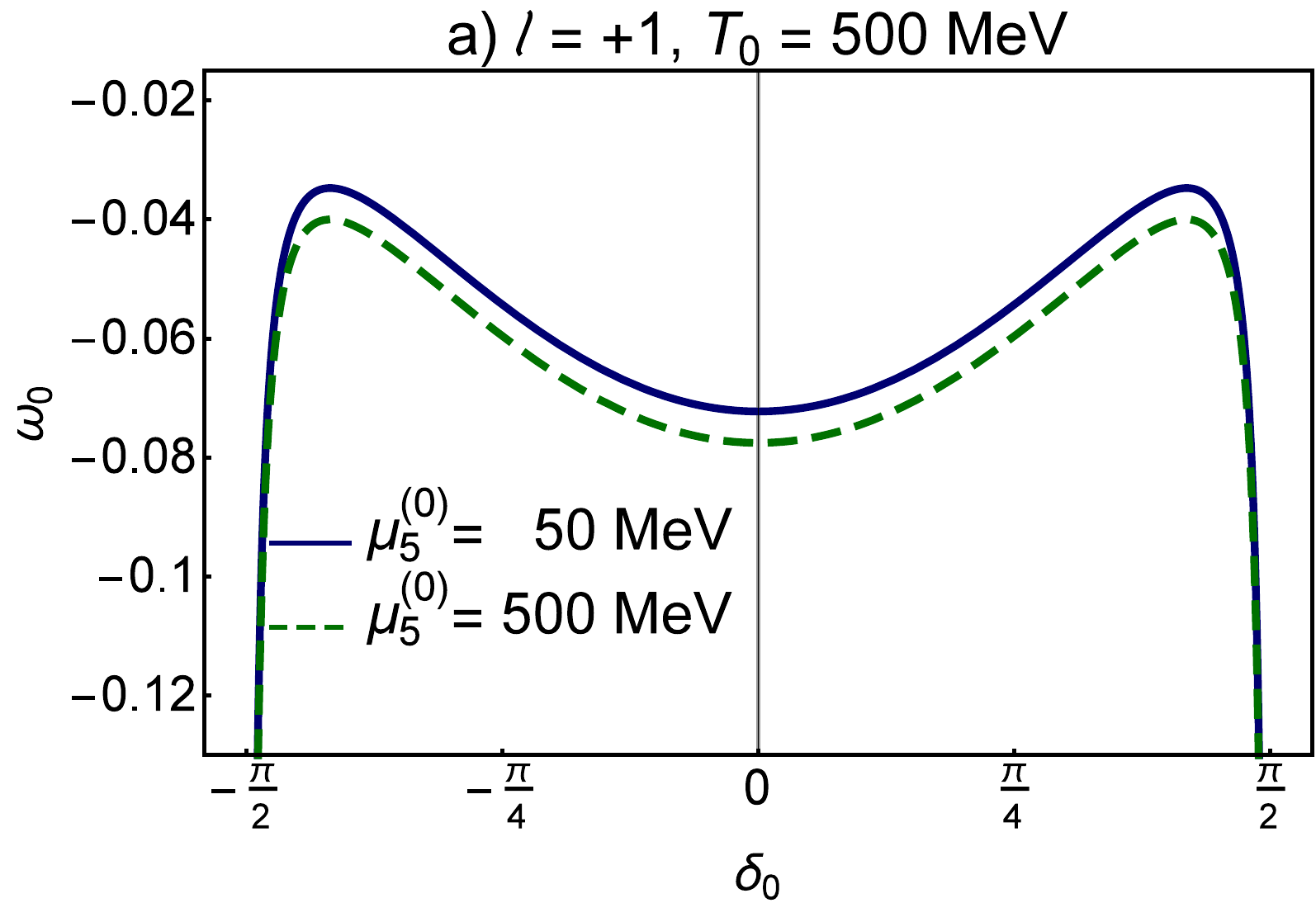}
	\includegraphics[width=8cm,height=6cm]{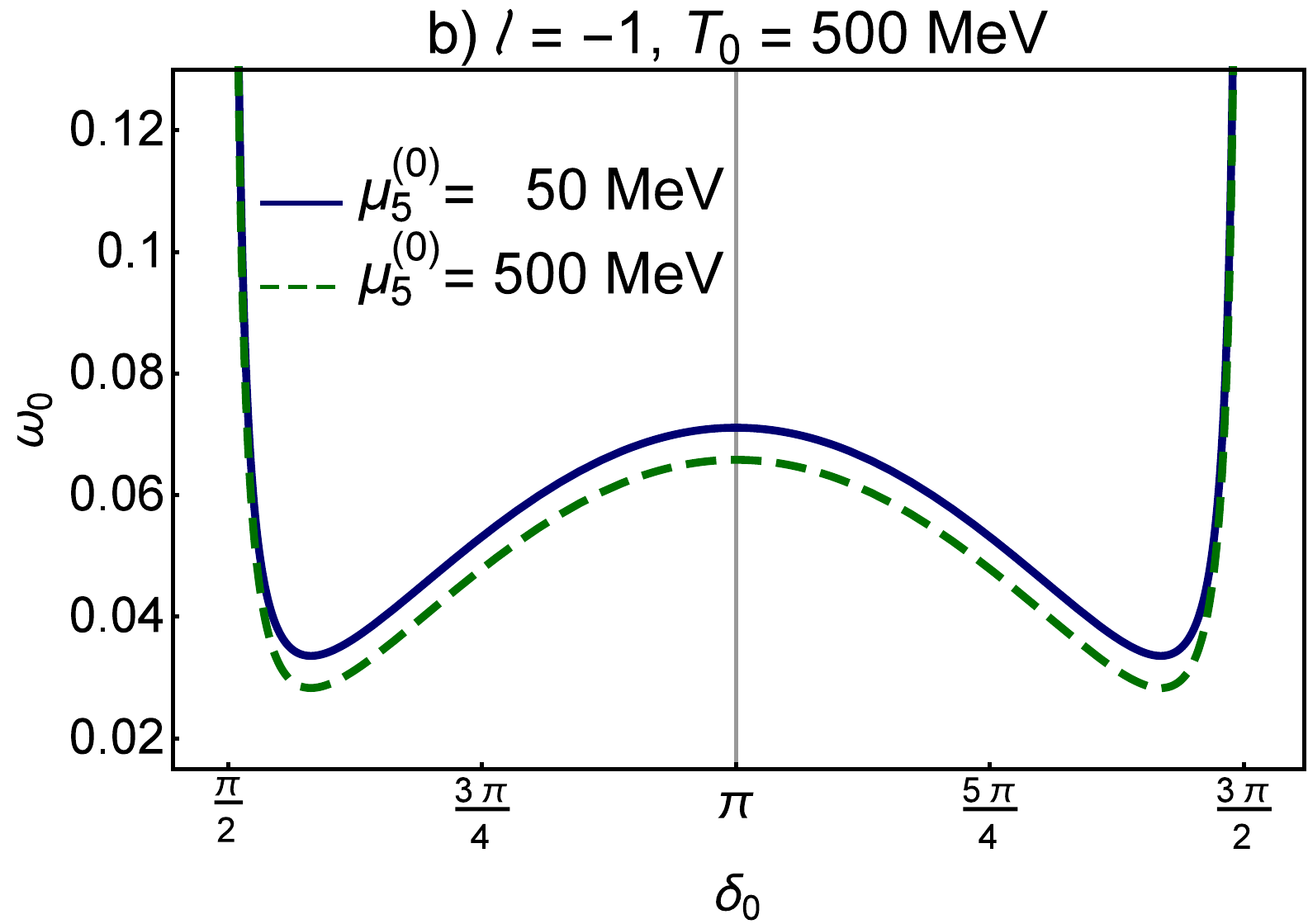}
	\caption{(color online).  The angular velocity $\omega_{0}$, appearing in (\ref{S7}), is plotted as a function of initial angle between the electric and magnetic field, $\delta_{0}$, for free parameters $\{\kappa,\tau_{0},\beta_{0},\chi_{e},\chi_{m}\}=\{3,0.5~\mbox{fm/c},0.1, 0,0\}$ and $\sigma_{0}=17.1$ MeVc, corresponding to $T_{0}=500$ MeV, as well as the initial axial chemical potential $\mu_{5}^{(0)}=50$ MeV (blue solid curves) and $\mu_{5}^{(0)}=500$ MeV (green dashed curves) in the interval $\delta_{0}\in\left(-\frac{\pi}{2},+\frac{\pi}{2}\right)$ (panel a) and $\delta_{0}\in\left(\frac{\pi}{2},\frac{3\pi}{2}\right)$ (panel b) where, by definition, $\ell=+1$ and $\ell=-1$, respectively.}\label{fig-1}
\end{figure}

\noindent
As indicated in the previous section, for $\tan\delta\neq 0$, the quantities ${\cal{M}}$ and ${\cal{N}}$ are determined analytically by making use of (\ref{E4}), (\ref{E9}) and (\ref{S41}). The results for ${\cal{M}}$ and ${\cal{N}}$ are given by (\ref{S12}) and (\ref{S13}), respectively. Plugging these results in the formal solutions of $B$ and $E$ from (\ref{E13}), we obtain the $\tau$-dependence of these fields in terms of aforementioned free parameters. Moreover, plugging the results for ${\cal{M}}$ and ${\cal{N}}$ into (\ref{S14}), the $\tau$-dependence of $\kappa_{B}$, and up to a constant numerical factor, the $\tau$-dependence of $\mu_{5}$ are also determined. To choose appropriate values for $\omega_{0}$ for fixed $\kappa_{B}^{(0)}$, let us consider (\ref{S14}). Setting $u=0$, we obtain
\begin{eqnarray}\label{D2}
\lefteqn{\hspace{-0.2cm}\kappa_{B}^{(0)}=\frac{1}{\ell\tau_{0}(1+\alpha_{0}^{2})\kappa}\bigg\{-\sqrt{1+\alpha_{0}^{2}}\beta_{0}\left(-1+\kappa+\tau_{0}\kappa\sigma_{0}\right.}\nonumber\\
&&\left.+\tau_{0}
\alpha_{0}^{2}\kappa\sigma_{0}
\right)-\ell\left(1+\alpha_{0}^{2}\right)\kappa\omega_{0}+\ell\left(1+\alpha_{0}^{2}
\right)\beta_{0}^{2}\kappa\omega_{0}
\bigg\}.\nonumber\\
\end{eqnarray}
Here, $\alpha_{0}=\tan\delta_{0},$ with $\delta_{0}$ the initial angle between the electric and magnetic fields. Using (\ref{D2}), it is possible to determine  $\omega_{0}$ in terms of free parameters $\{\kappa,\tau_0,\beta_0,\sigma_0,\kappa_B^{(0)},\delta_0,\chi_e,\chi_m\}$. In Fig. \ref{fig-1}, $\omega_0$ is plotted as a function of $\delta_{0}$ for $\delta_{0}\in\left(-\frac{\pi}{2},\frac{\pi}{2}\right)$ [Fig. \ref{fig-1}(a)] and $\delta_{0}\in\left(\frac{\pi}{2},\frac{3\pi}{2}\right)$ [Fig. \ref{fig-1}(b)].\footnote{Here, parentheses denote open intervals, i.e. $x\in(a,b)$ is equivalent with $a<x<b$.}  Free parameters are given by (\ref{D1}) and
\begin{eqnarray*}
	\{\sigma_{0},\kappa_B^{(0)},\chi_e,\chi_m\}=\{17.1~\mbox{MeVc},\kappa_{B}^{(0)},0,0\},
\end{eqnarray*}
with $\kappa_{B}^{(0)}=50 c$ MeV (blue solid curves) and $\kappa_{B}^{(0)}=500$ MeV (green dashed curves).
	\begin{figure}[hbt]
		\includegraphics[width=8cm,height=6cm]{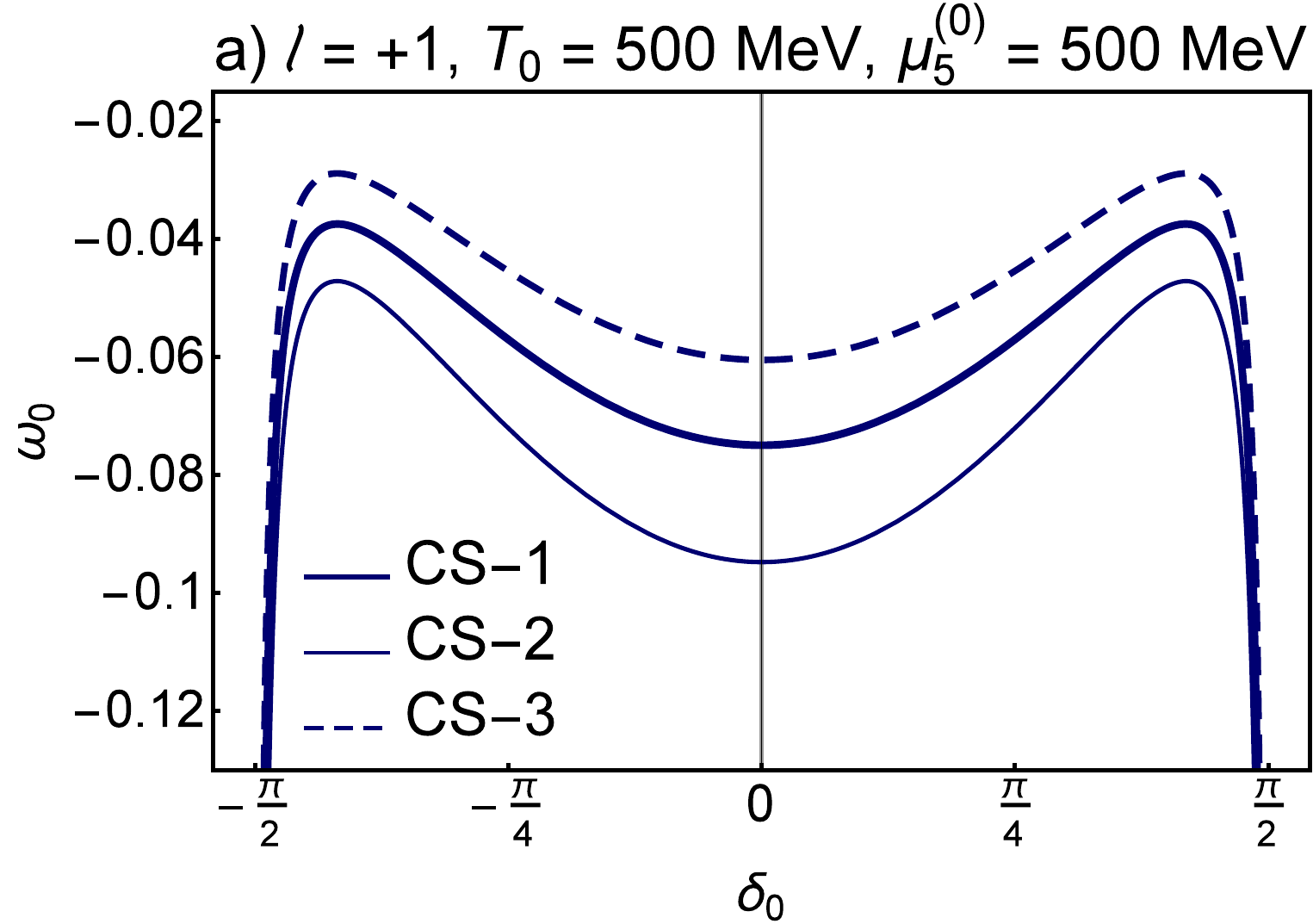}
		\includegraphics[width=8cm,height=6cm]{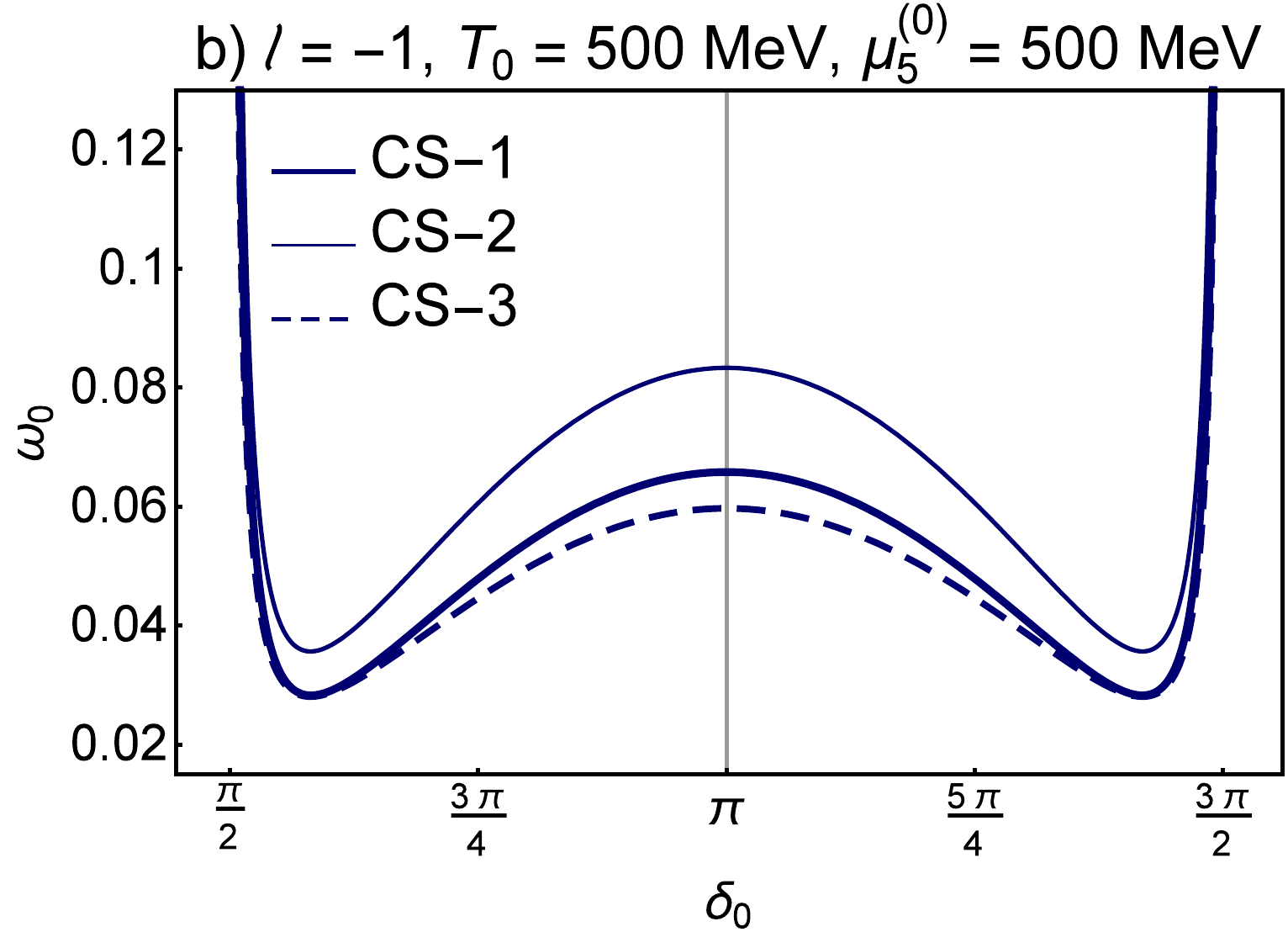}
		\caption{(color online). The angular velocity $\omega_{0}$ is plotted as a function of initial angle between the electric and magnetic field, $\delta_{0}$, for free parameters $\{\kappa,\tau_{0},\beta_{0},\mu_{5}^{(0)}\}=\{3,0.5,0.1,500~\mbox{MeV}\}$ and $\sigma_{0}=17.1$ MeVc, corresponding to $T_{0}=500$ MeV. Three different sets of $\chi_{e}$ and $\chi_{m}$, $\mbox{CS-i, i}=1,2,3$ from (\ref{D3}) are used. Panels (a) and (b) correspond to $\delta_{0}\in\left(-\frac{\pi}{2},+\frac{\pi}{2}\right)$ and $\delta_{0}\in\left(\frac{\pi}{2},\frac{3\pi}{2}\right)$, respectively. In both intervals of $\delta_{0}$, the curves for nonvanishing susceptibilities [CS-2 (thin solid curve) and CS-3 (dashed curve)] are slightly shifted relative to the case of vanishing $\chi_{e}$ and $\chi_{m}$ [CS-1 (thick solid curve)]. }\label{fig-2}
	\end{figure}
\par
Similar results arise for $\sigma_{0}$ corresponding to $T_{0}=250$ MeV.
According to these results, the range of $\omega_{0}$ does not vary too much by increasing $\mu_{5}^{(0)}$ from $\mu_{5}^{(0)}=50$ to $\mu_{5}^{(0)}=500$ MeV. Let us notice that, according to the definition of $\cos\delta_{0}=\ell\left(1+\alpha_{0}^{2}\right)^{1/2}$, we have $\ell=+1$ for $\delta_{0}\in\left(-\frac{\pi}{2}, +\frac{\pi}{2}\right)$ and $\ell=-1$ for $\delta_{0}\in\left(\frac{\pi}{2},\frac{3\pi}{2}\right)$. Moreover, as it turns out from Fig. \ref{fig-1}, in the intervals $\delta_{0}\in\left(-\frac{\pi}{2}, +\frac{\pi}{2}\right)$, $\omega_{0}$ is negative, while for $\delta_{0}\in\left(\frac{\pi}{2},\frac{3\pi}{2}\right)$, $\omega_{0}$ turns out to be positive. Hence, the product $\Omega_{0}=\ell\omega_{0}$ remains negative $\forall \delta_{0}$. In \cite{shokri-1}, we worked with positive and negative values of $\Omega_{0}$, and showed that for $\Omega_{0}>0$, the electric field becomes negative and thus unphysical. The above results confirm this observation in the case of nonvanishing $\alpha_{E}$.
\par
To study the effect of nonvanishing electric and magnetic susceptibilities on $\omega_{0}$, we have plotted in Fig. \ref{fig-2} the angular velocity $\omega_{0}$ as a function of $\delta_{0}$ for
\begin{eqnarray}\label{D3}
\begin{array}{ccrclcl}
\mbox{CS-1:}&&\{\chi_{e},\chi_{m}\}&=&\{0,0\},&&\mbox{thick solid curve},\nonumber\\
\mbox{CS-2:}&&\{\chi_{e},\chi_{m}\}&=&\{0.01,+0.2\},&&\mbox{thin solid curve},\nonumber\\
	\mbox{CS-3:}&&\{\chi_{e},\chi_{m}\}&=&\{0.01,-0.2\},&&\mbox{dashed curve,}\nonumber\\
	\end{array}\nonumber\\
	\end{eqnarray}
	\begin{table*}[t]
		\centering
		\begin{tabular}{|c|c|c|c|c|c||c|c|c|c|}
			\hline\hline
			\multicolumn{2}{|c|}{}&\multicolumn{4}{|c||}{Angles for}&\multicolumn{4}{|c|}{Angles for}\\
			\multicolumn{2}{|c|}{}&\multicolumn{4}{|c||}{$T_{0}=250~\mbox{MeV}$}
			&\multicolumn{4}{|c|}{$T_{0}=500~\mbox{MeV}$}\\
			\multicolumn{2}{|c|}{}&\multicolumn{2}{|c|}{$\mu_{5}^{(0)}=50~\mbox{MeV}$}&
			\multicolumn{2}{|c||}{$\mu_{5}^{(0)}=500~\mbox{MeV}$}&\multicolumn{2}{|c|}{$\mu_{5}^{(0)}=50~\mbox{MeV}$}&
			\multicolumn{2}{|c|}{$\mu_{5}^{(0)}=500~\mbox{MeV}$}
			\\
			&&$\chi_{e}=0$ &$\chi_{e}=0.01$ &$\chi_{e}=0$ &$\chi_{e}=0.01
			$ &$\chi_{e}=0$ &$\chi_{e}=0.01$ &$\chi_{e}=0$ &$\chi_{e}=0.01$
			\\
			$\ell$&$\omega_{0}$&$\chi_{m}=0$&$\chi_{m}=+0.2$&$\chi_{m}=0$&$
			\chi_{m}=+0.2$&$\chi_{m}=0$&$\chi_{m}=+0.2$&$\chi_{m}=0$&$
			\chi_{m}=+0.2$\\
			\hline
			$+1$&$-0.045$&$52.7^{\circ}$&$63.1^{\circ}$&$58.7^{\circ}$&$66.4^{\circ}$&
			$57.3^{\circ}$&$70.7^{\circ}$
			&$64.4^{\circ}$&$-$
			\\
			&&$86.9^{\circ}$&$86.0^{\circ}$&$86.5^{\circ}$&$85.4^{\circ}$&$83.2^{\circ}$&$78.9^{\circ}$
			&$81.5^{\circ}$&$-$
			\\
			$+1$&$-0.1$&$88.7^{\circ}$&$88.5^{\circ}$&$88.7^{\circ}$&$88.3^{\circ}$
			&$87.4^{\circ}$&$86.7^{\circ}$&$87.3^{\circ}$&$86.6^{\circ}$
			\\
			$-1$&$+0.1$&$268.7^{\circ}$&$268.4^{\circ}$&$268.8^{\circ}$&$268.5^{\circ}$
			&$267.5^{\circ}$&$267.2^{\circ}$
			&$267.6^{\circ}$&$266.9^{\circ}$
			\\
			$-1$&$+0.045$&$231.3^{\circ}$&$242.3^{\circ}$&$224.8^{\circ}$&$239.1^{\circ}$
			&$235.8^{\circ}$&$249.4^{\circ}$&$228.9^{\circ}$&$244.7^{\circ}$
			\\
			&&$267.1^{\circ}$&$266.5^{\circ}$&$267.4^{\circ}$&$266.5^{\circ}$&
			$263.4^{\circ}$&$259.4^{\circ}$
			&$264.4^{\circ}$&$261.4^{\circ}$
			\\
			\hline\hline
		\end{tabular}
		\caption{The angles $\delta_{0}$ corresponding to $\omega_{0}=\pm 0.045,\pm 0.01$ are listed for $T_{0}=250,500$ MeV, $\mu_{5}^{(0)}=50,500$ MeV and $\{\chi_{e},\chi_{m}\}=\{0,0\}$ and $\{\chi_{e},\chi_{m}\}=\{0.01,+0.2\}$. The values of $\delta_{0}$ for each fixed value of negative and positive $\omega_{0}$ are in the $\delta_{0}\in \left(0,\frac{\pi}{2}\right)$ and $\delta_{0}\in \left(\pi,\frac{3\pi}{2}\right)$ quadrants, respectively. The solutions for the other two quadrants are not presented here. It turns out that different properties of the medium, such as $T_{0},\beta_{0},\sigma_{0}, \mu_{5}^{(0)}, \chi_{e}$ and $\chi_{m}$, affect $\delta_{0}$ for each fixed value of $\omega_{0}$.}\label{table-1}
	\end{table*}
and $\sigma_{0}$ corresponding to $T_{0}=500$ MeV as well as $\mu_{5}^{(0)}=500$ MeV in two intervals $\delta_{0}\in\left(-\frac{\pi}{2},+\frac{\pi}{2}\right)$ [Fig. \ref{fig-2}(a)] and $\delta_{0}\in\left(\frac{\pi}{2},\frac{3\pi}{2}\right)$ [Fig. \ref{fig-2}(b)]. In both intervals of $\delta_{0}$, the curves for nonvanishing susceptibilities (CS-2 and CS-3) are slightly shifted relative to the case of vanishing $\chi_{e}$ and $\chi_{m}$ (CS-1). Let us notice that two cases CS-2 and CS-3 correspond to para- and diamagnetic fluids with $\chi_{m}>0$ and $\chi_{m}<0$, respectively.
\par
In what follows, we pick up a number of positive and negative $\omega_{0}=\pm 0.045,\pm 0.1$, and determine ${\cal{M}},{\cal{N}}$ and ${\cal{L}}$ using the method described before. To see more directly which $\delta_{0}$ corresponds to these $\omega_0$, we use the information arising in Fig. \ref{fig-1}, together with the corresponding results to $T_{0}=250$ MeV, and present in Table \ref{table-1} a list of initial angles $\delta_{0}$ corresponding to these $\omega_{0}$s for the case $\mu_{5}^{(0)}=50$ and $\mu_{5}^{(0)}=500$ MeV as well as $\{\chi_{e},\chi_{m}\}=\{0,0\}$ and $\{\chi_{e},\chi_{m}\}=\{0.1,+0.2\}$. We are, in particular, interested in $\chi_{m}>0$, because, according to lattice QCD results, the QGP created at the RHIC and LHC is paramagnetic \cite{latticepara}. Let us notice that for $\ell=+1$ and $\ell=-1$, we present only angles in $\delta_{0}\in \left(0,\frac{\pi}{2}\right)$ and $\delta_0\in \left(\pi,\frac{3\pi}{2}\right)$ quadrants, respectively. In the case of nonvanishing AH coefficient $\kappa_{E}$, $\delta_{0}$ is determined by the initial value of $\alpha_{0}$ through $\tan\delta_{0}=\alpha_{0}$. According to the above scenario, affected by $\delta_{0},\beta_{0},\sigma_{0},\chi_{e}$ and $\chi_{m}$ of the fluid, $\omega_{0}$ is fixed. It has a crucial role, in particular, on the evolution of the axial chemical potential $\mu_{5}$ (or equivalently the CM conductivity $\kappa_{B}$).
\subsubsection{Evolution of $B,E$ and $T$}\label{sec6A2}
\begin{figure}[hbt]
		\includegraphics[width=8cm,height=6.5cm]{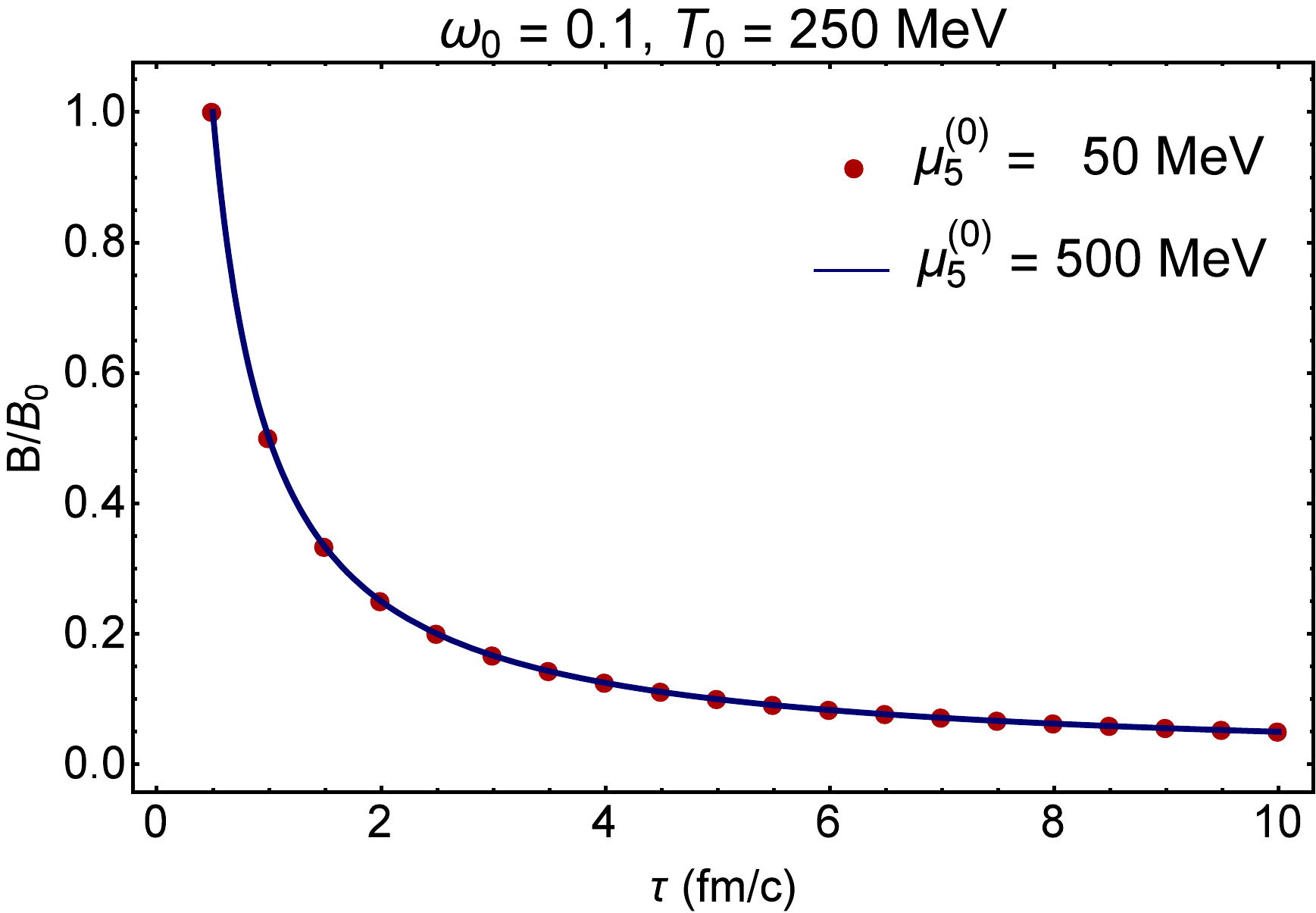}
		\caption{(color online). The $\tau$-dependence of $B/B_{0}$ is demonstrated in the case $\alpha_{E}\neq 0$ for $\omega_{0}=0.1$, $T_{0}=250$ MeV and $\mu_{5}^{(0)}=50$ MeV (red dots) and $\mu_{5}=500$ MeV (black curve). Other parameters are given in (\ref{D1}). As it turns out, the effect of different initial axial chemical potential on the evolution of the magnetic field is negligible.}\label{fig-3}
	\end{figure}

In Fig. \ref{fig-3}, we have plotted $B/B_{0}$ as a function of $\tau\in[0.5,10]$ fm/c for $\omega_{0}=+0.1$ and zero susceptibilities, and compared its evolution for two initial axial chemical potentials $\mu_{5}^{(0)}=50$ MeV (red dots) and $\mu_{5}^{(0)}=500$ MeV (black solid curve). To do this, we first consider (\ref{D2}). Then, plugging $\kappa_{B}^{(0)}=50c$ MeV and $\kappa_{B}^{(0)}=500c$ MeV, as well as $\{\kappa,\tau_{0},\beta_{0}\}$ from (\ref{D1}) and
\begin{eqnarray*}
\{\ell,\omega_{0},\sigma_{0}\}=\{ -1,0.1,8.6~\mbox{MeVc}\},
\end{eqnarray*}
into (\ref{D2}), we determine the corresponding $\alpha_{0}$ to $\mu_{5}^{(0)}=50$ MeV and $\mu_{5}^{(0)}=500$ MeV. We arrive at $\alpha_{0}=45.66$ and $\alpha_{0}=47.03$, respectively. Plugging these quantities into (\ref{S12}), and bearing in mind that $\delta_{0}=\tan^{-1}\alpha_{0}$, we arrive at ${\cal{M}}(\tau)$. This leads to $B/B_{0}$ once the formal ansatz (\ref{E13}) is used. The results presented in Fig. \ref{fig-3} shows that the effect of initial axial chemical potential on the proper time dependence of the magnetic field is negligible. We also plotted $B/B_{0}$ for $T_{0}=500$ MeV with $\sigma_{0}=17.1$, and arrived at the same conclusion. This shows that the effect of different initial electric conductivity on $B/B_{0}$ is negligible.
\par
\begin{figure}[hbt]
		\includegraphics[width=8cm,height=6.5cm]{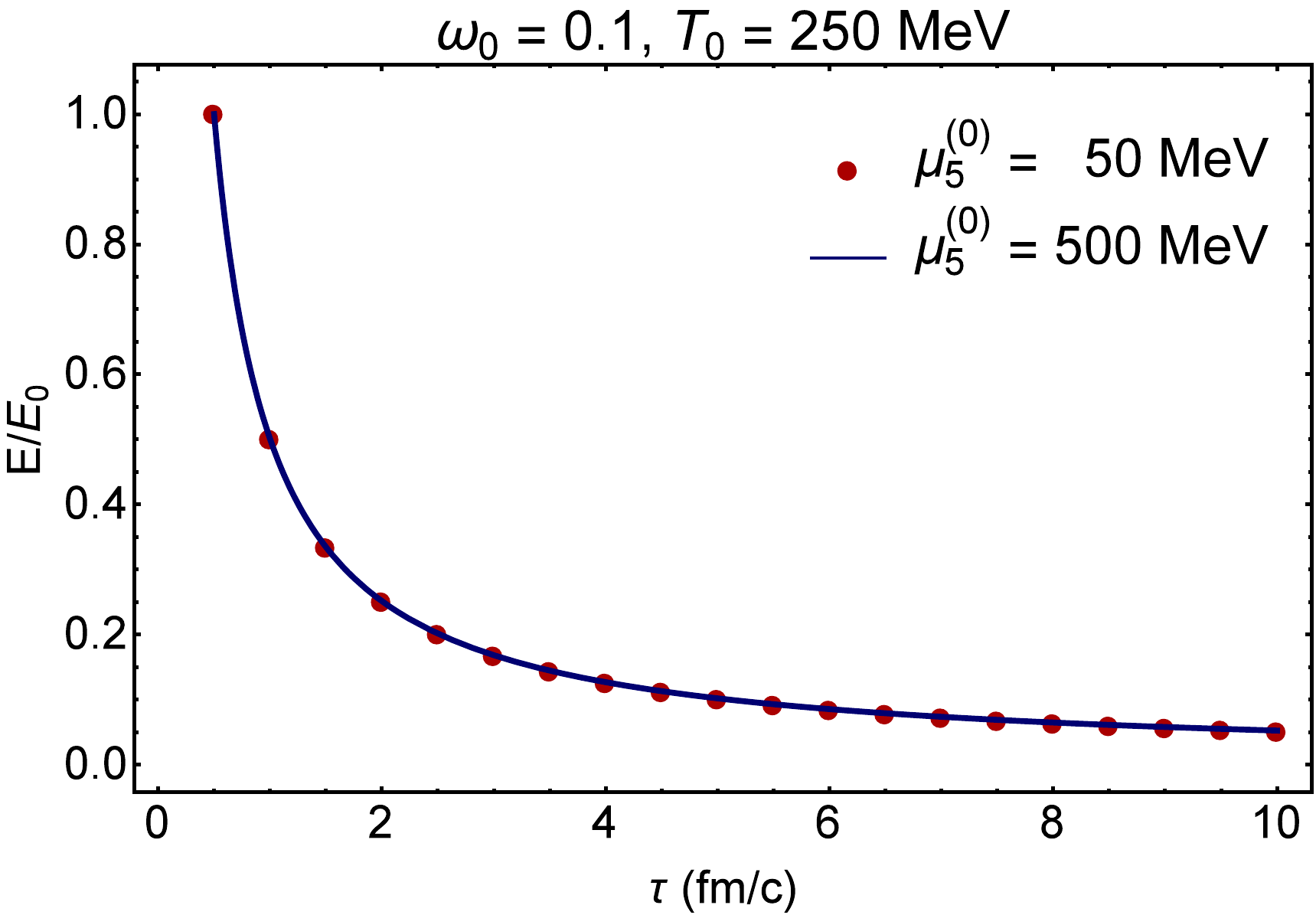}
		\caption{(color online). The $\tau$-dependence of $E/E_{0}$ is demonstrated in the case $\alpha_{E}\neq 0$ for $\omega_{0}=0.1$, $T_{0}=250$ MeV and $\mu_{5}^{(0)}=50$ MeV (red dots) and $\mu_{5}=500$ MeV (black curve). Other parameters are given in (\ref{D1}). It turns out that the effect of different initial axial chemical potential on the evolution of the electric field is negligible. }\label{fig-4}
	\end{figure}

Plugging the same $\alpha_{0}=45.66$ and $\alpha_{0}=47.03$, corresponding to $\mu_{5}^{(0)}=50$ MeV and $\mu_{5}^{(0)}=500$ MeV, together with free parameters (\ref{D1}) into (\ref{S13}), we arrive at ${\cal{N}}$. Using the formal ansatz (\ref{E13}), we then obtain $E/E_{0}$. In Fig. \ref{fig-4}, $E/E_{0}$ is plotted as a function of $\tau\in [0.5,10]$ fm/c for $\omega_{0}=0.1$ and the same axial chemical potentials $\mu_{5}^{(0)}=50$ MeV (red dots) as well as $\mu_{5}^{(0)}=500$ MeV (black solid curve) as above. Similar to the case of $B/B_{0}$, the effect of different initial axial chemical potential on $E/E_{0}$ turns out to be negligible. The same is also true for the effect of initial electric conductivity.
	\par
\begin{figure}[hbt]
			\includegraphics[width=8cm,height=6.5cm]{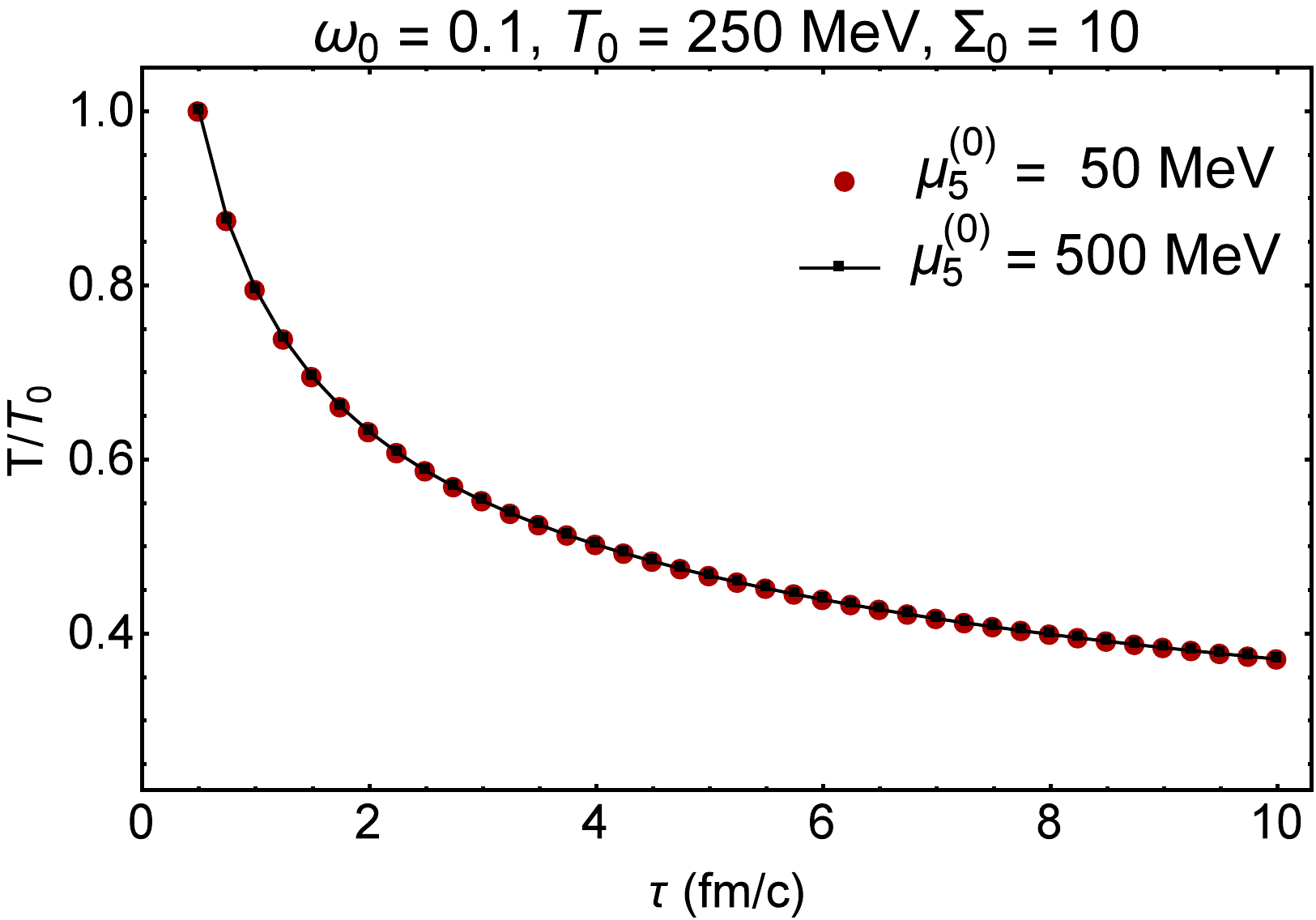}
			\caption{(color online). The $\tau$-dependence of $T/T_{0}$ is demonstrated in the case $\alpha_{E}\neq 0$ for $\omega_{0}=0.1$, $T_{0}=250$ MeV and $\mu_{5}^{(0)}=50$ MeV (red dots) and $\mu_{5}=500$ MeV (black dotted curve). Other parameters are given in (\ref{D1}). Similar to $B$ and $E$, the effect of different initial axial chemical potential on the evolution of the temperature is also negligible.}\label{fig-5}
		\end{figure}

	To determine $T/T_{0}$ from (\ref{E13}), let us consider $e^{{\cal{L}}/\kappa}$ from (\ref{S18}), and define a new parameter $\Sigma_{0}\equiv \frac{B_{0}^{2}}{\epsilon_{0}}$. Plugging also $\sigma(\tau)$ from (\ref{E18}) into (\ref{S18}), and bearing in mind that $\cos\delta=\ell\left(1+\alpha_{E}^{2}\right)^{1/2}$ with $\alpha_{E}$ from (\ref{S3}), we arrive at a more appropriate expression for $e^{{\cal{L}}/\kappa}$. In the case of zero susceptibilities only the first two terms in (\ref{S18}) contribute. Using the same free parameters as in the case of $B/B_{0}$ and $E/E_{0}$ from Figs. \ref{fig-3} and \ref{fig-4} together with $\Sigma_{0}=10$, we arrive at the proper time dependence of $T/T_{0}$. This is demonstrated in Fig. \ref{fig-5}. Similar to previous examples, the effect of different initial axial chemical potential on the evolution of the temperature is negligible. Moreover, as it turns out, different values of initial electric conductivity does not affect the temperature too much.
			\par\noindent
We also checked the effect of negative $\omega_{0}$ on $B,E$ and $T$, and arrived at the same conclusions. Choosing various $\Sigma_{0}$ does not change the results presented in Fig. \ref{fig-5} as well.
\subsubsection{Evolution of $\mu_{5}$}\label{sec6A3}
\begin{figure}[hbt]
	\includegraphics[width=8cm,height=6.5cm]{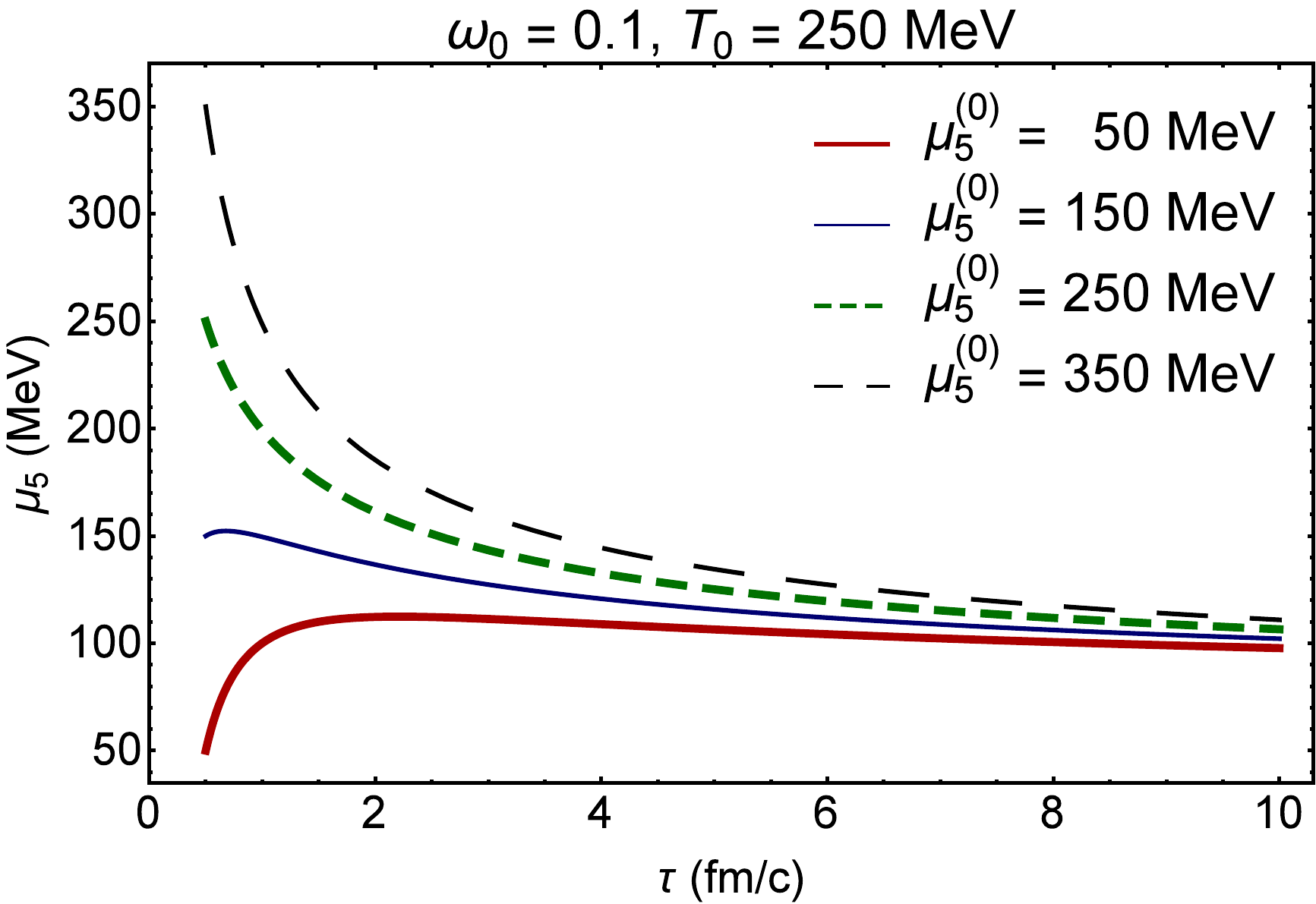}
	\caption{(color online).  The $\tau$-dependence of $\mu_{5}=c^{-1}\kappa_{B}$ is demonstrated in the case $\alpha_{E}\neq 0$ for $T_{0}=250$ MeV and various initial axial chemical potentials $\mu_{5}^{(0)}=50,150,250,350$ MeV, denoted by red thick and blue thin solid curves as well as green thick and black thin dashed curves. Depending on its initial value, $\mu_{5}$ increases or decreases in the first $\tau=2-4$ fm/c, approaches asymptotically to $\mu_{5}\simeq 100-120$ MeV, and remains almost constant afterwards.}\label{fig-6}
\end{figure}

In contrast to $E,B$ and $T$, the proper time dependence of $\mu_{5}$ is strongly affected by the initial axial chemical potential $\mu_{5}^{(0)}$, the angular velocity $\omega_{0}$ and the initial electric conductivity $\sigma_{0}$. Let us first consider the effect of various $\mu_{5}^{(0)}$ on the proper time dependence of $\mu_{5}$ for fixed  $\{\ell,\omega_{0},\sigma_{0},\chi_{e},\chi_{m}\}$. To determine it, we compute the corresponding $\alpha_{0}$ to $\mu_{5}^{(0)}=50,150,250,350$ MeV by plugging
\begin{eqnarray*}
\lefteqn{\{\kappa,\tau_{0},\beta_{0},\ell, \omega_{0},\sigma_{0},\chi_{e},\chi_{m}\}
}\\
&=&\{3,0.5~\mbox{fm/c},0.1,-1,0.1,8.6~\mbox{MeVc},0,0\}
\end{eqnarray*}
and $\kappa_{B}^{(0)}=\mu_{5}^{(0)}c=50c,150c,250c,350c$ MeV into (\ref{D2}).
We arrive at $\alpha_{0}=45.66, 45.96,  46.27,  46.57$ for these $\kappa_{B}^{(0)}$. Using these $\alpha_{0}$ and aforementioned free parameters, we then determine ${\cal{M}}, {\cal{N}}$ and their derivatives with respect to $u$ from (\ref{S12}) and (\ref{S13}). Plugging all these quantities in (\ref{S14}), the proper time dependence of $\kappa_{B}=\mu_{5}c$ is determined. In Fig. \ref{fig-6}, the effect of various initial axial chemical potential $\mu_{5}^{(0)}=50,150,250,350$ MeV on the evolution of $\mu_{5}$ is plotted for $T_{0}=250$ MeV. Depending on its initial value, $\mu_{5}$ either increases (small $\mu_{5}^{(0)}$) or decreases (large $\mu_{5}^{(0)}$) with increasing $\tau$. According to these and several other results with different $\omega_{0}$ and $T_{0}$, $\mu_{5}$ approaches asymptotically to a certain value $\mu_{5}\simeq 100-120$ MeV at $\tau=2-4$ fm/c, and remains almost constant afterwards. This can be interpreted as the production of an approximately constant  CM current, independent of the initial value of $\mu_{5}$, at $\tau\geq 6$ fm/c.
\par
In a more realistic model, where the pressure $p$ and the axial chemical potential $\mu_{5}$ are related,\footnote{In our Bjorkenian setup, $p/p_{0}= \left(\tau_{0}/\tau\right)^{1+1/\kappa}$. Together with $T/T_0=\left(\tau_{0}/\tau\right)^{1/\kappa}$ and $\kappa=3$, it thus leads to $p\propto T^{4}$. In this setup $\mu_{5}$ does not appear in $p$.} it is possible to relate a finite change in $\mu_{5}$ to a difference in the axial number density $n_{5}$, using $n_{5}=\frac{\partial p}{\partial\mu_{5}}$. As we have demonstrated in Fig. \ref{fig-6}, at $\tau\sim 2$ fm/c, $\mu_{5}$ increases from $\mu_{5}^{(0)}=50$ MeV to $\mu_{5}\sim 100$ MeV, and decreases from $\mu_{5}^{(0)}=350$ MeV to $\mu_{5}\sim 200$ MeV, respectively. On the other hand, according to Fig. \ref{fig-5}, at the same $\tau\sim 2$ fm/c, the temperature decreases nearly $40\%$ from $T_{0}=250$ MeV to $T\sim 150$ MeV. Using
$$
p\left(T,\mu_{5}\right)=\frac{g_{\mbox{\tiny{QGP}}}\pi^{2}}{90}T^{4}+\frac{N_{c} N_{f}}{6}\mu_{5}^{2}T^{2}+\frac{N_{c}N_{f}}{12\pi^{2}}\mu_{5}^{4},
$$
from \cite{hirono2014},\footnote{We neglect $\mu$ in $p$ from \cite{hirono2014}.} with $g_{\mbox{\tiny{QGP}}}=g_{gl}+\frac{7}{8}g_{q}$ the number of degrees of freedom with $g_{gl}=(N_{c}^{2}-1)N_{s}$ and $g_{q}=2N_{c}N_{f}N_{s}$ and $N_{f}=3,N_{c}=3$ as well $N_{s}=2$ the number of flavors, colors as well as spin states of quarks and transverse gluons, we arrive for $\mu_{5}^{(0)}=50$ MeV and $\mu_{5}^{(0)}=350$ MeV to $\Delta n_{5}\sim \left(0.13~\mbox{GeV}\right)^{3}$ and $\Delta n_{5}\sim \left(0.29~\mbox{GeV}\right)^{3}$, respectively.
In general, defining $\mu_{5}=\mu_{R}-\mu_{L}$ as being the difference of right- and left-handed chemical potential, $\Delta\mu_{5}>0$ and $\Delta\mu_{5}<0$ are related to a chirality flip in favor of right- and left-handed quarks, respectively.
\par
\begin{figure}[hbt]
			\includegraphics[width=8cm,height=6.5cm]{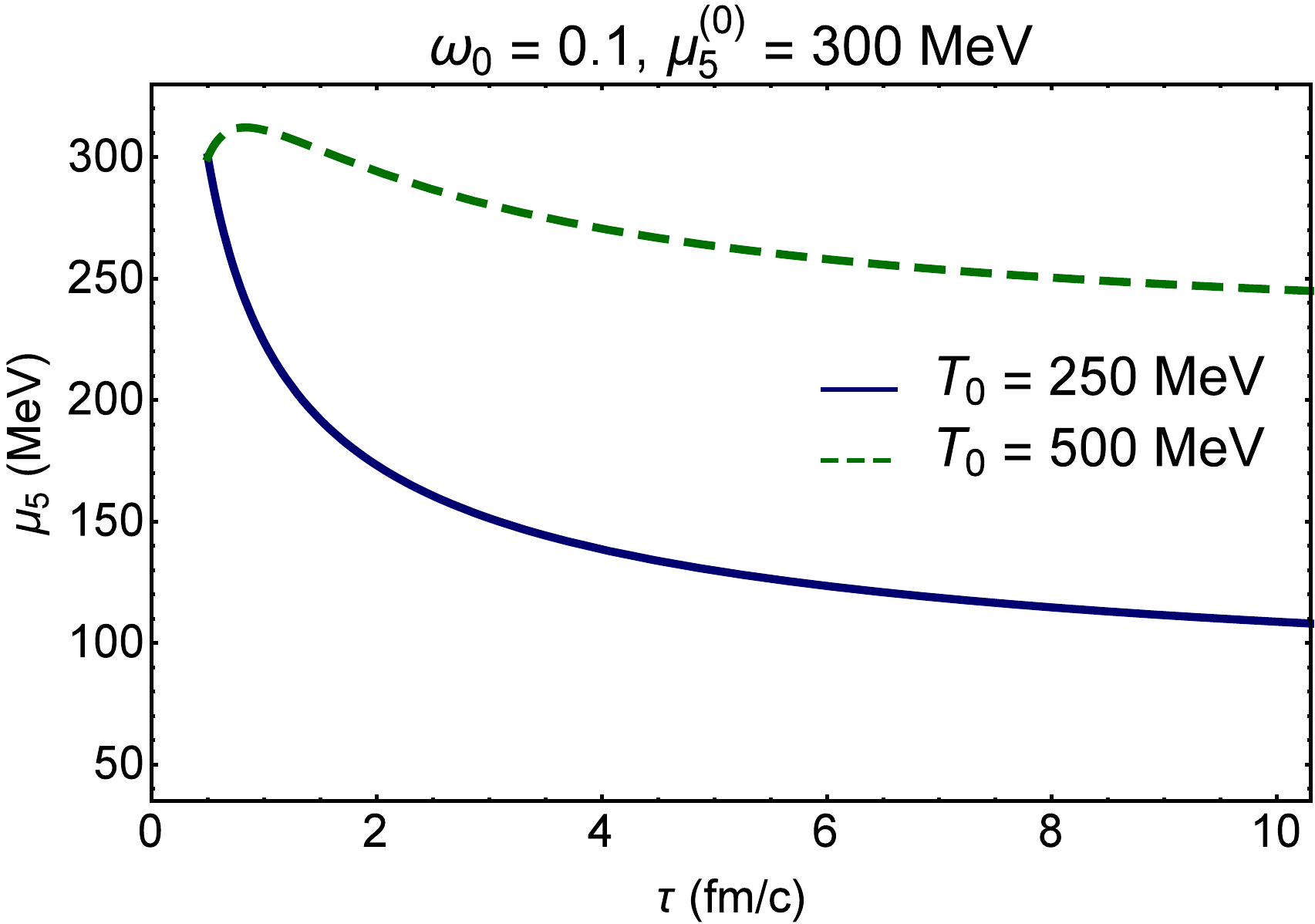}
			\caption{(color online). The $\tau$-dependence of $\mu_{5}$ is plotted for $\{\ell=-1,\omega_{0}=0.1\}$ and $\mu_5^{(0)}=300$ MeV at two different temperatures $T_{0}=250$ MeV (blue solid curve) and $T_{0}=500$ MeV (green dashed curve), giving rise to $\sigma_{0}=8.6$ MeVc and $\sigma_{0}=17.1$ MeVc, respectively. It turns out that the axial chemical potential $\mu_{5}$, or equivalently CM conductivity $\kappa_{B}$, decays slower for larger values of $T_{0}$, or equivalently larger initial electric conductivity of the medium.}\label{fig-7}
		\end{figure}

The evolution of $\mu_{5}$ is also affected by the initial value of the electric conductivity, $\sigma_{0}$. Following the procedure described above, we determine $\mu_{5}$ for $\omega_{0}=0.1$ and $\mu_{5}^{(0)}=300$ MeV at two different initial temperatures $T_{0}=250$ MeV and $T_{0}=500$ MeV, giving rise to $\sigma_{0}=8.6$ MeVc and $\sigma_{0}=17.1$ MeVc, respectively. The $\tau$-dependence of the corresponding $\mu_{5}$ is plotted in Fig. \ref{fig-7}. Here, blue solid and green dashed curves correspond to $T_{0}=250$ MeV and $T_{0}=500$ MeV. According to this result, the axial chemical potential $\mu_{5}$, or equivalently the CM conductivity $\kappa_{B}$, decays slower for larger initial temperatures, or equivalently larger initial electric conductivities of the medium. We have repeated this computation for various positive $\omega_{0}$ as well as initial values of $\mu_{5}$, and arrived at the same conclusion.
\par
\begin{figure}[hbt]
	    	\includegraphics[width=8cm,height=6.5cm]{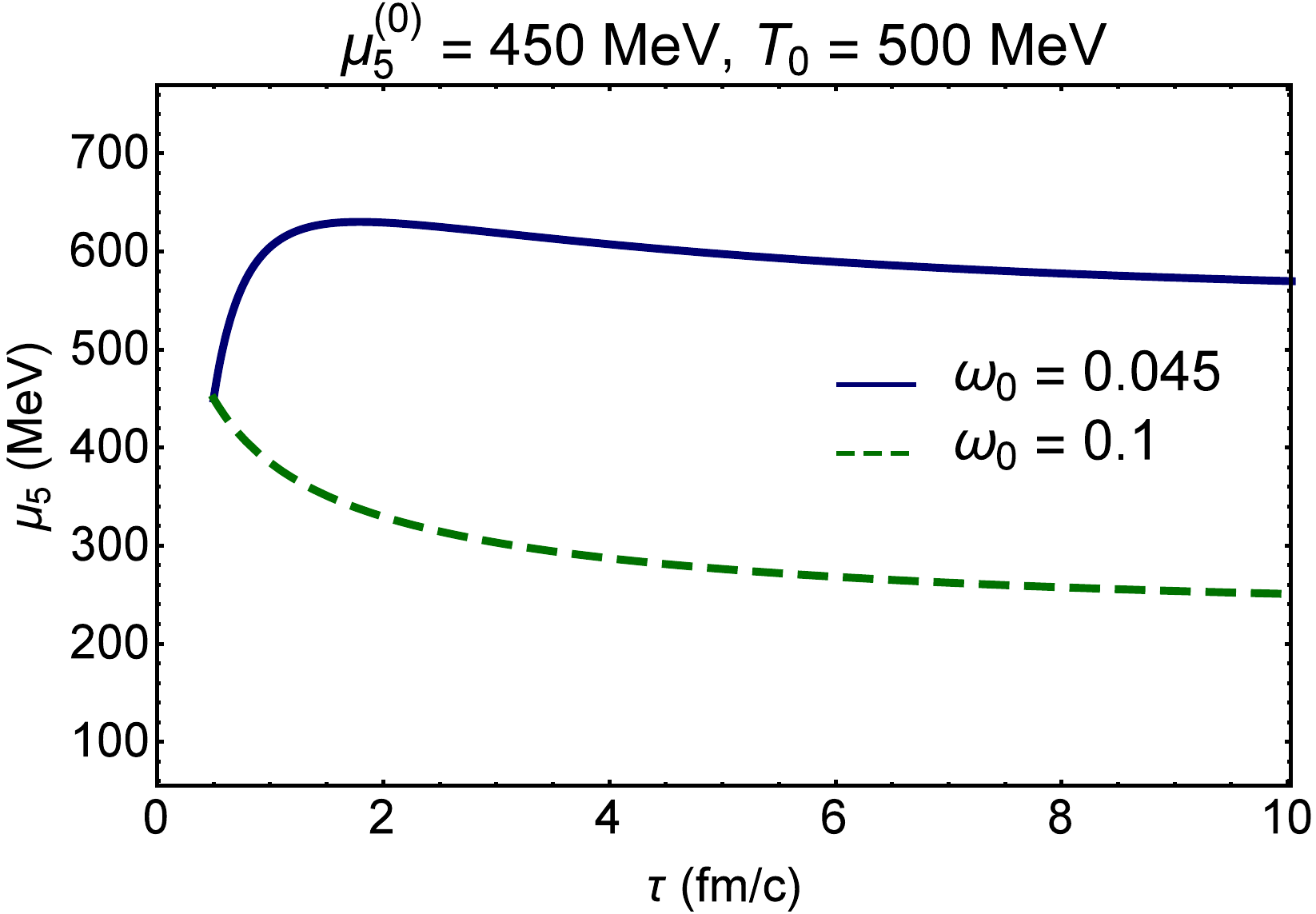}
			\caption{(color online). The $\tau$-dependence of $\mu_{5}$ is plotted for $\ell=-1$ and $\mu_5^{(0)}=450$ MeV and $T_{0}=500$ MeV for two different angular velocity $\omega_{0}=0.045$ (blue solid curve) and $\omega_{0}=0.1$ (green dashed curve). Other free parameters are given in (\ref{D4}). It turns out that the axial chemical potential $\mu_{5}$ decays slower for smaller values of $\omega_{0}$.}\label{fig-8}
			\end{figure}

In Fig. \ref{fig-8}, the $\tau$-dependence of $\mu_{5}$ is plotted for two different values of $\omega_{0}=0.045$ (blue solid curve) and $\omega_{0}=0.1$ (green dashed curve). The set of free parameters corresponding to this plot are given by
\begin{eqnarray}\label{D4}
\lefteqn{
\{\kappa,\tau_{0},\beta_{0},\ell,\sigma_{0},\kappa_{B}^{(0)},\chi_{e},\chi_{m}\}
}\nonumber\\
&=&\{3,0.5~\mbox{fm/c},0.1,-1,17.1~\mbox{MeVc}, 450~\mbox{MeV}, 0,0\}.\nonumber\\
\end{eqnarray}
			\par\noindent
According to these results, $\mu_{5}$ decays faster for larger values of positive $\omega_{0}$.
As aforementioned, positive $\omega_{0}$s correspond to initial angles $\delta_{0}$ in the second and third quadrants, i.e. $\delta_{0}\in\left(\frac{\pi}{2},\frac{3\pi}{2}\right)$.
\par
\begin{figure*}[hbt]
              	\includegraphics[width=5.7cm,height=4.5cm]{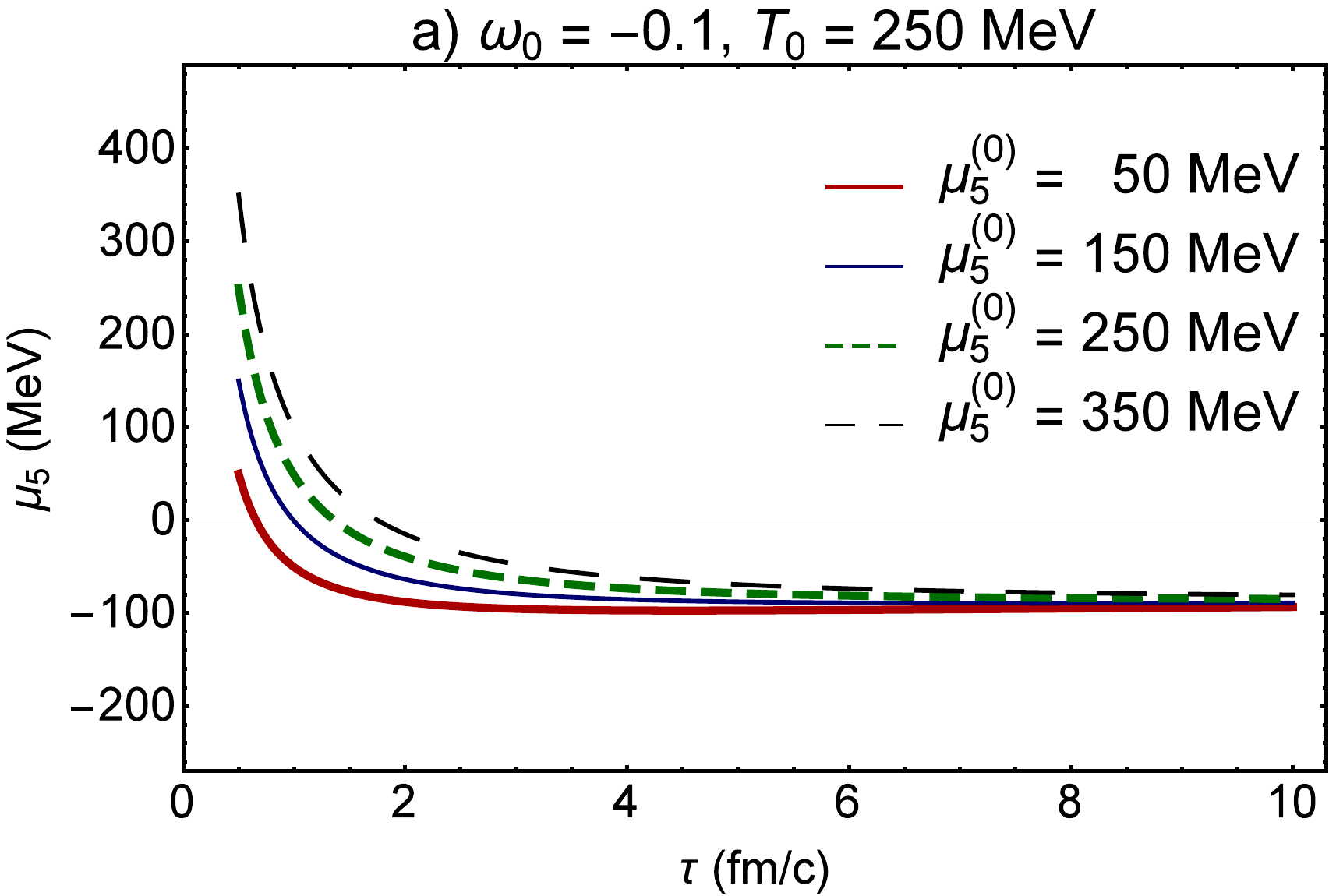}
              	\includegraphics[width=5.7cm,height=4.5cm]{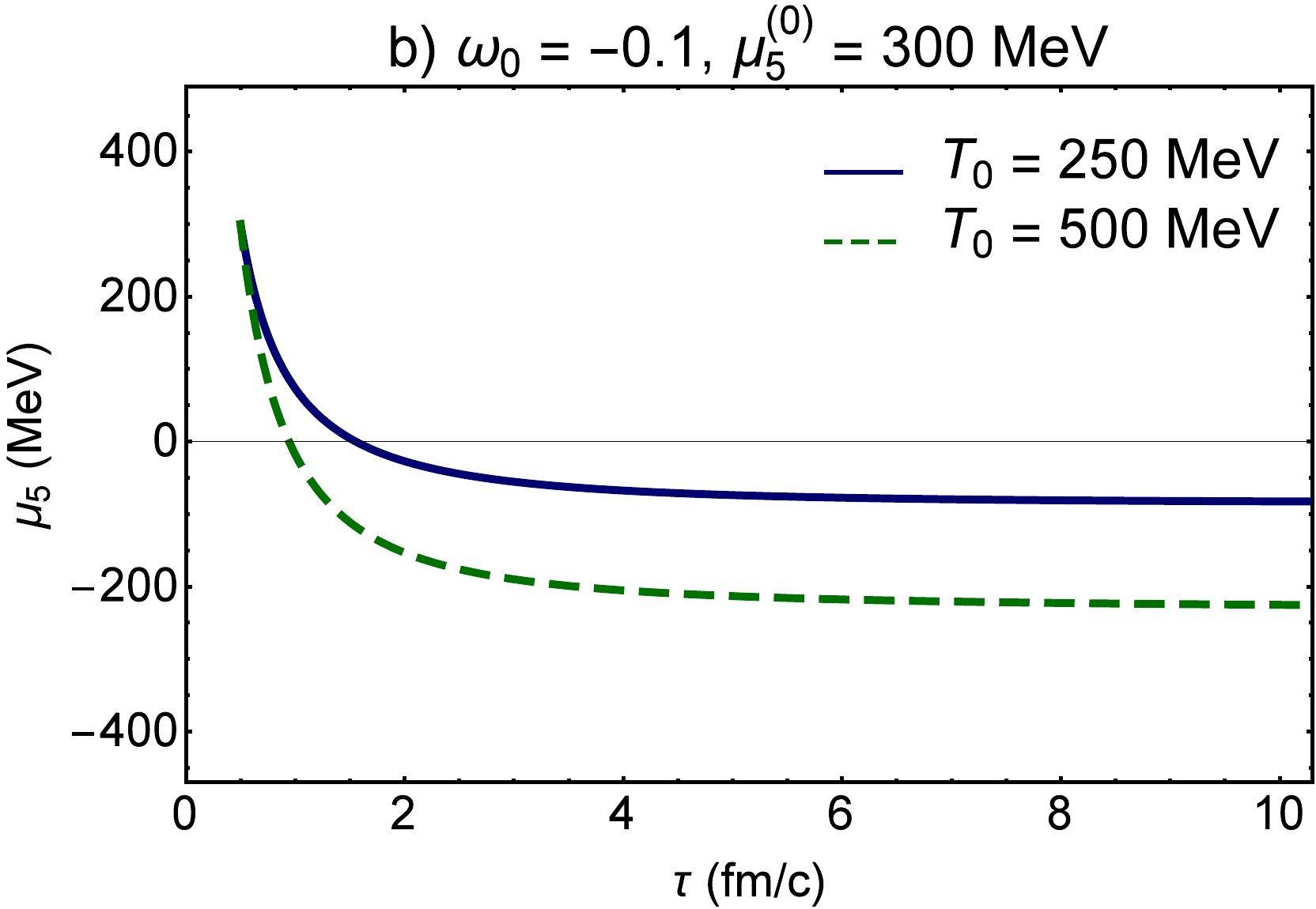}
              	\includegraphics[width=5.7cm,height=4.5cm]{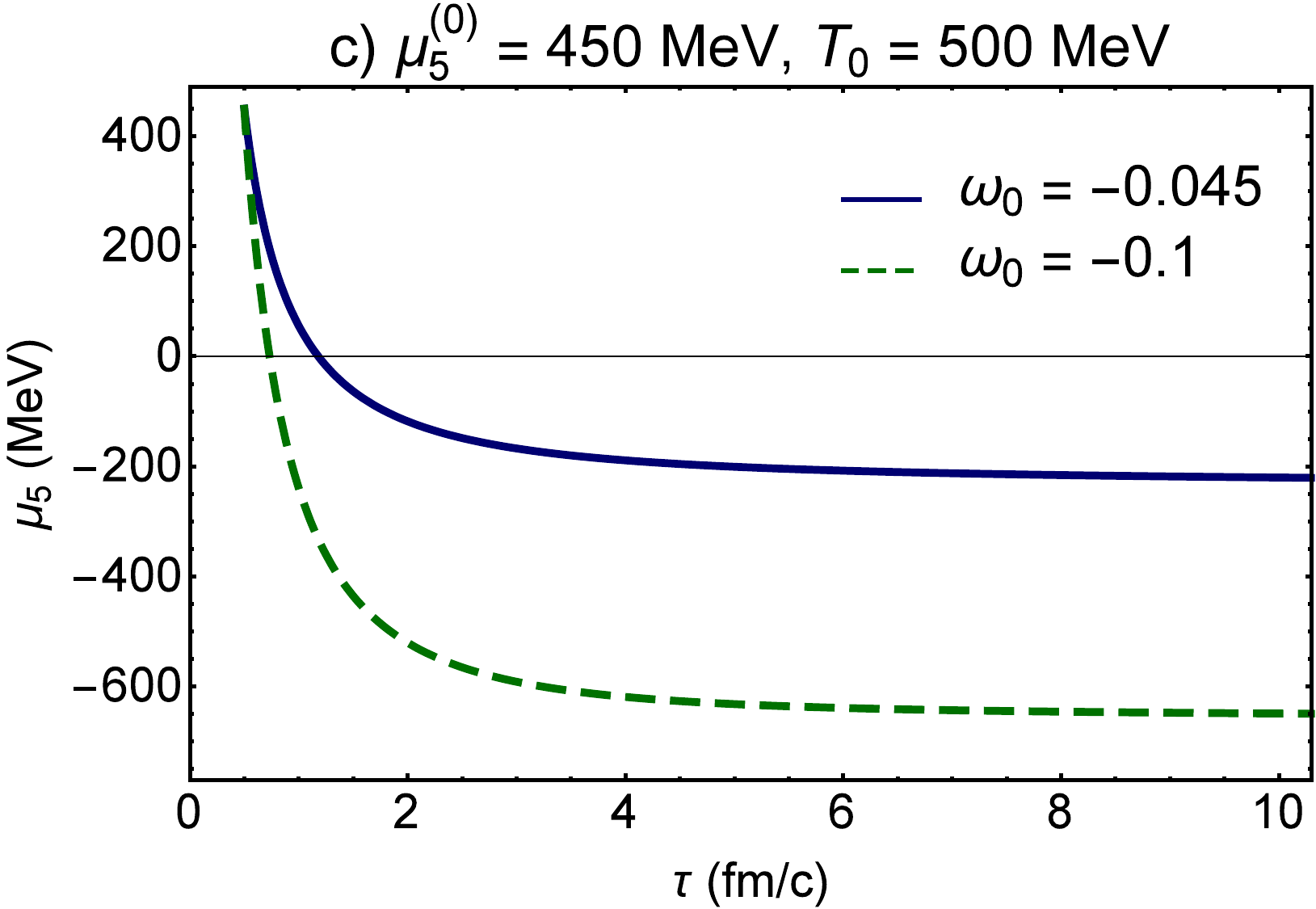}
              	\caption{(color online). Using the set of free parameters (\ref{D5}), we have plotted in panel (a) the $\tau$-dependence of $\mu_{5}$ for fixed  $\omega_{0}=-0.1$, $T_{0}=250$ MeV and $\mu_{5}^{(0)}=50,150,250,350$ MeV. The smaller the initial value of axial chemical potential is, the faster $\mu_{5}$ decays. In panel (b), same free parameters are used, and the $\tau$-dependence of $\mu_{5}$ is plotted for fixed $\omega_{0}=-0.1$, $\mu_{5}^{(0)}=300$ MeV and $T_{0}=250$ MeV (blue solid curve) and $T_{0}=500$ MeV (green dashed curve). As it turns out, $\mu_{5}$ increases faster for larger values of initial electric conductivity $\sigma_{0}$. In panel (c), the $\tau$-dependence of $\mu_{5}$ is plotted for fixed $\mu_{5}^{(0)}=450$ MeV, $T_{0}=500$ MeV and $\omega_{0}=-0.045$ (blue solid curve) and $\omega_{0}=-0.1$ (green dashed curve), using the same set of free parameters. According to these results, $\mu_{5}$ decays much slower for larger values of negative $\omega_{0}$.
                The results demonstrated in panels (a)-(c) for negative values of $\omega_{0}$ are to be compared with the results from Figs. \ref{fig-6}-\ref{fig-8} for positive values of $\omega_{0}$. In contrast to those results, for negative $\omega_{0}$, $\mu_{5}$ changes its sign as time evolves. The sign flip of $\mu_{5}$ can be interpreted as a change in the direction of the CM current which is proportional to $\kappa_{B}=\mu_{5}c$.}\label{fig-9}
              \end{figure*}

 As concerns the effect of negative values of $\omega_{0}$, we have repeated the above computations for negative $\omega_{0}$, and arrived partly at different results. In particular, the conclusions concerning the evolution of $\mu_{5}$ are different from those corresponding to positive $\omega_{0}$. In Fig. \ref{fig-9}, we have chosen negative $\omega_{0}$, and plotted the counterparts of Figs. \ref{fig-6}-\ref{fig-8}. Apart from the free parameters $\{\omega_{0},\mu_{5}^{(0)},T_{0}\}$, which are indicated in the figures, following choice of remaining parameters is made:
  \begin{eqnarray}\label{D5}
  	  	\{\kappa,\tau_{0},\beta_{0},\ell,\chi_{e},\chi_{m}\}=\{3,0.5~\mbox{fm/c},0.1,+1, 0,0\}.\nonumber\\
  \end{eqnarray}
\par\noindent
Let us first compare the results from Fig. \ref{fig-9}(a) with corresponding results from Fig. \ref{fig-6}. As it turns out, for negative $\omega_{0}$, $\mu_{5}$ decreases for all values of $\mu_{5}^{(0)}=50,150,250,350$ MeV. This is in contrast to the evolution of $\mu_{5}$ for positive $\omega_{0}$, demonstrated in Fig. \ref{fig-6}. Moreover, for negative $\omega_{0}$, $\mu_{5}$ decays faster for smaller values of the initial axial chemical potential.  Independent of its initial value, however, $\mu_{5}$ becomes negative for $\tau\gtrsim 2$ fm/c. For smaller values of $\mu_{5}^{(0)}$, $\mu_{5}$'s sign flips earlier than $\tau=2$ fm/c. Bearing in mind that $\mu_{5}=\kappa_{B}c^{-1}$, a sign flip of $\mu_{5}$ can be interpreted as a change in the direction of the CM current.
\par
In Fig. \ref{fig-9}(b), the effect of $T_{0}$, or equivalently $\sigma_{0}$, on the evolution of $\mu_{5}$ is explored for $\omega_{0}=-0.1$ and $\mu_{5}^{(0)}=300$ MeV at $T_{0}=250$ MeV (blue solid curve) and $T_{0}=500$ MeV (green dashed curve). In contrast to the result demonstrated in Fig.  \ref{fig-7} for $\omega_{0}=+0.1$, for $\omega_{0}=-0.1$, $\mu_{5}$ increases faster for larger values of initial electric conductivity $\sigma_{0}$. This can be regarded as one of the differences between effects associated with positive and negative $\omega_{0}$. Apart from this,  whereas the axial chemical potential for $\omega_{0}>0$ remains positive during its evolution, its sign flips for $\omega_{0}<0$. According to the results demonstrated in Fig. \ref{fig-9}(b), the (proper) time at which $\mu_{5}$'s sign is flipped becomes smaller the larger the initial value of electric conductivity $\sigma_{0}$ is.
\par
To study the effect of different negative $\omega_{0}$ on the $\tau$-dependence of $\mu_{5}$, we have plotted $\mu_{5}(\tau)$ in Fig. \ref{fig-9}(c) for fixed value of $\mu_{5}^{(0)}=450$ MeV and $T_{0}=500$ MeV and for two different values of $\omega_{0}=-0.045$ (blue solid curve) and $\omega_{0}=-0.1$ (green dashed curve). In contrast to the results demonstrated in Fig. \ref{fig-8} for positive $\omega_{0}$, it turns out that apart from the fact that for negative $\omega_{0}$ a sign flip of $\mu_{5}$ occurs at an early proper time, the axial chemical potential $\mu_{5}$ decays slower for larger values of negative $\omega_{0}$. Moreover, $\mu_{5}$ decay to larger negative values is more emphasized than for smaller values of $\omega_{0}$.
\par
We finally notice that the above conclusions, arising from the plots demonstrated in Fig. \ref{fig-9}, are independent of the choice of $\omega_{0}, \mu_{5}^{(0)}$ and $T_{0}$ (or $\sigma_{0}$). We have repeated the above computations for the case of nonvanishing electric and magnetic susceptibilities, and arrived at the same qualitative results and conclusions. The interplay between these susceptibilities and the angular velocity $\omega_{0}$, and their effects on the evolution of electromagnetic and hydrodynamic fields are already studied in \cite{shokri-1}.
\subsection{Case 2: Vanishing AH coefficient}\label{sec6B}
                \begin{figure}[hbt]
                	\includegraphics[width=8cm,height=6.5cm]{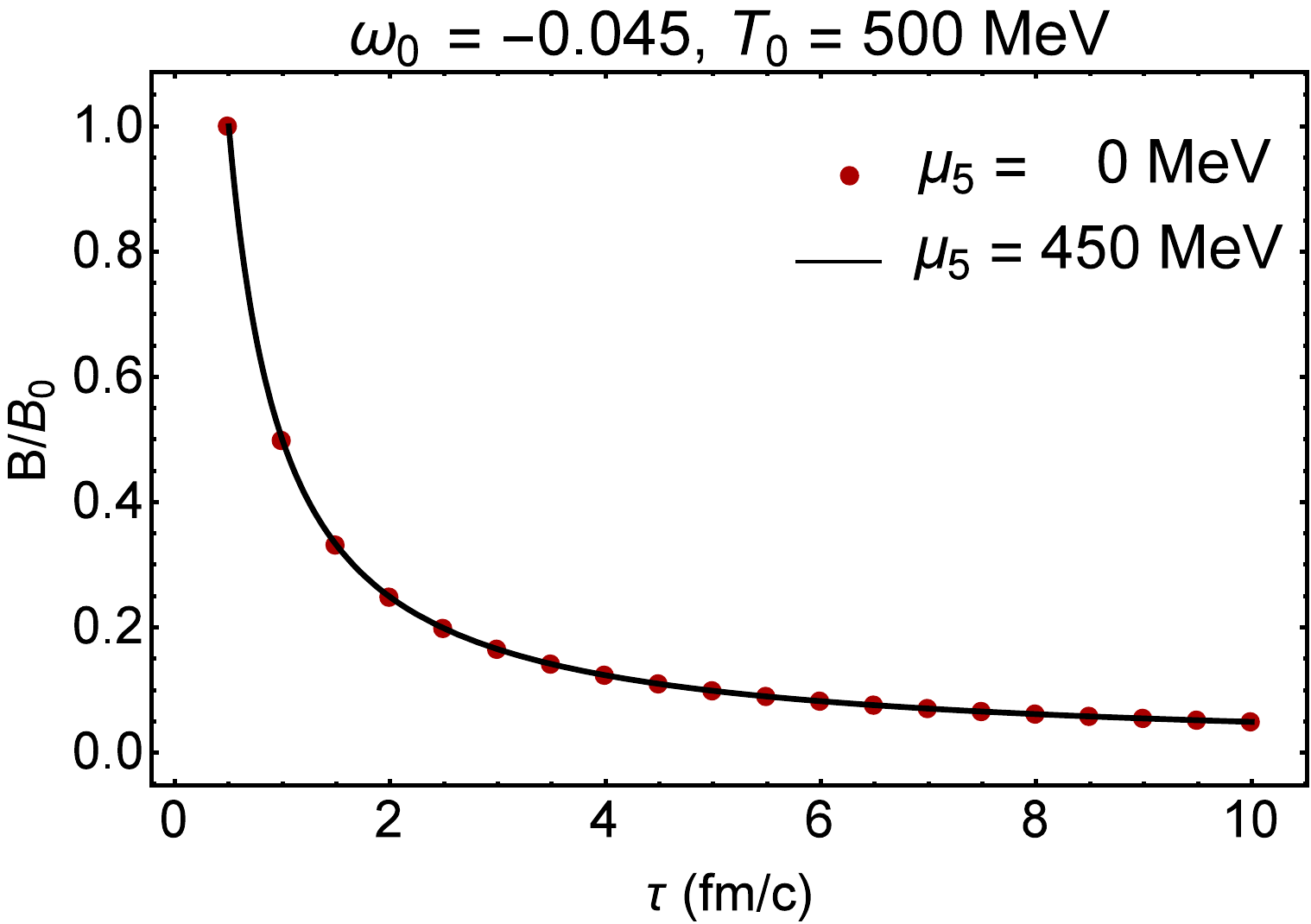}
                	\caption{(color online). The $\tau$-dependence of $B/B_{0}$ is demonstrated in the case of $\alpha_{E}=0$ for the set of free parameters (\ref{D6}) and $\mu_{5}=0,450$ MeV. As it turns out, the effect of axial chemical potential on the evolution of the magnetic field is negligible. }\label{fig-10}
                \end{figure}
                \begin{figure}[hbt]
                	\includegraphics[width=8cm,height=6.5cm]{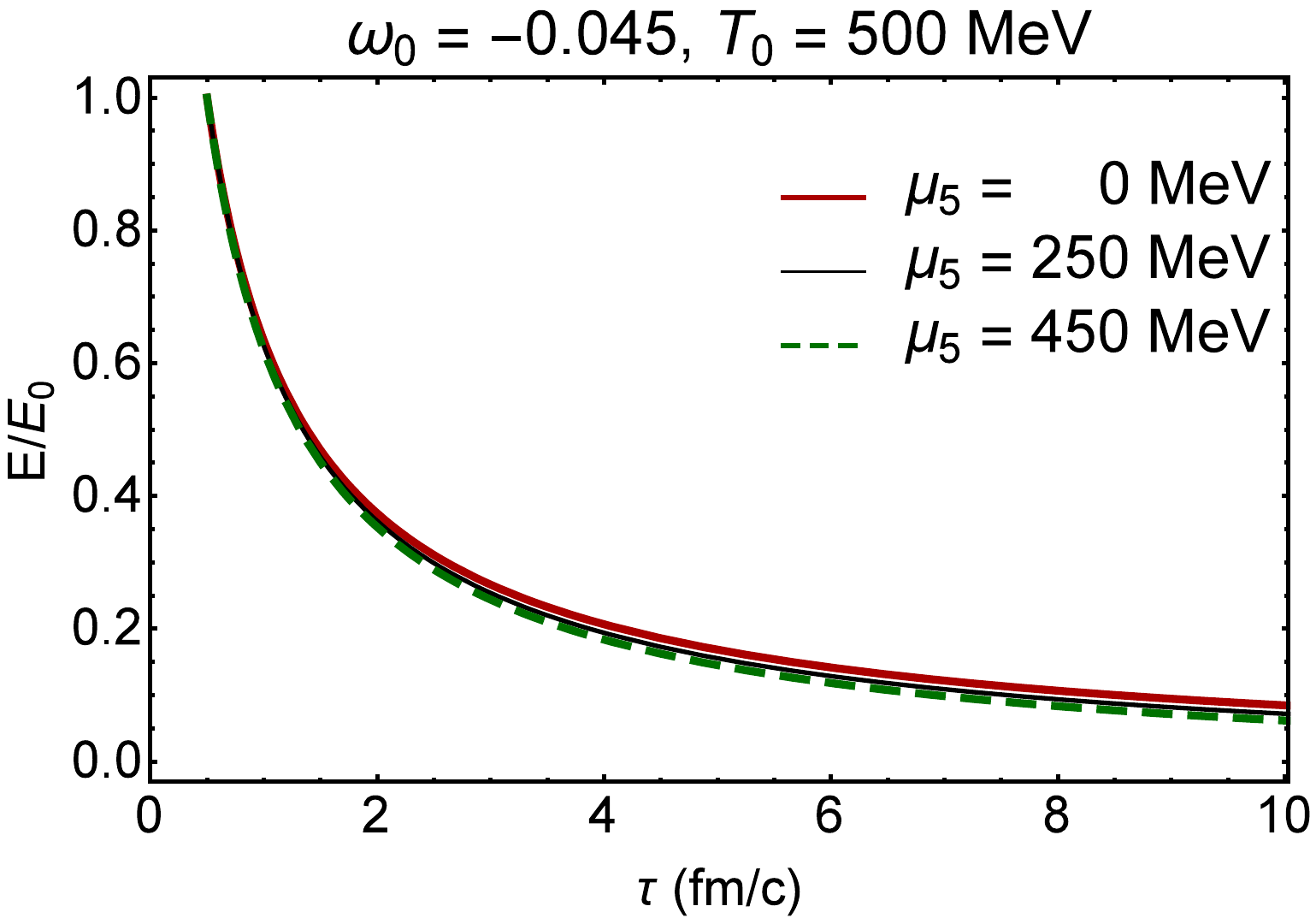}
                	\caption{(color online). The $\tau$-dependence of $E/E_{0}$ is demonstrated in the case of $\alpha_{E}=0$ for $\mu_{5}=0$ (red thick solid curve), $\mu_{5}=250$ MeV (black thin solid curve) as well as $\mu_{5}=450$ MeV (green dashed curve). Here, the set of free parameters (\ref{D6}) is used. In the case of $\omega_{0}<0$, $E/E_{0}$ decays faster for larger values of $\mu_{5}$.}\label{fig-11}
                \end{figure}
                \begin{figure}[hbt]
                	\includegraphics[width=8cm,height=6.5cm]{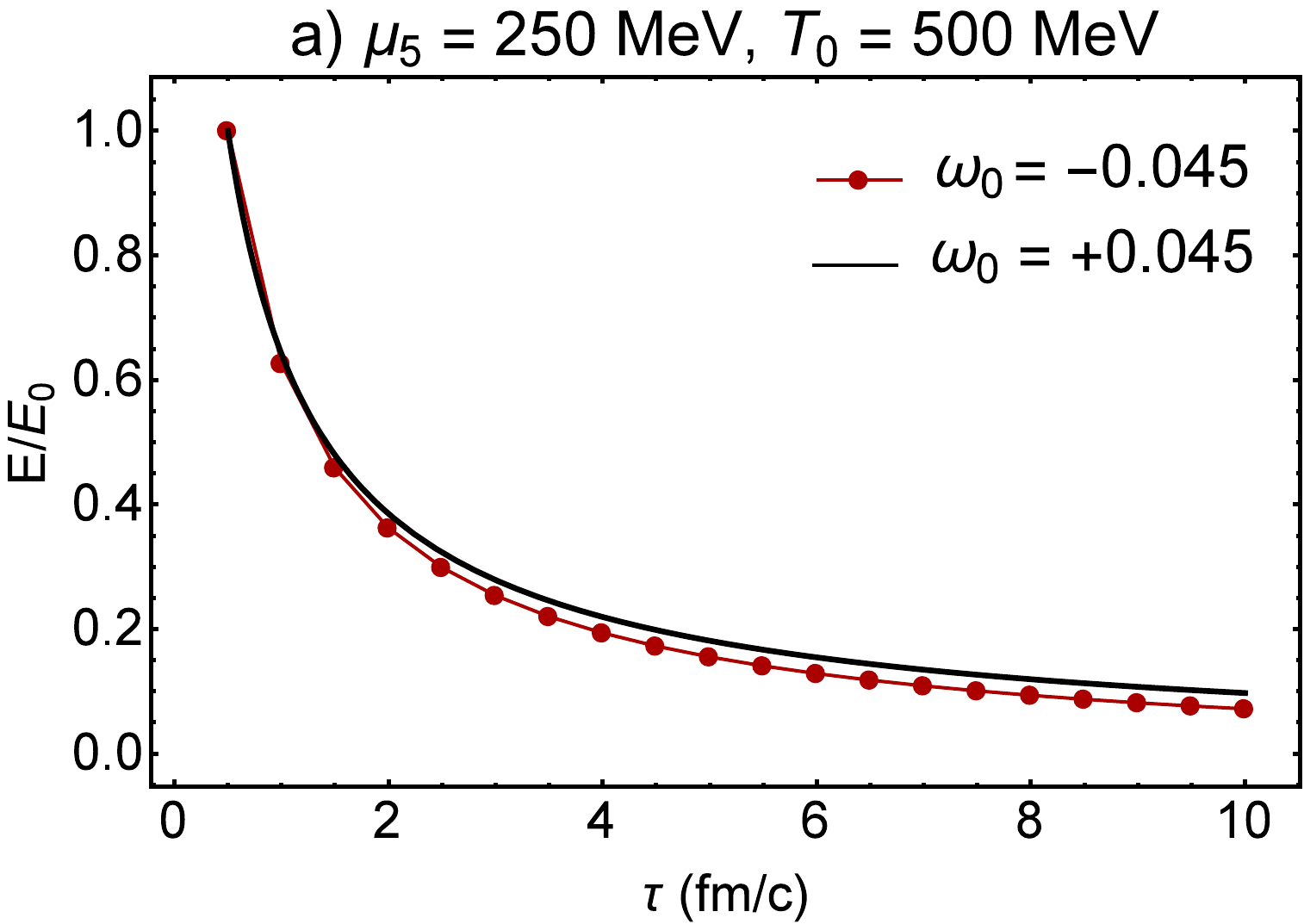}
                	\includegraphics[width=8cm,height=6.5cm]{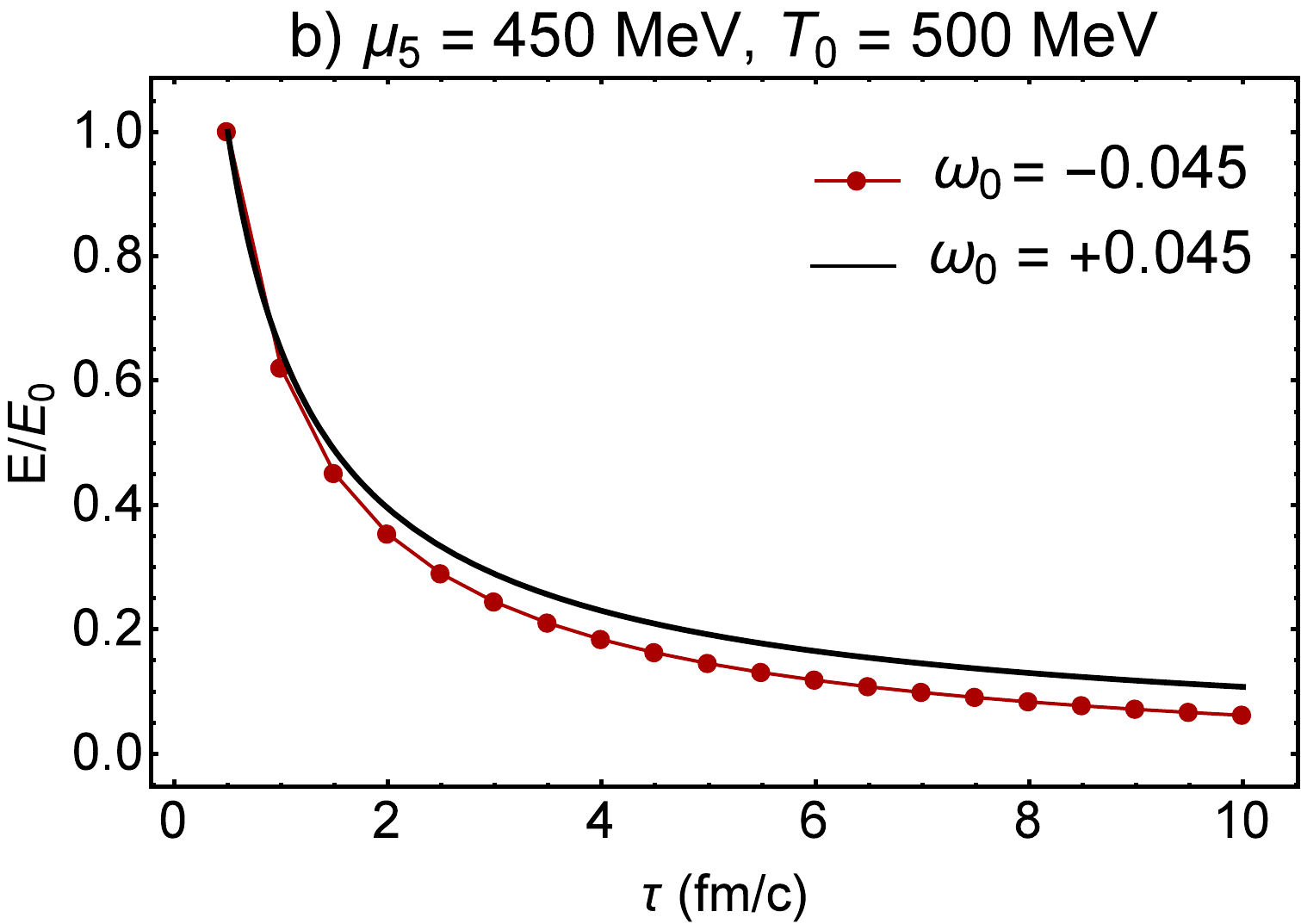}
                	\caption{(color online). The effect of positive and negative angular velocity on the evolution of the electric field is demonstrated. To do this, $\{\ell,\omega_{0}\}$ is chosen to be $\{+1,-0.045\}$ (red dotted curve) and $\{-1,+0.045\}$ (black curve). Panels (a) and (b) correspond to two different axial chemical potential, $\mu_{5}=250$ MeV (panel a) and $\mu_{5}=450$ MeV (panel b). The rest of parameters are given in (\ref{D6}). A comparison between these two panels shows that the difference between the effect of positive and negative $\omega_{0}$ on the decay rate of $E$ increases with increasing $\mu_{5}$.}\label{fig-12}
                \end{figure}
                \begin{figure}
                	\includegraphics[width=8cm,height=6.5cm]{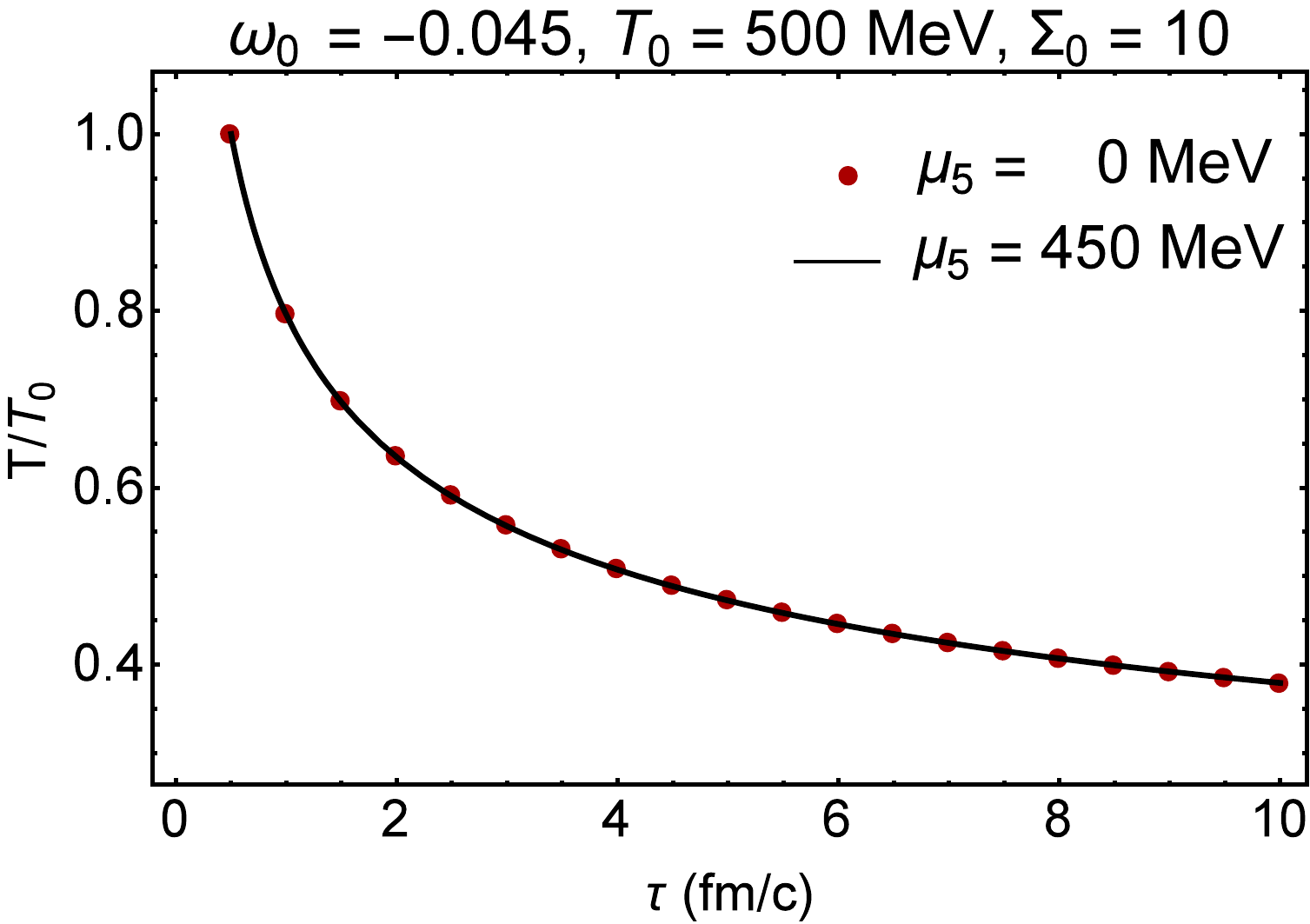}
                	\caption{(color online). The $\tau$-dependence of $T/T_{0}$ is demonstrated in the case $\alpha_{E}=0$ for the set of free parameters (\ref{D6}), $\Sigma_{0}=10$ and $\mu_{5}=450$ MeV. According to these results, the effect of axial chemical potential on the evolution of the temperature is negligible.}\label{fig-13}
                \end{figure}
                \begin{figure}[hbt]
                	\includegraphics[width=8cm,height=6.5cm]{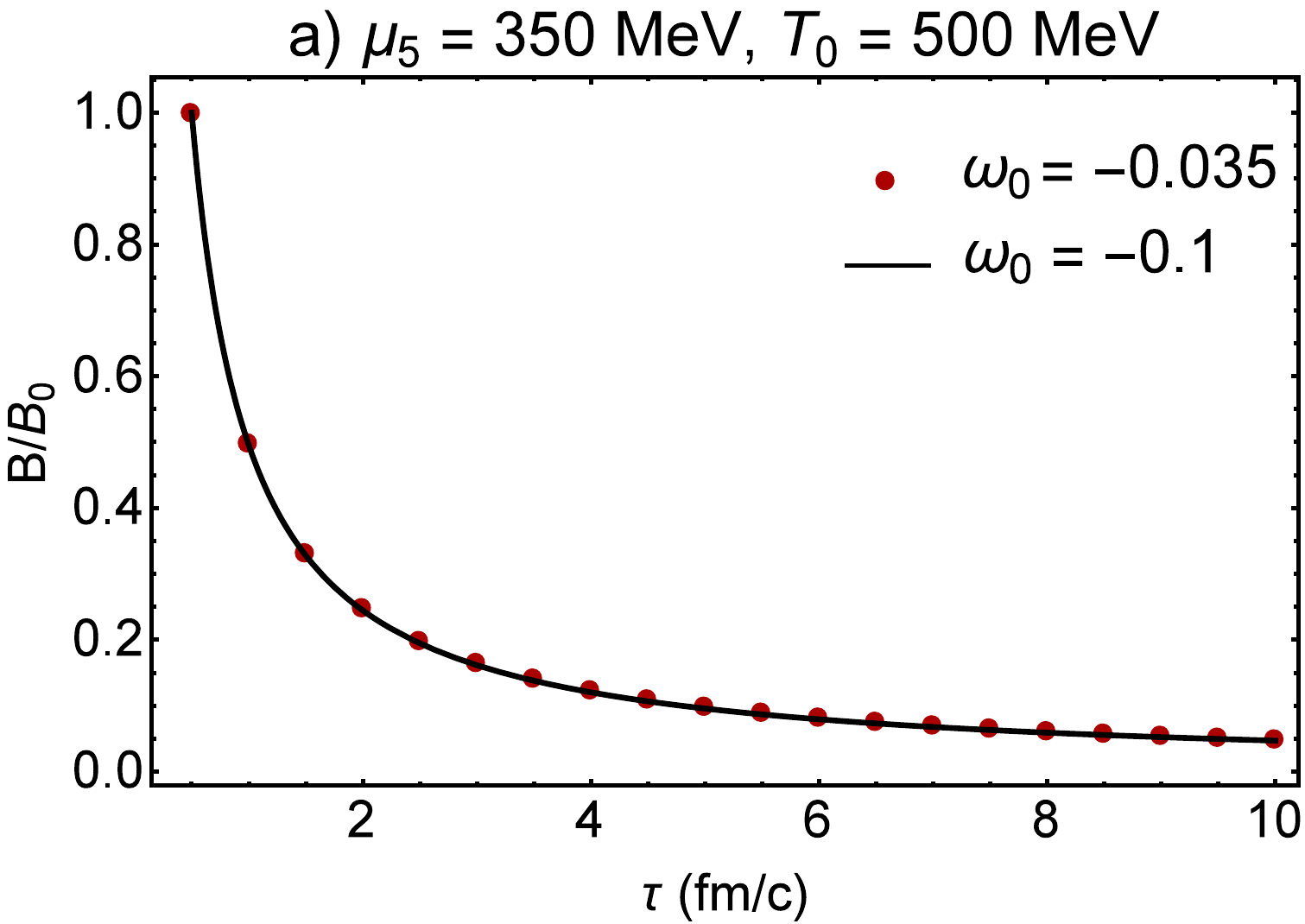}
                	\includegraphics[width=8cm,height=6.5cm]{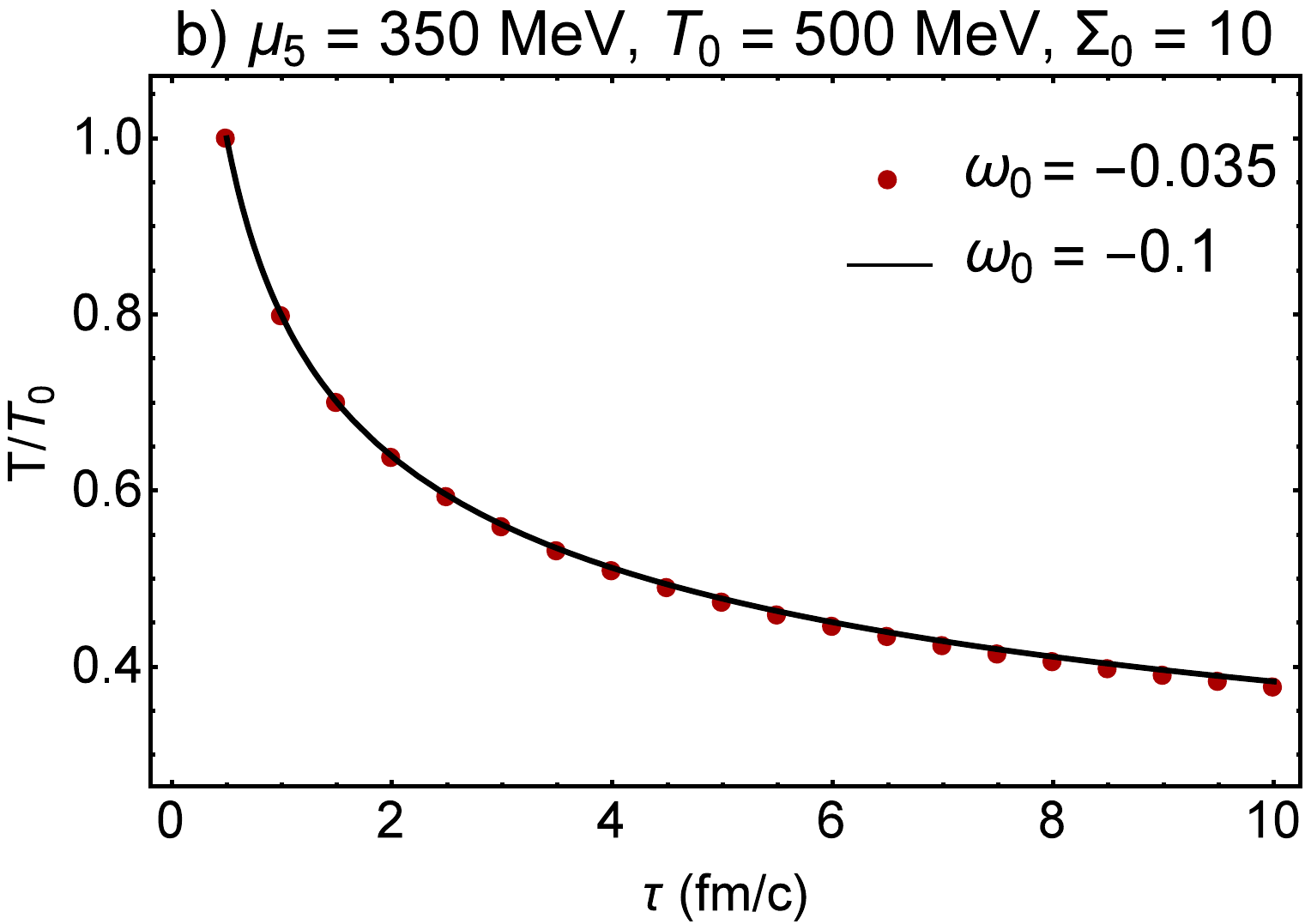}
                	\caption{(color online). (a) The $\tau$-dependence of $B/B_{0}$ is plotted for fixed $\mu_{5}=350$ MeV and different $\omega_{0}=-0.035$ (red dots) and $\omega_{0}=-0.1$ (black curve). (b) The $\tau$-dependence of $T/T_{0}$ is plotted for fixed $\mu_{5}=350$ MeV and different $\omega_{0}=-0.035$ (red dots) and $\omega_{0}=-0.1$ (black curve). It turns out that the evolution  of $B$ and $T$ is not significantly affected by different choices for $\omega_{0}$.}\label{fig-14}
                \end{figure}
                \begin{figure*}[hbt]
                	\includegraphics[width=8cm,height=6.5cm]{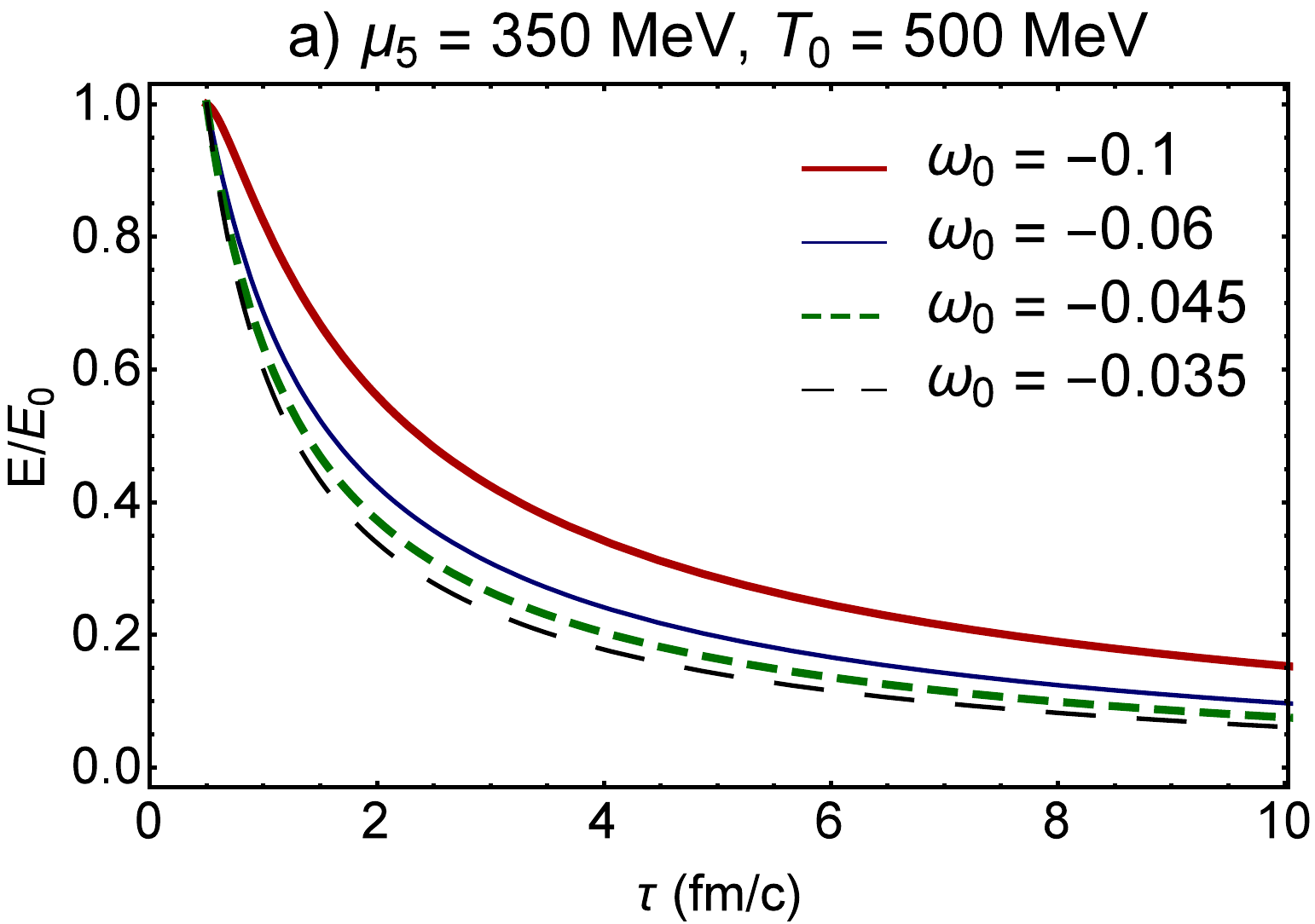}
                	\includegraphics[width=8cm,height=6.5cm]{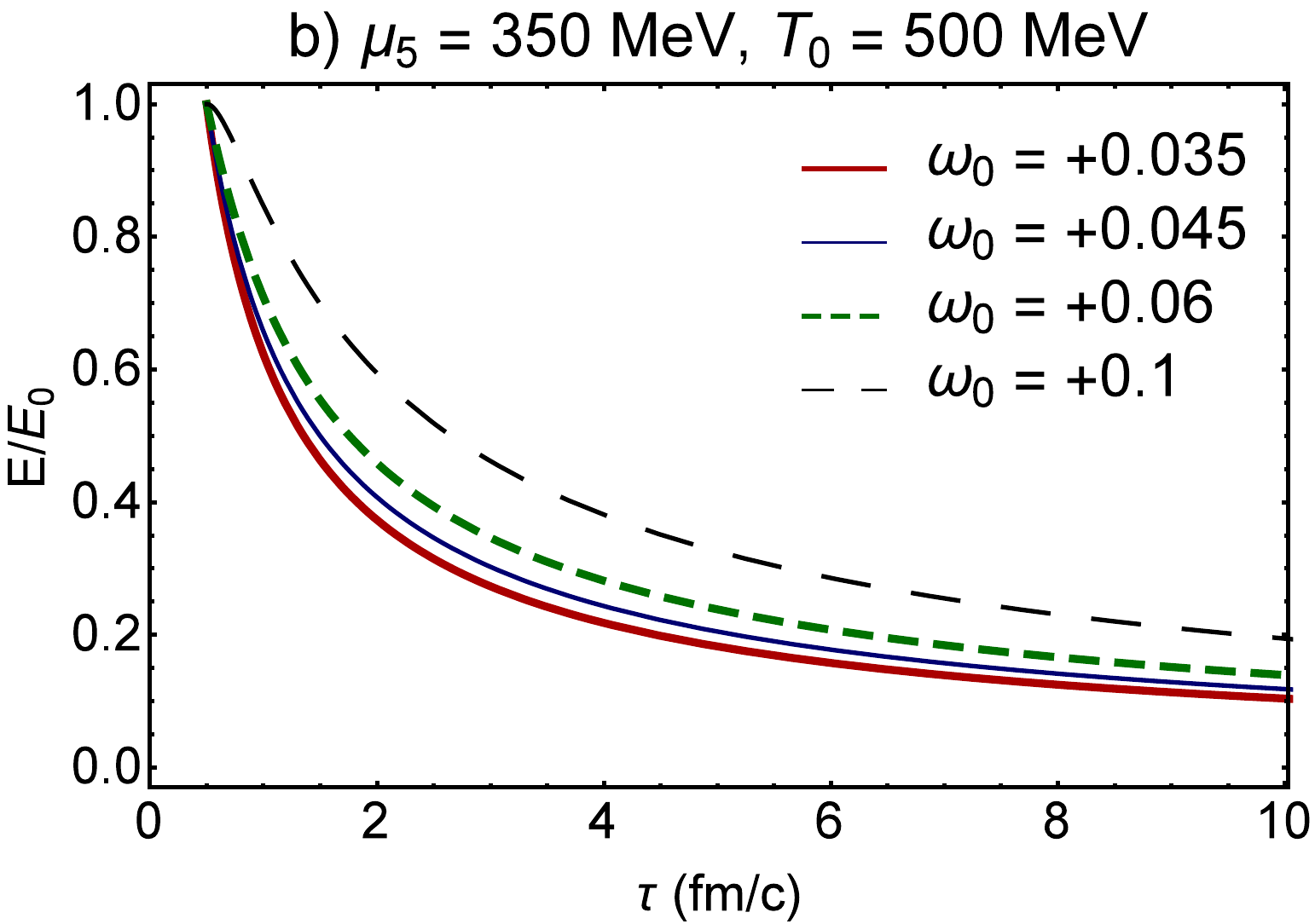}
                	\caption{(color online). (a) The $\tau$-dependence of $E/E_{0}$ is plotted for fixed $\mu_{5}=350$ MeV and different $\omega_{0}=-0.1$ (red thick solid curve), $\omega_{0}=-0.06$ (blue thin solid curve), $\omega_{0}=-0.045$ (green thick dashed curve), $\omega_{0}=-0.035$ (black thin dashed curve). For negative $\omega_{0}$, corresponding to parallel electric and magnetic fields, the electric field decays faster for larger values of $\omega_{0}$.  (b) The $\tau$-dependence of $E/E_{0}$ is plotted for fixed $\mu_{5}=350$ MeV and different $\omega_{0}=+0.035$ (red thick solid curve), $\omega_{0}=+0.045$ (blue thin solid curve), $\omega_{0}=+0.06$ (green thick dashed curve), $\omega_{0}=+0.1$ (black thin dashed curve). For positive $\omega_{0}$, corresponding to antiparallel electric and magnetic fields, the electric field decays faster for smaller values of $\omega_{0}$. }\label{fig-15}
                \end{figure*}
As explained at the beginning of this section, in the case of vanishing AH coefficient $\alpha_{E}$,\footnote{This assumption is equivalent with the assumption of vanishing $\alpha_{0}$.} we have to work with a constant value of $\kappa_{B}$. For simplicity, we consider only the case of vanishing electric and magnetic susceptibilities (see \cite{shokri-1} for the results corresponding to nonvanishing susceptibilities). We mainly focus on the evolution of $B,E$ and $T$. Two different aspects of the effect of $\mu_{5}$ (or equivalently $\kappa_{B}$) and $\omega_{0}$ on the $\tau$-dependence of $B,E$ and $T$ are scrutinized:
\begin{enumerate}
\item The effect of different constant $\mu_{5}$ and a fixed $\omega_{0}$.
\item The effect of different constant $\omega_{0}$ and a fixed $\mu_{5}$.
\end{enumerate}
In both cases, we arrive at the conclusion that the effect of $\omega_{0}$ and $\mu_{5}$ on the evolution of $B$ and $T$ can be neglected, while the $\tau$-dependence of the electric field is affected by different choices of $\omega_{0}$ and $\mu_{5}$. To show this, let us start by studying the effect of different constant $\mu_{5}$ and a fixed $\omega_{0}$ on the evolution of $B,E$ and $T$. In Fig. \ref{fig-10}, the $\tau$-dependence of $B/B_{0}$ is plotted for
\begin{eqnarray}\label{D6}
\lefteqn{
	\{\kappa,\tau_{0}, \beta_{0},\sigma_{0},\ell,\omega_{0}, \chi_{e},\chi_{m}\}}\nonumber\\
&=&\{3,0.5~\mbox{fm/c}, 0.1,17.1~\mbox{MeVc},+1,-0.045,0,0\},\nonumber\\
\end{eqnarray}
and $\mu_{5}=0,450$ MeV. The formal solution of $B/B_{0}$ is given in (\ref{E13}) in term of ${\cal{M}}$. To determine ${\cal{M}}$, we numerically solved the master equation (\ref{S16}) with $\sigma$ from (\ref{E18}) and the above set of free parameters (\ref{D6}) and $\mu_{5}=0,450$ MeV. As it is shown in Fig. \ref{fig-10}, the effect of $\mu_{5}$ on the evolution of the magnetic field is negligible. Same conclusion arises if we replace $\{\ell,\omega_{0}\}=\{+1,-0.045\}$ with $\{\ell,\omega_{0}\}=\{+1,-0.045\}$. Other choices of free parameters lead also to the same qualitative behavior.
\par
To determine the $\tau$-dependence of $E/E_{0}$, the formal solution of $E$ from (\ref{E13}) is used. Here, ${\cal{N}}$ is determined from (\ref{S17}), where, in particular, the previously determined ${\cal{M}}$ for the set (\ref{D6}) is used to find the $\tau$-dependence of $\frac{d{\cal{M}}}{du}$. This gives rise to the $\tau$-dependence of ${\cal{N}}$ and eventually to the evolution of $E/E_{0}$. The latter is  demonstrated in Fig. \ref{fig-11} for $\mu_{5}=0$ (red thick solid curve), $\mu_{5}=250$ MeV (black thin solid curve) and $\mu_{5}=450$ MeV (green dashed curve). According to these results, for $\omega_{0}<0$, the larger the axial chemical potential $\mu_{5}$ is, the faster $E$ decreases with $\tau$. Replacing $\{\ell,\omega_{0}\}=\{+1,-0.045\}$ in (\ref{D6}) with $\{\ell,\omega_{0}\}=\{-1,+0.045\}$, and following same steps as described before, we arrive at the $\tau$-dependence of $E$ for positive $\omega_{0}$. In contrast to the case of $\omega_{0}<0$, for $\omega_{0}>0$, $E/E_{0}$ decays slower for larger values of $\mu_{5}$.
\par
The opposite effect of positive (negative) and negative (positive) $\omega_{0}$ ($\ell$) on the evolution of the electric field is demonstrated in Fig. \ref{fig-12}. Here, red dotted and black curves correspond to $\{\ell,\omega_{0}\}=\{+1,-0.045\}$ and $\{\ell,\omega_{0}\}=\{-1,+0.045\}$, respectively. Other free parameters are given in (\ref{D6}). A comparison between Figs. \ref{fig-12}(a)  with $\mu_{5}=250$ MeV and \ref{fig-12}(b) with $\mu_{5}=450$ MeV shows that the difference between the effect of positive and negative $\omega_{0}$ on the decay rate of $E$ increases with increasing $\mu_{5}$. Same conclusions arise by using other sets of free parameters and positive as well as negative $\omega_{0}$.
\par
In Fig. \ref{fig-13},  the proper time dependence of $T/T_{0}$ is demonstrated for the set of parameters (\ref{D6}) and $\Sigma_{0}=B_{0}^{2}/\epsilon_{0}=10$. To do this, we used the formal solution of $T$ from (\ref{E13}) with $\exp\left({\cal{L}}/\kappa\right)$ given in (\ref{S18}).  For vanishing electric and magnetic susceptibilities, only the first three terms in (\ref{S18}) contribute to $\exp\left({\cal{L}}/\kappa\right)$. To determine them, previous results for ${\cal{M}}$ and ${\cal{N}}$ with free parameters (\ref{D6}) and $\mu=0,450$ MeV are used. One should bear in mind that in the case of $\alpha_{E}=0$, the CM conductivity $\kappa_{B}$ is constant and $\cos\delta=\ell=\pm 1$. According to the results demonstrated in Fig. \ref{fig-13}, the effect of axial chemical potential on the evolution of $T$ is negligible. Same conclusion arises if one replaces $\{\ell,\omega_{0}\}=\{+1,-0.045\}$ with $\{\ell,\omega_{0}\}=\{+1,-0.045\}$. Other choices of free parameters lead also to the same qualitative behavior.
\par
Let us finally study the effect of different $\omega_{0}$ and a fixed $\mu_{5}$ on the evolution of $B,E$ and $T$. To do this, we used the same method as described above, and determined the $\tau$-dependence of $B, E$ and $T$ for various $\omega_{0}=-0.035,-0.045,-0.06, -0.1$ and a fixed $\mu_{5}=350$ MeV. Other free parameters are given in (\ref{D6}). In Figs. \ref{fig-14}(a) and \ref{fig-14}(b) the proper time dependence of $B/B_{0}$ and $T/T_{0}$ are plotted for $\omega_{0}=-0.035$ and $\omega_{0}=-0.1$. The results demonstrated in these figures confirm our previous conclusion stating that different choices of $\omega_{0}$ do not affect the evolution of $B$ and $T$ significantly. The same conclusion arises for other sets of free parameters.
\par
As aforementioned, the evolution of the electric field is strongly affected by $\omega_{0}$ for a fixed $\mu_{5}$. In Fig. \ref{fig-15}(a), the $\tau$-dependence of $E/E_{0}$ is plotted for fixed $\mu_{5}=350$ MeV and $\omega_{0}=-0.1$ (red thick solid curve), $\omega_{0}=-0.06$ (blue thin solid curve), $\omega_{0}=-0.045$ (green thick dashed curve), $\omega_{0}=-0.035$ (black thin dashed curve). Other free parameters are given in (\ref{D6}). According to these results, the electric field decays faster for larger values of negative $\omega_{0}$. This is, however, in contrast to the effect of positive $\omega_{0}$ on the decay rate of the electric field. This is demonstrated in Fig. \ref{fig-15}(b), where $\mu_{5}=350$ MeV is fixed, and angular velocities are given by $\omega_{0}=+0.035$ (red thick solid curve), $\omega_{0}=+0.045$ (blue thin solid curve), $\omega_{0}=+0.06$ (green thick dashed curve) and $\omega_{0}=+0.1$ (black thin dashed curve). As it turns out, the electric field decays faster for smaller values of positive $\omega_{0}$.
\section{Concluding remarks}\label{sec7}
\setcounter{equation}{0}
Building on our prior results from \cite{shokri-1}, we explored, in the present paper, the physical features of the nonrotating and rotating solutions for the electric and magnetic fields $E^{\mu}$ and $B^{\mu}$ by extending the previously considered Lagrangian of the Maxwell theory with an additional ${\cal{CP}}$ violating Chern-Simons term $F^{\mu\nu}\tilde{F}_{\mu\nu}$, which is proportional to a pseudo-scalar axionlike field $\vartheta(x)$. Using this MCS Lagrangian, we arrived, in particular, at the corresponding equation of motion and energy-momentum tensor to the MCS theory. Combining the latter with the energy-momentum tensor of a nonviscous hydrodynamics, using the homogeneous and inhomogeneous MCS equations, and making the same assumptions as in \cite{shokri-1}, we arrived at an appropriate formulation for the nonideal transverse CSMHD. We emphasized that the specific feature of the current appearing in the inhomogeneous MCS equation is the presence of two nondissipative currents, the chiral magnetic and the anomalous Hall currents.  Denoting the CM and AH conductivities by $\kappa_{B}$ and $\kappa_{E}$, respectively, we showed that in a transverse CSMHD, these coefficients are the Lorentz boost transformed of the time and space derivatives of the $\vartheta$ field, $P_{0}=\partial_{0}\vartheta$ and $P_{3}=\partial_{3}\vartheta$. We were, in particular, interested in the effect of these anomalous currents on the evolution of electromagnetic and hydrodynamic fields.
\par
Following the same steps as in \cite{shokri-1}, we arrived at the constitutive equations of nonideal transverse CSMHD. Comparing these equations with the constitutive equations of transverse MHD, there appears additional terms proportional to $\kappa_{B}$ and $\kappa_{E}$ [see, in particular, (\ref{E3}), (\ref{E4}), (\ref{E10}) and (\ref{E11})]. Same inhomogeneous continuity equations as in \cite{shokri-1} with the generic form $\partial_{\mu}\left(fu^{\mu}\right)=fD\lambda$ and $f\in\{B,E,T^{\kappa}\}$\footnote{ Here, $\kappa=c_s^{-2}$ arises in the equation of state $\epsilon=\kappa p$.} as well as $\lambda\in \{{\cal{M}},{\cal{N}},{\cal{L}}\}$ characterize the nonideal transverse CSMHD. The formal solutions to these differential equations are presented in (\ref{E13}).
\par
In Sec. \ref{sec4}, we presented a number of results arising from the solution of the constitutive equations of CSMHD.  One of the most remarkable ones was that the relative angle between $\boldsymbol{E}$ and $\boldsymbol{B}$ is given in terms of the AH coefficient $\kappa_{E}$ and the electric conductivity of the fluid $\sigma$ through $\delta=\tan^{-1}\alpha_{E}$ with $\alpha_{E}=\kappa_{E}/\sigma$.
This result is consistent with our findings  for transverse MHD from \cite{shokri-1}, as for vanishing $\kappa_{E}$ and nonvanishing $\sigma$, $\tan\delta$ vanishes, and $\boldsymbol{E}$ and $\boldsymbol{B}$ fields become either parallel or antiparallel as in \cite{shokri-1}. Similar results were also found in \cite{ads}, using gauge/gravity duality.
The angle $\delta$ was then shown to be boost-invariant ($\eta$-independent). Its $\tau$-dependence, however, was given by the $\tau$-dependence of $\kappa_{E}$ from (\ref{S2}) and $\sigma$ from (\ref{E18}). We considered two cases of vanishing and nonvanishing AH coefficient, and determined separately the $\tau$-dependence of $\vartheta, B,E$ and $T$ for these cases. For the case of nonvanishing $\kappa_{E}$, we were able to determine analytical solutions for ${\cal{M}}$ and ${\cal{N}}$, which eventually led to the $\tau$-dependence of $B=|\boldsymbol{B}|$ and $E=|\boldsymbol{E}|$. For $\kappa_{E}=0$, in contrast, ${\cal{M}}$ is determined by two distinct differential equations (\ref{S15}) and (\ref{S16}), corresponding to nonrotating and rotating solutions for $B$. Once ${\cal{M}}$ is determined, ${\cal{N}}$ and ${\cal{L}}$ could also be determined. They eventually led to nonrotating and rotating solutions for $E$ and $T$ in the nonideal CSMHD. We noticed that for nonvanishing AH coefficient, (\ref{E14}) was the key relation, that, once combined with other constitutive equations, revealed analytical solutions for ${\cal{M}}$. For vanishing AH coefficient, this equation is trivially satisfied.
\par
As concerns the angles $\zeta$ and $\phi$, for nonvanishing $\kappa_{E}$, they are, as in \cite{shokri-1}, linear in $\eta$, and depend, in contrast to the $\kappa_{E}=0$ case, explicitly on $\tau$. Hence, although the relative angle of $\boldsymbol{E}$ and $\boldsymbol{B}$ fields is $\eta$-independent, the angles $\zeta$ and $\phi$ change uniformly with $\eta$. The corresponding angular velocity $\omega_{0}$ turned out to be constant. We showed that in $\kappa_{E}\neq 0$ case, $\omega_0$ is given in terms of the initial conditions for $E,B,\sigma$ and the CM as well as AH coefficients $\kappa_B$ and $\kappa_{E}$ at $\tau_0$. This is in contrast to \cite{shokri-1}, where for $\kappa_{E}=0$ the angular velocity $\omega_{0}$ was part of initial conditions.
\par
Using constitutive equations, we also showed that in the $\kappa_{E}\neq 0$ case, the $\tau$-dependence of the CM conductivity $\kappa_{B}$ can be completely determined in terms ${\cal{M}}, {\cal{N}}$ and their derivatives with respect to $u=\ln\frac{\tau}{\tau_{0}}$ as well as a number of free parameters $\{E_{0},B_{0}, \sigma_{0},\alpha_{0},\chi_{e},\chi_{m}\}$ [see (\ref{S14})]. For $\kappa_{E}=0$, however, $\kappa_{B}$ is constant and, similar to $\omega_{0}$, part of initial conditions. Bearing in mind that $\kappa_{B}$ is proportional to the axial chemical potential $\mu_{5}$, the evolution of $\kappa_{B}$ in the case of $\kappa_{E}\neq 0$ led automatically to the $\tau$-dependence of $\mu_{5}$. Starting with different initial values of $\mu_{5}$, we explored the evolution of $\mu_{5}$ in Sec. \ref{sec6}. We were, in particular, interested in the effect of $\omega_{0}$ on this evolution. We considered two different cases of positive and negative $\omega_{0}$ in Figs. \ref{fig-6}-\ref{fig-8} as well as Fig. \ref{fig-9}, and showed that for positive $\omega_{0}$,  under certain circumstances, $\mu_{5}$ increases during the evolution of the chiral fluid, whereas for negative $\omega_{0}$, it always decreases, and at some point even changes its sign from positive to negative [compare Fig. \ref{fig-6} with Fig. \ref{fig-9}(a)]. This sign flip in $\mu_{5}$ indicates a change in the direction of the CM current, which is proportional to $\kappa_{B}\propto \mu_{5}$. In Sec. \ref{sec6}, we quantified the relation between $\Delta \mu_{5}$ with a change in the axial number density $n_{5}$ in a more realistic model, where the pressure $p$ depends, apart from $T$, on $\mu_{5}$. We notice that the  (proper) time dependence of $\Delta \mu_{5}$ can also be brought into relation to $\Delta{\cal{H}}$, where ${\cal{H}}\equiv \frac{1}{V}\int d^{3}x \boldsymbol{A}\cdot\boldsymbol{B}$ is the magnetic helicity. Here,  $\boldsymbol{B}=\boldsymbol{\nabla}\times \boldsymbol{A}$. For $T\gg \mu_{5}$ the corresponding relation is given by \cite{manuel2015}
$$
\frac{d\mu_{5}}{dt}=-\Gamma_{f}\mu_{5}-\frac{c}{\pi\chi_{5}}\frac{d{\cal{H}}}{dt},
$$
where $\Gamma_{f}$ is the rate of helicity-flipping, $c=\sum_{f}q_{f}^{2}\frac{e^{2}}{2\pi}$ is defined before and  $\chi_{5}\equiv \frac{\partial n_{5}}{\partial\mu_{5}}$ is the chiral susceptibility of the medium. It would be interesting to further scrutinize the results arisen in Sec. \ref{sec6} for the $\tau$-dependence of $\mu_{5}$ with regard to the helicity flip in the QGP with a chirality imbalance. The corresponding backreaction is supposed to affect the lifetime of the magnetic field, because helical magnetic fields are apparently more long lived \cite{manuel2015, brandenburg2011}.
\par
As concerns the effect of different initial values of electric conductivities $\sigma$ on $\mu_{5}$, it turned out that for positive (negative) $\omega_{0}$, larger (smaller) values of $\sigma_{0}$ inhibit the rapid decay of $\kappa_{B}$ as well as $\mu_{5}$ [compare Figs. \ref{fig-7} with Fig. \ref{fig-9}(b)]. For a fixed $\sigma_{0}$, however, larger (smaller) values of positive (negative) $\omega_{0}$ enhance the decay rate of $\kappa_{B}$ as well as $\mu_{5}$ [compare Figs. \ref{fig-8} with Fig. \ref{fig-9}(c)]. Let us remind that positive and negative signs for $\omega_{0}$ is indirectly related to whether $\delta$ is from $\left(\frac{\pi}{2}, \frac{3\pi}{2}\right)$ or $\left(-\frac{\pi}{2},+\frac{\pi}{2}\right)$ intervals.
\par
The results presented in this paper can be extended in many ways. As aforementioned, the Bjorken flow is mainly characterized by a uniform longitudinal
expansion of an ideal relativistic fluid. Although it is able to describe the early time dynamics of the QGP created in HICs, various experimental results,
in particular, the transverse momentum of final hadrons signals a significant radial expansion apart from the longitudinal one. There are many attempts to overcome
this specific shortcoming of Bjorken flow, among others, the Gubser \cite{gubser2010} and $3+1$ dimensional self-similar flow \cite{csorgo2003}. In \cite{shokri-3}, we
present a generalization of these flows to relativistic ideal MHD. Extending the derivations in \cite{shokri-3} to nonideal MHD, the resulting model can be used
as a basis to a computation similar to that which is carried out in the present paper. In particular, the role of chiral vortical current can be explored in this setup,
as the vorticity vanishes in a $1+1$ dimensional setup.  Another open question is the inclusion of dissipative terms, both in the energy-momentum tensor and electromagnetic
currents, as the evolution of magnetic fields, in particular the primordial ones, is usually described by the system of nonrelativistic Maxwell and Navier-Stokes equations
\cite{boyarski2011}. Hydrodynamic dissipations modify the constitutive equations, and, in this way, the proper time dependence of the electric and magnetic fields may also be affected.
\par
The above results, in particular, the rotation of electric and magnetic fields, the evolution of the axionlike field $\vartheta$, and the $\tau$ dependence of the CM and AH conductivities, $\kappa_{B}$ and $\kappa_{E}$ may have important and not yet explored effects not only on various observables in HIC experiments, like the axial charge and photon production rates, but also on various transport properties of electrons in Weyl semimetals. A consistent hydrodynamical description of Weyl semimetals is recently presented in \cite{shovkovy2017}. In \cite{shovkovy2017,shovkovy2018-1}, it is shown that Chern-Simons contributions, including CME and AHE, strongly modify the dispersion relation of the collective modes in Weyl semimetals.
The role played by the Chern-Simons terms on the hydrodynamical flow of chiral electrons in a Weyl semimetal slab is studied in \cite{shovkovy2018-2}. It would be interesting to study the application of our results, mainly resulted from the assumption of a uniform and longitudinal expansion of the fluid, in the physics of Weyl semimetals, and to compare the corresponding findings with the results in \cite{shovkovy2017,shovkovy2018-1,shovkovy2018-2}.
\section{Acknowledgments}\label{sec8}
\setcounter{equation}{0}
\par\noindent
The authors thank S.M.A. Tabatabaee for useful discussions, and S. Kundu for introducing the references \cite{ads} to us. N.S. thanks M. Kargarian for enlightening discussions
on Anomalous Hall current in condensed matter physics. This work is supported by Sharif University of Technology's Office of Vice President for Research under Grant No: G960212/Sadooghi.  
\begin{appendix}
\section{Maxwell-Chern-Simons energy-momentum tensor}\label{appA}
\setcounter{equation}{0}
To derive the MCS energy-momentum tensor $F_{\mu\nu}$ from (\ref{N3}), let us start with
\begin{eqnarray}\label{appA1}
f_{\mu}\equiv F_{\mu\nu}J^{\nu},
\end{eqnarray}
with ${\cal{J}}^{\nu}=J^{\nu}-cP_{\mu}\tilde{F}^{\mu\nu}$, as defined in Sec. \ref{sec2}. Using $P_{\mu}=\partial_{\mu}\vartheta$ and the homogeneous Maxwell equation $\partial_{\mu}\tilde{F}^{\mu\nu}=0$, we arrive first at
\begin{eqnarray}\label{appA2}
\partial_{\mu}{\cal{F}}^{\mu\nu}=J^{\nu},
\end{eqnarray}
with ${\cal{F}}_{\mu\nu}=F_{\mu\nu}+c\vartheta\tilde{F}_{\mu\nu}$. Then, plugging (\ref{appA2}) into (\ref{appA1}), we obtain
\begin{eqnarray}\label{appA3}
f_{\mu}=F_{\mu\nu}\partial_{\rho}{\cal{F}}^{\rho\nu}.
\end{eqnarray}
Performing a number of straightforward algebraic manipulations, where, in particular, the homogeneous Maxwell equation in the form
\begin{eqnarray}\label{appA4}
\partial_{\rho}F_{\mu\nu}+\partial_{\nu}F_{\rho\mu}+\partial_{\mu}F_{\nu\rho}=0,
\end{eqnarray}
is used, we arrive at
\begin{eqnarray}\label{appA5}
f_{\mu}=\partial_{\rho}\left(F_{\mu\nu}{\cal{F}}^{\rho\nu}\right)+\frac{1}{2}(\partial_{\mu}F_{\nu\rho}){\cal{F}}^{\rho\nu}.
\end{eqnarray}
Plugging, at this stage, the definition of ${\cal{F}}_{\mu\nu}$ into the second term on the rhs of (\ref{appA5}), we get
\begin{eqnarray}\label{appA6}
f_{\mu}&=&\partial_{\rho}\left(F_{\mu\nu}{\cal{F}}^{\rho\nu}\right)+\frac{1}{2}\partial_{\mu}(F_{\nu\rho}{\cal{F}}^{\rho\nu})-\frac{1}{2}F_{\nu\rho}\partial_{\mu}F^{\rho\nu}\nonumber\\
&&-\frac{c}{2}F_{\nu\rho}P_{\mu}\tilde{F}^{\rho\nu}-\frac{c\vartheta}{2}F_{\nu\rho}\partial_{\mu}\tilde{F}^{\rho\nu}.
\end{eqnarray}
Using (\ref{appA4}), the third and last terms on the rhs of (\ref{appA6}) are given by
\begin{eqnarray}\label{appA7}
F_{\nu\rho}\partial_{\mu}F^{\rho\nu}&=&-\frac{1}{2}\partial_{\mu}\left(F_{\nu\rho}F^{\nu\rho}\right), \nonumber\\
c\vartheta F_{\nu\rho}\partial_{\mu}\tilde{F}^{\rho\nu}&=&-\frac{1}{2}\partial_{\mu}\left(c\vartheta\tilde{F}^{\alpha\beta}F_{\alpha\beta}\right)+\frac{c}{2}
P_{\mu}\tilde{F}^{\alpha\beta}F_{\alpha\beta}.\nonumber\\
\end{eqnarray}
Plugging these expressions into (\ref{appA6}), we arrive after some algebraic manipulations at
\begin{eqnarray}\label{appA8}
J^{\nu}F_{\nu\mu}&=&\partial_{\rho}\left({\cal{F}}^{\rho\nu}F_{\nu\mu}-\frac{1}{4}g^{\rho}_{~\mu}F_{\nu\sigma}{\cal{F}}^{\nu\sigma}\right)+\frac{c}{4}P_{\mu}F_{\nu\rho}\tilde{F}^{\nu\rho},\nonumber\\
\end{eqnarray}
where (\ref{appA1}) is used. The expression arising in the total derivative can be identified as the MCS energy-momentum tensor $T^{\mu\nu}_{\mbox{\tiny{MCS}}}$. We therefore have
\begin{eqnarray}\label{appA9}
\partial_{\mu}T^{\mu\nu}_{\mbox{\tiny{MCS}}}=J_{\mu}F^{\mu\nu}+\frac{c}{4}P^{\nu}F_{\mu\rho}\tilde{F}^{\mu\rho},
\end{eqnarray}
with
\begin{eqnarray}\label{appA10}
T^{\mu\nu}_{\mbox{\tiny{MCS}}}={\cal{F}}^{\mu\rho}F_{\rho}^{~\nu}+\frac{1}{4}g^{\mu\nu}
F_{\rho\sigma}{\cal{F}}^{\rho\sigma},
\end{eqnarray}
as claimed.
\section{$\boldsymbol{\tau}$ and $\boldsymbol{\eta}$ dependence of the longitudinal components of the electric and magnetic fields}\label{appB}
\setcounter{equation}{0}
As it is explicitly stated in Sec. \ref{sec2},  the longitudinal components of $E^{\mu}$ and $B^{\mu}$ vanish because of symmetry properties of the transverse MHD. Using, in particular, the definition of $B^{\mu}$ and $E^{\mu}$ in terms of $F^{\mu\nu}$ in the paragraph below (\ref{N7}), we have
\begin{eqnarray}\label{appB1}
B_{0}=-\sinh\eta F_{12},\qquad B_{z}=-\cosh\eta F_{12},
\end{eqnarray}
and
\begin{eqnarray}\label{appB2}
E_{0}=\sinh\eta F^{30},\qquad E_{z}=\cosh\eta F^{30}.
\end{eqnarray}
For $B_{0}=B_{z}=0$ and $E_{0}=E_{z}=0$, we have, in particular, $F_{12}=0$ and $F^{30}=0$.
In this Appendix, we first show that $F_{12}$ and $F^{30}$ do not evolve with $\tau$ and $\eta$, i.e.,
\begin{eqnarray}
\frac{\partial B_{i}}{\partial\tau}=\frac{\partial B_{i}}{\partial\eta}=0,\qquad i=0,z,\label{appB3}\\
\frac{\partial E_{i}}{\partial\tau}=\frac{\partial E_{i}}{\partial\eta}=0,\qquad i=0,z,\label{appB4}
\end{eqnarray}
as stated in (\ref{N12}). To prove (\ref{appB3}), let us start with the homogeneous Maxwell equation in the form (\ref{appA4}). For $\left(\mu,\nu,\rho\right)=\left(0,1,2\right)$ and $\left(\mu,\nu,\rho\right)=\left(3,1,2\right)$, we have
\begin{eqnarray}\label{appB5}
\partial_{2}F_{01}+\partial_{1}F_{20}+\partial_{0}F_{12}=0,
\end{eqnarray}
and
\begin{eqnarray}\label{appB6}
\partial_{2}F_{31}+\partial_{1}F_{23}+\partial_{3}F_{12}=0.
\end{eqnarray}
Because of the assumed translational invariance in the $x$-$y$ plane, all terms in (\ref{appB5}) and (\ref{appB6}) including $\partial_{1}$ and $\partial_{2}$ vanish. As concerns the remaining terms, $\partial_{0}F_{12}$ in (\ref{appB5}) and $\partial_{3}F_{12}$ in (\ref{appB6}), they are given by
\begin{eqnarray}\label{appB7}
\hspace{-0.5cm}\frac{\partial F_{12}}{\partial t}&=&\left(\cosh\eta\frac{\partial}{\partial\tau}-\frac{1}{\tau}\sinh\eta\frac{\partial}
{\partial\eta}\right)F_{12}=0,\nonumber\\
\hspace{-0.5cm}\frac{\partial F_{12}}{\partial z}&=&\left(-\sinh\eta\frac{\partial}{\partial\tau}+\frac{1}{\tau}\cosh\eta\frac{\partial}
{\partial\eta}\right)F_{12}=0.
\end{eqnarray}
Here, (\ref{N14}) is used. Combining there two relations, we first obtain
\begin{eqnarray}\label{appB8}
\frac{\partial F_{12}}{\partial \tau}=0,\qquad \frac{\partial F_{12}}{\partial\eta}=0.
\end{eqnarray}
Using, at this stage, (\ref{appB1}) and, in particular,  $F_{12}=0$, we finally arrive at
(\ref{appB3}).
\par
As concerns the $\tau$- and $\eta$-dependence of the longitudinal components of $E^{\mu}$, we start with the inhomogeneous MCS equation of motion from (\ref{N2}), with ${\cal{J}}^{\mu}$ from (\ref{N23}). For $\nu=0,3$, we have
\begin{eqnarray}\label{appB9}
\frac{\partial F^{30}}{\partial t}=-{\cal{J}}^{3},\qquad \frac{\partial F^{30}}{\partial z}={\cal{J}}^{0}.
\end{eqnarray}
Using (\ref{N23}) and $B_{i}=E_{i}=0, i=0,z$, we arrive at
\begin{eqnarray}\label{appB10}
{\cal{J}}^{3}=\chi_{e}\frac{\partial E^{3}}{\partial\tau},\qquad {\cal{J}}^{0}=\chi_{e}\frac{\partial E^{0}}{\partial\tau}.
\end{eqnarray}
Plugging (\ref{appB10}) into (\ref{appB9}), and using (\ref{appB2}) as well as the definitions of $\partial_t$ and $\partial_z$ from (\ref{N14}), we obtain
\begin{eqnarray}\label{appB11}
\hspace{-0.8cm}\left(\left(1+\chi_{e}\right)\cosh\eta\frac{\partial}{\partial\tau}-\frac{1}{\tau}\sinh\eta\frac{\partial}
{\partial\eta}\right)F^{30}&=&0,\nonumber\\
\hspace{-0.8cm}\left(-\left(1+\chi_{e}\right)\sinh\eta\frac{\partial}{\partial\tau}+\frac{1}{\tau}\cosh\eta\frac{\partial}
{\partial\eta}\right)F^{30}&=&0.	
\end{eqnarray}
Combining these two relations, we first obtain
\begin{eqnarray}\label{appB12}
\frac{\partial F^{30}}{\partial\tau}=0,\qquad \frac{\partial F^{30}}{\partial \eta}=0.
\end{eqnarray}
Using, at this stage, (\ref{appB2}) and, in particular, $F^{30}=0$, we finally arrive at (\ref{appB4}). A comparison with the proof of the same claims (\ref{appB3}) and (\ref{appB4}) in \cite{shokri-1}, where no ${\cal{CP}}$ violating term was considered in the Lagrangian density of the Maxwell theory, we observe that the additional terms in ${\cal{J}}^{\mu}$ proportional to $c$ have no effects on the evolution of the longitudinal components of $B^{\mu}$ and $E^{\mu}$.
\end{appendix}

\end{document}